\documentclass{article}

\usepackage{PRIMEarxiv}

\usepackage{amsmath}
\usepackage{caption}
\usepackage[colorlinks=true]{hyperref}
\usepackage{float}
\usepackage{subcaption}
\usepackage{multirow}
\usepackage{array}
\usepackage{soul}
\usepackage[capitalise]{cleveref}
\usepackage{graphicx,xcolor}
\usepackage[title]{appendix}

\definecolor{pink1}{RGB}{219, 48, 122}
\definecolor{asparagus}{rgb}{0.53, 0.66, 0.42}
\definecolor{amber}{rgb}{1.0, 0.35, 0.0}
\definecolor{applegreen}{rgb}{0.55, 0.71, 0.0}
\definecolor{nodiffusion}{rgb}{0.6, 0.0, 0.0}
\definecolor{diffusion}{rgb}{0.0 0.6, 0.0}
\definecolor{br}{rgb}{0.0, 0.0, 1.0}

\newcommand\Tstrut{\rule{0pt}{2.6ex}}         
\newcommand\Bstrut{\rule[-0.9ex]{0pt}{0pt}}   

\newcommand{\specialcell}[2][c]{%
  \begin{tabular}[#1]{@{}c@{}}#2\end{tabular}}

\title{Scalable computational algorithms for geo-spatial COVID-19 spread in high performance computing
}

\author{
   Sudhi P V \\
  Department of Civil and Environmental Engineering\\
  Carleton University\\
  Ottawa,Ontario\\
  Canada. \\
   \And
  Victorita Dolean \\
  Department of Mathematics and Statistics\\
  University of Strathclyde, Glasgow, Scotland and \\
  Laboratoire J.A. Dieudonn\'e, CNRS, Universit\'e C\^ote d'Azur \\
  Nice, France.
   \AND
   Pierre Jolivet \\
   Institut de Recherche en Informatique de Toulouse\\
   CNRS, ENSEEIHT\\
   Toulouse\\
   France.
   \And
   Brandon Robinson \\
   Department of Civil and Environmental Engineering\\
  Carleton University\\
  Ottawa,Ontario\\
  Canada. \\
   \And
   Jodi D. Edwards \\
   School of Epidemiology and Public Health\\
   University of Ottawa,\\
   University of Ottawa Heart Institute and \\
   ICES,
   Ottawa, ON, Canada. 
   \And
   Tetyana Kendzerska \\
   Department of Medicine, Faculty of Medicine, \\
   Division of Respirology, University of Ottawa,\\
   The Ottawa Hospital Research Institute and \\
   ICES\\
   Ottawa, ON, Canada.
   \And
   Abhijit Sarkar\\
   Department of Civil and Environmental Engineering\\
  Carleton University\\
  Ottawa,Ontario\\
  Canada.\\
}

\begin{document}
\maketitle

\begin{abstract}
A nonlinear partial differential equation (PDE) based compartmental model of COVID-19 provides a continuous trace of infection over space and time. Finer resolutions in the spatial discretization, the inclusion of additional model compartments and model stratifications based on clinically relevant categories contribute to an increase in the number of unknowns to the order of millions. We adopt a parallel scalable solver allowing faster solutions for these high fidelity models. The solver combines domain decomposition and algebraic multigrid preconditioners at multiple levels to achieve the desired strong and weak scalability. As a numerical illustration of this general methodology, a five-compartment susceptible-exposed-infected-recovered-deceased (SEIRD) model of COVID-19  is used to demonstrate the scalability and effectiveness of  the proposed solver for a large geographical domain (Southern Ontario). It is possible to predict the infections up to three months for a system size of 92 million (using 1780 processes) within 7 hours saving months of computational effort needed for the conventional solvers.
\end{abstract}


\section{Introduction}

After the emergence of the coronavirus disease 2019 (COVID-19), many policy decisions directly affecting personal gatherings, business operations, and healthcare utilization have been predicated on forecasted case counts and hospitalization statistics. Many such predictions use compartmental models based on ordinary differential equations (ODEs), which capture temporal variation for a population. Compartmental models derived from the susceptible-infected-removed (SIR) model, have been used extensively. These models often include additional compartments and stratifications based on age, comorbidity, sex etc., to account for complex disease dynamics \cite{Tuite,Robinsone052681}. These approaches are based on an aggregated population for a given geographical domain and are well suited for individual population centres and cities or for studying global trends in broader regions. Agent and network-based models \cite{agent,bertaglia2021hyperbolic} are also popular particularly for studying localized virus spread by employing micro-scale data concerning disease transmission and population structure, which are difficult to acquire for large geographical domains. However, when the geographical domain of interest is a province/state/region or a country, spatial variations in disease dynamics become critical for accurate predictions \cite{GIS,google_mobility}. Critically compartmental models based on partial differential equations (PDEs) \cite{rabies_spatial,seird_1,oden_seird,seird_2} capture the continuous spread of virus both in space and time, providing a more complete description of disease dynamics over a large geographical domain.  Their outputs indicate highly contagious zones and the evolution of disease dynamics among other clinically relevant information which can inform the decision makers about preventative measures and hospital preparedness.

The numerical solution of the problem involves spatial and temporal discretizations leading to nonlinear system of equations, typically with millions of unknowns. A parallel iterative solver with an appropriate preconditioner can be used to achieve faster solutions for the linearized system derived from the nonlinear system. Domain decomposition (DD) methods refer to approaches for solving linear (or nonlinear) systems arising from PDEs that rely on dividing the domain into smaller sub problems and concurrently iterating to find a converged solution. They are generally used as preconditioners to Krylov subspace based solvers because of the inherent parallelism and ability to adapt to complex problems providing faster convergence. Development of DD has parallelly evolved into two branches of overlapping and non-overlapping decompositions  \cite{victorita_book,dd_TFchan,dd_tarek,dd_toselli,dd_smith}. We utilize an overlapping Schwarz DD framework to develop efficient preconditioners for the nonlinear system. Newton-Krylov methods linearize the system using Newton's iteration and then employ a preconditioned Krylov solver for the linear system obtained at each step \cite{NK_shadid,Nk_knoll}. Other nonlinear preconditioning techniques available in the literature \cite{PC_nonlinear1,PC_nonlinear2,PC_nonlinear3} solve nonlinear problems in subdomains with modified preconditioners. These are mainly suitable when localized nonlinearity cannot be effectively handled by global linearization. In this paper we use Picard linearization for our nonlinear system followed by a DD-based iterative solver at each step for simplicity. We find our linearization and preconditioning strategy satisfactory for the current five-compartment model problem to demonstrate the importance and suitability of DD methods to compartmental models of COVID-19.

Now, we note below the compelling reasons for applying a DD-based solver for COVID modelling. 
\begin{itemize}
    \item Like their ODE-based counterparts, PDE-based compartmental models can be very fine grained, accounting for accurate dynamics among population (see \cref{ac}). This can further be extended to consider different age groups, socio-economic status, vaccination status, and co-morbidity (e.g., \cite{Robinsone052681}). We note that corresponding spatial and temporal data for these stratifications could be obtained from private healthcare databases (e.g. \cite{ices}). This highly complex coupled model can be solved efficiently using an iterative solver equipped with parallel preconditioners offered by domain decomposition methods. 
    
    \item The application of the above mentioned five-compartment model with finer spatial and temporal discretizations covering large geographical areas (including many public health units) involves system sizes above millions and the associated computational cost.
    
    \item The inclusion of uncertainty in the model (as in \cite{bertaglia2021hyperbolic,kinetic_uncertain}) by considering parameters as random variables or random fields leads to stochastic PDEs which can be solved using sampling (Monte Carlo or quadrature) methods or sampling-free stochastic Galerkin methods \cite{knio_uq}. Slow convergence of Monte Carlo or large stochastic dimensions can increase the number of deterministic sample evaluations which in turn increases the computational cost for sampling approaches. Even though DD-based solvers can be utilized for each deterministic evaluation, parallel overhead may reduce its efficiency. Sampling-free stochastic Galerkin methods demand solutions of a large linear/nonlinear system of equations for the stochastic PDEs. Depending on the stochastic discretization (number of input random variables/order of output expansion), the size of this linear system may grow exponentially which requires scalable parallel solvers built on DD-based methods (e.g. \cite{sarkar_stochasticdd}).
    
    \item In the Bayesian inference framework, inverse problems for the estimation of model parameters requires the forward model to be evaluated numerous times for the computation of the likelihood function \cite{oden_seird}. For a high resolution model with many compartments where a single forward evaluation takes hours, this computation could take months to complete. With a highly scalable and efficient solver developed here, this computational cost can be reduced to days or even hours.
    
    \item For time-dependant nonlinear system of PDEs such as compartmental model of COVID-19, different solution strategies can be adopted as: ($i$) discretize in time, linearize and adopt a linear preconditioner for the iterative solver for non-symmetric system, ($ii$) discretize in time and apply a nonlinear preconditioner \cite{PC_nonlinear1,PC_nonlinear2,PC_nonlinear3}, ($iii$) apply a parallel in time method \cite{paratime}. The first strategy is adopted in this paper for which  an efficient preconditioner for the GMRES iterative solver is described. In contrast to the conjugate gradient  (CG) method used for symmetric and positive definite systems, the convergence criteria of the GMRES algorithm is difficult to establish using the spectral information on the coefficient matrix. It is thus important to identify a problem-specific preconditioner that expedites the convergence  and improves scalabilities of the iterative solver.

\end{itemize}

To this end, the main contributions of the paper are as follows.
\begin{itemize}
	\item Development of a parallel scalable solver for the solution of complex compartmental models of COVID-19 for large geographical areas.
	\item Scalability studies of the one-level restricted additive Schwarz (RAS) and two-grid Schwarz preconditioner variants in terms of the iteration count and solution time.
    \item Development of an efficient solver using a two-grid restricted additive Schwarz (RAS) preconditioner for the nonlinear coupled PDEs: The adapted solver with an  algebraic multigrid preconditioning for the coarse problem reduces the execution time and improves scalabilities. The comprehensive numerical experiments demonstrate the superior performance of this two-grid RAS solver against the other variant of two-grid RAS preconditioner. 
    \item Numerical illustration of a SEIRD model with spatio-temporally varying infection rate parameters that capture the realistic trends of infection through the region. 
   \item Application of the two-grid RAS-based solver to a large geographical domain of Southern Ontario with over 92 million unknowns demonstrating the efficiency of the solver in a realistic setting. 
   \item Verification studies for one and two-dimensional compartmental models of COVID-19 using the method of manufactured solutions (MMS).
\end{itemize}

The paper is organized as follows. Section \ref{methodology} introduces the PDE-based compartmental model and formulation of weak form on which the solver is applied. Section \ref{preconditioner} briefly covers the basics of overlapping Schwarz method and coarse corrections. Section \ref{numerical} applies the different preconditioners and assesses their relative effectiveness. The selected approach is then applied to a realistic model of Southern Ontario. The weak form of the equations are given in Appendix \ref{aa}, the model validation is shown in Appendix \ref{ab}, and details on a more complex $22$-compartment PDE-based model are provided in Appendix \ref{ac}.

\section{Methodology}\label{methodology}
\subsection{Compartmental model}

A compartmental model consisting of susceptible-exposed-infected-recovered-deceased (SEIRD) states is considered. The susceptible compartment is the population density of individuals who are vulnerable to infection, but have not yet been exposed. This state acts as the feeding state for infection to spread. The current model does not consider the possibility of reinfection after recovery (as we will only consider a single wave of infection), and thus the susceptible compartment decreases monotonically until a steady-state is reached. The exposed and infected compartments consist of people who are the carriers of virus and who may infect others. We distinguish between these two compartments by defining exposed as cases that are asymptomatic or cases that end in recovery without being detected, whereas infected accounts for cases that are positively identified. The spread of infected people is thus considered to be lower due to quarantine/isolation measures after detection. The deceased compartment accounts for people who die of acute COVID-19, the recovered compartment accounts for all other possible outcomes associated with COVID-19 infection where the individual survives.

\begin{figure}[htbp!]
    \centering
\includegraphics[width=110mm]{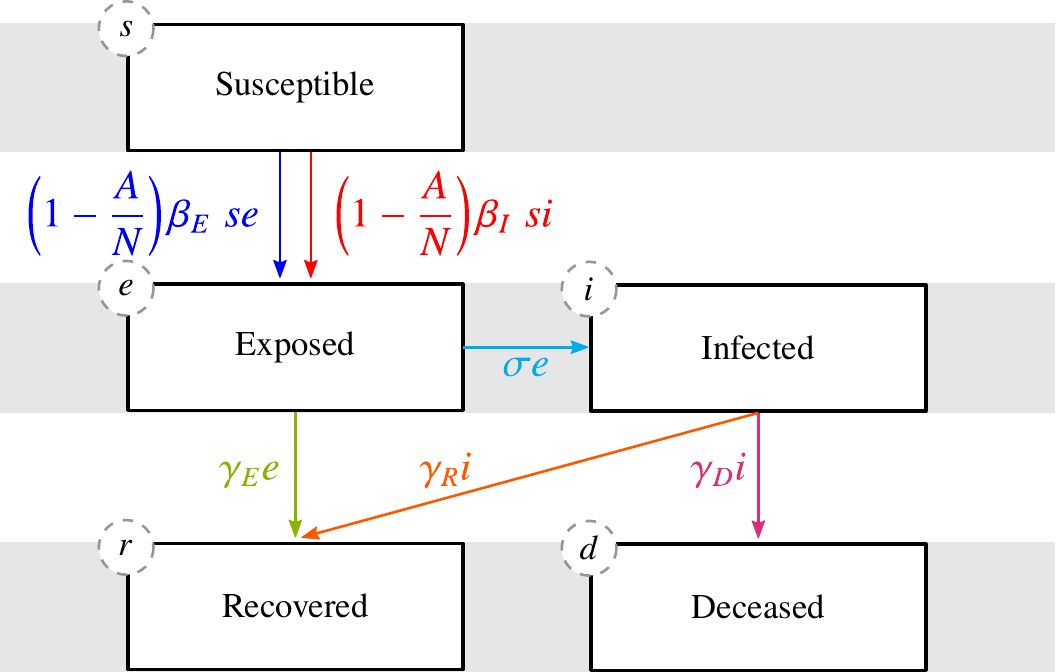}
    \caption{five-compartment model (adapted from \cite{seird_1,seird_2,oden_seird})}
    \label{fig:5state}
\end{figure}

The movement of population is generally based on assumptions on the behaviour of the host and their interaction with the surrounding environment \cite{pde_ecology}. A randomly mixing population at the micro-scale, can be modeled as a diffusion process in the macro-scale \cite{murray_1}. The current model assumes a heterogeneous population-dependent diffusion term for the movement of population in space. Along with diffusion in space, at each time step, the population changes states according to the interaction between the respective compartments. All model states are functions of space $\mathbf{x}$ and time $t$, however we generally omit these in our notations for brevity. The spatio-temporal evolution of the densities of the susceptible $s(\mathbf{x},t)$, exposed $e(\mathbf{x},t)$, infected $i (\mathbf{x},t)$, recovered $r(\mathbf{x},t)$, and deceased $d(\mathbf{x},t) $ compartments are described by the following coupled nonlinear partial differential equations (PDEs) \cite{seird_1,seird_2,oden_seird}.
\begin{align}
\label{Eq.seird1} \bigg.\partial _t s &= \nabla \cdot \left( N \bar{\nu}_S \nabla s \right) + \alpha N  -  { \left(1- \frac{A}{N}\right) \beta_I\; s i } - { \left(1- \frac{A}{N}\right) \beta_E \; s e } - \mu s  \\
\label{Eq.seird2} \bigg.\partial _t e &= \nabla \cdot \left(N \bar{\nu}_E \nabla e \right) +  { \left(1- \frac{A}{N}\right) \beta_I \; s i } + { \left(1- \frac{A}{N}\right) \beta_E \; s e } - {\sigma e} - {\gamma_E e} - \mu e  \\
\label{Eq.seird3} \bigg.\partial _t i &= \nabla \cdot \left(N \bar{\nu}_I \nabla i \right) + {\sigma e} - {\gamma_R i} -  {\gamma_D i} - \mu i  \\
\label{Eq.seird4}\bigg. \partial _t r &= \nabla \cdot \left(N \bar{\nu}_R \nabla r \right) + {\gamma_E e}  + {\gamma_R i} - \mu r  \\
\label{Eq.seird5}\bigg. \partial _t d &={\gamma_D i},
\end{align}

where $N$ is the total population density defined as 
\begin{equation}\label{Eq.N}
N(\mathbf{x},t) = s(\mathbf{x},t)  + e(\mathbf{x},t)  + i (\mathbf{x},t) + r(\mathbf{x},t) +d(\mathbf{x},t),
\end{equation}
and $\bar{\nu}_{S},\bar{\nu}_E,\bar{\nu}_I$ and $\bar{\nu}_R$ are the scalar diffusion coefficients dictating the spread of infection in space. \cref{fig:5state} shows the flow of individuals through compartments and the associated parameters. Note that the five-compartment model \cite{seird_1,seird_2,oden_seird} in  \cref{fig:5state} is a simplified version of the 22-compartment model shown in \cref{ac}. As an initial investigation, this five-compartment model is used to demonstrate the usefullness of the domain decomposition based solvers.

In the equations above: ($i$) susceptible individuals flow from $s$ to $e$ after being exposed to the disease by coming in contact with individuals in either the $e$ or $i$ compartments according to the rate parameters $\beta_E$, $\beta_I$, ($ii$) individuals in the exposed compartment either move to the infected compartment $i$ if they are detected according to the rate parameter $\sigma$, or if they go undetected, they remain in the $e$ compartment for the duration of their infectious period before proceeding to the recovered compartment at the rate $\gamma_E$, ($iii$) individuals in the $i$ compartment will proceed to one of either the recovered $r$ or deceased compartment $d$ according to rate parameters $\gamma_R$ and $\gamma_D$, respectively. The susceptible and exposed compartments contain a term $ (1- A/N)$, called the \textit{Allee effect}. For a given value of $A$, this term enforces higher transmission rates in locations of higher population density and conversely lower transmission rates for regions with lower population density. The terms containing $\alpha$ and $\mu$ are the population vital dynamics, representing the birth rate and death rates (excluding deaths caused by COVID-19). Given the time-scale, both are considered zero for the present study. 

In general, precise values of these parameters are not known but can be obtained by calibrating the model using data and prior clinical knowledge. When the domain is isolated (as is the case here), the total population over the whole domain remains constant in time, but the density may vary through space and time. 
Nonlinearities arise in these equations from the diffusion and the reaction terms. The diffusion coefficients $\bar{\nu}_{S},\bar{\nu}_E,\bar{\nu}_I, \bar{\nu}_R$ and infection rates $\beta_E, \beta_I$ control the spatial distribution of infection and their relative magnitudes make the process diffusion or reaction controlled. The nonlinear coupled PDEs are discretized using the finite element method, converting them to a system of linear algebraic equations as explained in next section.

\subsection{Weak form}
The states and reaction terms are represented as vectors and matrices as follows \cite{seird_2},

\begin{equation}\label{Eq.vec_component1}
\mathbf{u} = \begin{bmatrix}
     s\\e\\i\\r\\d
\end{bmatrix} \;\;\;
\boldsymbol{\nu}(\mathbf{u}) = N(\mathbf{x},t) \begin{bmatrix}
    \bar{\nu}_S & 0 & 0 & 0 & 0 \\
    0 & \bar{\nu}_E & 0 & 0 & 0 \\
    0 & 0 & \bar{\nu}_I & 0 & 0 \\
    0 & 0 & 0 & \bar{\nu}_R & 0 \\
    0 & 0 & 0 & 0 & 0 \\
    \end{bmatrix},
\end{equation}

\begingroup
\begin{equation}\label{Eq.vec_component2}
\medmuskip = 0.5mu
\setlength{\arraycolsep}{1pt}
 \mathbf{B(\mathbf{u})} =
{\begin{bmatrix}
   \alpha-\mu -{ (1- \frac{A}{N}) \beta_E  e } -  { (1- \frac{A}{N}) \beta_I i  } & \alpha & \alpha & \alpha & \alpha \\
    { (1- \frac{A}{N}) \beta_I\; i } & { (1- \frac{A}{N}) \beta_E s  } -\mu -{\sigma } -{\gamma_E } & 0 & 0 & 0 \\
    0 & {\sigma} & {-\gamma_R } -  {\gamma_D } - \mu  & 0 & 0 \\
     0 & {\gamma_E } & {\gamma_R } & -\mu & 0 \\
    0 & 0  & {\gamma_D } & 0 & 0 \\
\end{bmatrix}}.
\end{equation}

\endgroup

\normalsize
Hence the aforementioned SEIRD PDE system can be written concisely as \cite{seird_2},
\begin{equation}
    \partial_t \mathbf{u} - \nabla \cdot ( \boldsymbol{\nu} (\mathbf{u}) \nabla \mathbf{u} ) = \mathbf{B(\mathbf{u})}\mathbf{u}.
\end{equation}

A time discrete version of this PDE system for the state vector $\mathbf{u}$ can be written using a backward Euler/implicit method as,
\begin{equation}\label{Eq.timediscrete}
\mathbf{u}^{n+1} - \Delta t (\nabla \cdot ( \boldsymbol{\nu} (\mathbf{u}^{n+1}) \nabla \mathbf{u}^{n+1})) = \mathbf{u}^{n} + \Delta t \; (\mathbf{B}(\mathbf{u}^{n+1})\mathbf{u}^{n+1} ),
\end{equation}
where $\mathbf{u}^{n+1}$ is the solution at time step $n+1$ computed from the solution $\mathbf{u}^{n}(\mathbf{x}) \approx \mathbf{u}(\mathbf{x},t^n) $ for the previous time step. When $n = 0$, $ \mathbf{u}^{0}(\mathbf{x})$ is the initial condition to the problem. The weak form of \cref{Eq.timediscrete} becomes,

\begin{equation}\label{Eq.weakform}
(\mathbf{u}^{n+1}, \mathbf{v}) + \Delta t ( \boldsymbol{\nu} (\mathbf{u}^{n+1}) \nabla \mathbf{u}^{n+1}, \nabla \mathbf{v})  - \Delta t \; ( \mathbf{B}(\mathbf{u}^{n+1})\mathbf{u}^{n+1}, \mathbf{v}) \\
- \Delta t \int_{\Gamma_N} \boldsymbol{\nu} (\mathbf{u}^{n+1}) \nabla \mathbf{u}^{n+1} \cdot \hat{\mathbf{n}}  
= (\mathbf{u}^{n}, \mathbf{v}),
\end{equation}
where $(u,v) = \int_{\Omega} u\;v\;d\mathbf{x}$ , $\Omega$ is the spatial domain, $\Gamma_N$ denotes the boundary where Neumann boundary condition is specified and $\hat{\mathbf{n}}$ denotes the unit normal vector to the boundary.
A homogeneous Neumann boundary for the domain enforces a no-flux condition which prevents the migration of population across the boundary of the domain. This means complete isolation of the region, which is representative of restrictions on international/interprovincial travel. 

This nonlinear system must be solved using either Picard iterations or Newton's method inside the time discretizations. For simplicity, we chose to apply Picard iteration directly to \cref{Eq.weakform} as below.
\begin{equation}\label{Eq.weak_vec_implicit}
\begin{split}
(\mathbf{u}^{n+1,k+1}, \mathbf{v}) + \Delta t ( \textcolor{red} { \boldsymbol{\nu} (\mathbf{u}^{n+1,k}) } \nabla \mathbf{u}^{n+1,k+1}, \nabla \mathbf{v})  &- \Delta t \; 
( \textcolor{red} {  \mathbf{B}(\mathbf{u}^{n+1,k}) } \mathbf{u}^{n+1,k+1}, \mathbf{v}) 
= (\mathbf{u}^{n}, \mathbf{v}).
\end{split}
\end{equation}
Consider the Picard iteration at the time step $n+1$ with the current iteration number $k+1$. Initial guess for each iteration at $\mathbf{u}^{n+1,k=0}$ is chosen as the solution at the previous time step $\mathbf{u}^{n}$.
In the Picard iteration, the nonlinear coefficient terms are treated explicitly, using the solution from the previous iteration, resulting in a linear approximation at each iteration. Note that the colour-coded coefficient terms in \cref{Eq.weak_vec_implicit} depend only on the previous iterate $k$. Thus the solution to the nonlinear coupled PDE system is calculated at each Picard iteration inside each time step through \cref{Eq.weak_vec_implicit}.

For the current model, a fully coupled approach involves solving the system of five PDEs in \cref{Eq.weak_vec_implicit} together as a vector of compartments. However, this approach does not significantly improve the efficiency of the solver due to the weak coupling observed among compartments after applying Picard iteration.   Note that the introduction of additional compartments, model stratifications, the inclusion of uncertainty etc., can affect the system properties. In such cases, it may be necessary to adopt a fully coupled solution approach  and use Newton's method.
The detailed equations involving the formulation for each compartment is shown in Appendix \ref{aa}. We use triangular elements with linear interppolation functions for discretization. The linear system assembled from the weak form can be solved using Krylov subspace solvers with appropriate preconditioners inside each Picard iteration. The  Picard iteration is assumed to be converged  when the error, $\epsilon$ reaches a specified tolerance defined as,

\begin{align}\label{Eq:picardtol}
    \epsilon = \frac{\parallel \mathbf{u}^{n+1,k+1} _h - \mathbf{u}^{n+1,k} _h \parallel _2 }{\parallel \mathbf{u}^{n} _h \parallel _2},
\end{align}
where $\mathbf{u} _h$ represents the discretized solution vector.
The next section introduces domain decomposition concepts and the associated preconditioners to be applied to the above model.

\section{Overlapping Schwarz method and two-grid preconditioners}\label{preconditioner}

The domain decomposition method is a \textit{divide and conquer} algorithm that provides fast solutions to computational models involving PDEs \cite{dd_TFchan}. Independent treatment of each part of a complex domain provides a naturally parallel formulation for multiphysics and heterogeneous domains. Generally, these methods are applied as preconditioners to Krylov solvers at the discrete level.

\begin{figure}[H]
      \centering
        \includegraphics[width=0.5\textwidth]{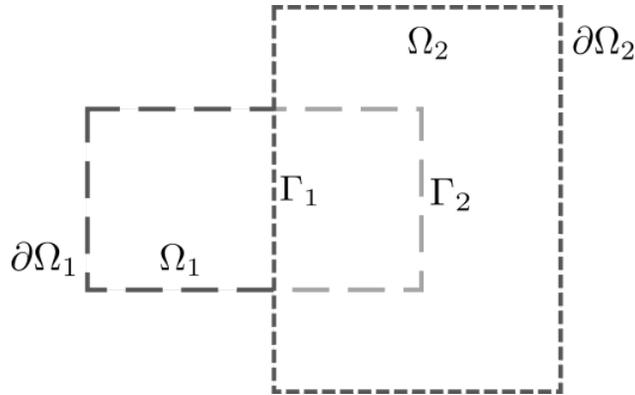} 
        \caption{Overlapping domain decomposition}\label{fig:overlap}
\end{figure}

The two main branches of DD methods, namely overlapping and non-overlapping, are fundamentally distinguished based on whether or not the adjacent subdomains overlap one another. Consider the domain $\Omega$, illustrated in Figure \ref{fig:overlap} as the union of two overlapping subdomains, $\Omega = \Omega_1 \cup \Omega_2$. The subdomain boundaries are $\partial \Omega_1$ and $\partial \Omega_2$ and  the artificial boundary created inside each subdomains are $\Gamma_1$ and $\Gamma_2$. The overlapping DD method considers subdomains on which the solution is sought independently using artificial boundary values from adjacent subdomain. The overlapped regions are then updated by combining the solutions from each subdomain and the process is repeated to reach convergence of the solution. The method can be sequential or parallel. We present here briefly the iterative version of an overlapping method for a decomposition into two subdomains. 

Consider the Poisson problem of finding the solution $u$ over the domain $\Omega$ with homogeneous boundaries. This problem can be split into independent sub-problems in each subdomain as below \cite{dd_smith,victorita_book,dd_toselli,dd_tarek},
\begin{equation}
\label{Eq.Parallel Schwarz}
\begin{split}
    - \Delta u_i^{n+1} &= f \quad \rm{in} \quad \Omega_i \\
     u_i^{n+1} &= 0 \quad \rm{on} \quad \partial \Omega_i \cap \partial \Omega \\
     u_i^{n+1}&= u^{n}_{3-i} \quad \rm{on} \quad  \Gamma_{3-i}.
\end{split}
\end{equation}

In the parallel case, finding the solution at the iteration $n+1$ on the artificial boundary of the $i^{th}$ subdomain $u_{i=1,2}$ involves the previous solution iterate from the adjacent subdomains which permits independent solve. A sequential version on the other hand alternately solves the subdomains $i=1,2$ using the updated solution at the current step. The implementation of overlapping methods are more straightforward than their non-overlapping counterparts, wherein it is necessary to solve the combined interface problem before tackling the interior nodes of each subdomain. Also, the definition of interface points can be quite involved especially in higher dimensions where cross points can appear. Overlapping methods only require consecutive solutions of the original problem in smaller subdomains and the exchange of artificial boundary solutions to neighbouring subdomains. While Schwarz methods can be used as solvers or as preconditioners for accelerated convergence of Krylov solvers, they are seldom used as solvers due to their slow convergence compared to Krylov solvers. Preconditioners are operators applied to coefficient matrix transforming them to have favourable properties of convergence for iterative solvers \cite{precondioner_benzi,preconditioning_wathen}. This transformation, in general alters the spectral properties and conditions the coefficient matrix. 
We briefly discuss two main types of preconditioners used in the literature in a discrete framework \cite{victorita_book}.

The additive Schwarz method (ASM) relies on local solutions in subdomains which are then assembled globally as \cite{victorita_book},
    \begin{align}
        \mathbf{M}^{-1}_{\rm{ASM}} = \sum_{i=1}^N \mathbf{R}_i^T (\mathbf{R}_i \mathbf{A} \mathbf{R}_i^T)^{-1} \mathbf{R}_i,
         \label{Eq.ASM}
    \end{align}

where $\mathbf{A}$ represents the linearized coefficient matrix assembled inside each Picard iteration for each time step, $\mathbf{R}_i$ represents the restriction operator transferring the solution vector on global mesh to the local subdomain level, and $\mathbf{R}_i^T$ the extension matrix reversing the operation. The ASM exchanges information between subdomains without taking into account the redundancies in the overlap. For this reason, this can only be used as a preconditioner, since its iterative counterpart doesn't converge to the true solution. It can be noted that the preconditioner formed is symmetric.
The restricted additive Schwarz (RAS) preconditioner is defined as \cite{victorita_book},
\begin{align}
        \mathbf{M}^{-1}_{\rm{RAS}} = \sum_{i=1}^N \mathbf{R}_i^T \mathbf{D}_i (\mathbf{R}_i \mathbf{A} \mathbf{R}_i^T)^{-1} \mathbf{R}_i,
        \label{Eq.RAS}
    \end{align}
where $\mathbf{D}_i$ are the Boolean square matrices called partition of unity matrices such that $\mathbf{I}_d = \sum_{i=1}^N \mathbf{R}_i^T \mathbf{D}_i \mathbf{R}_i $.
The partition of unity matrices scale the residuals so that consistent contributions from each subdomain is only added. These preconditioners are not assembled explicitly, but a series of steps replicating their action are applied, i.e., matrix-vector computations and local linear solves. The global residual computed is distributed to the local subdomain and a local Dirichlet problem is solved. This solution is combined using the partition of unity matrices for all subdomains. Both RAS and ASM solve for the increment/correction to the solution for all subdomains before combining them appropriately. ASM and RAS differ only in the way these corrections are combined. We use RAS-based preconditioners since it provides faster convergence than its counterpart.

\subsection{Coarse corrections}

\begin{figure}[H]
    \centering
    \begin{subfigure}[b]{0.3\textwidth}
      \centering
        \includegraphics[width=\textwidth]{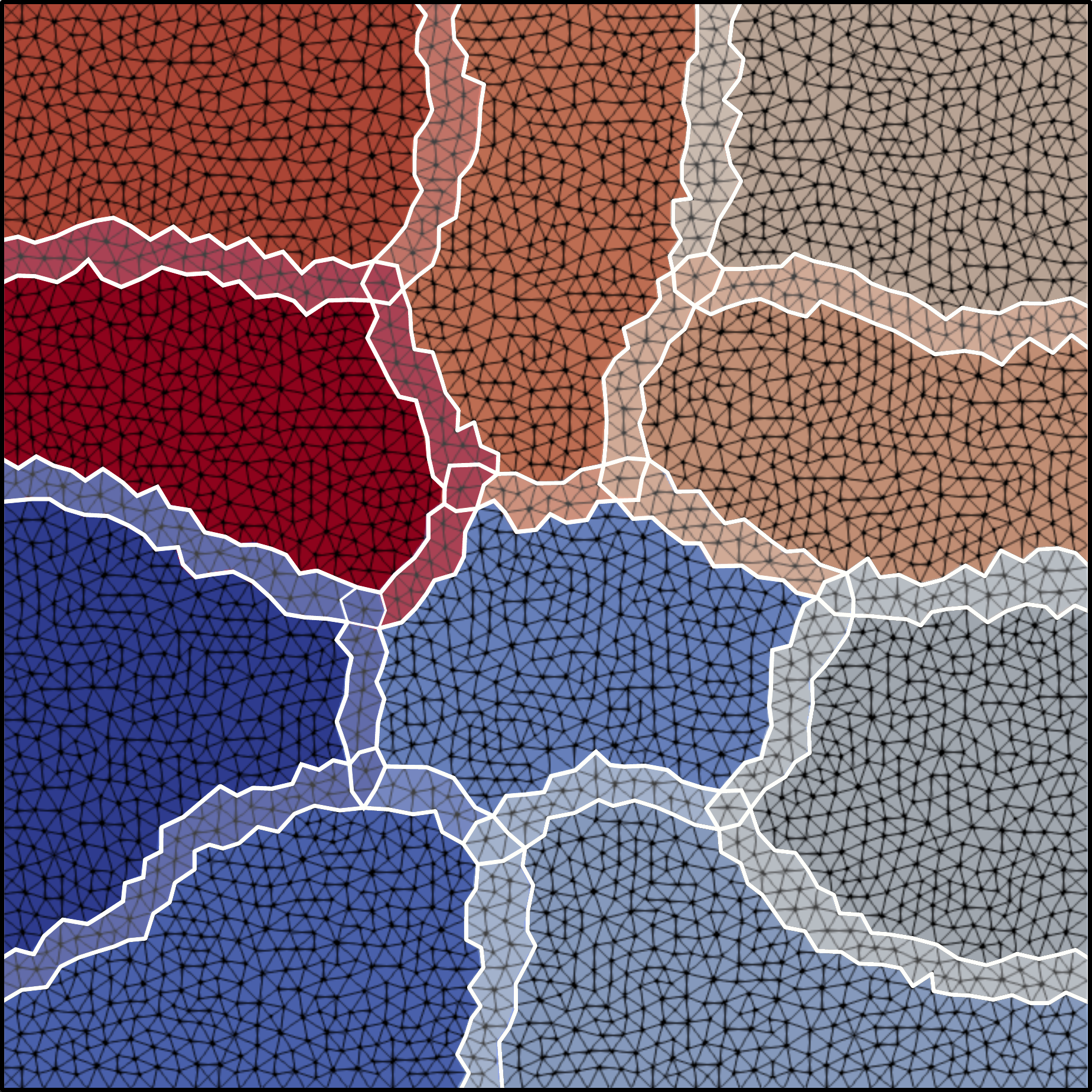} 
        \caption{Fine grid with $10$ subdomains}
    \end{subfigure}
    \qquad
    \begin{subfigure}[b]{0.3\textwidth}
      		 \centering
        \includegraphics[width=\textwidth]{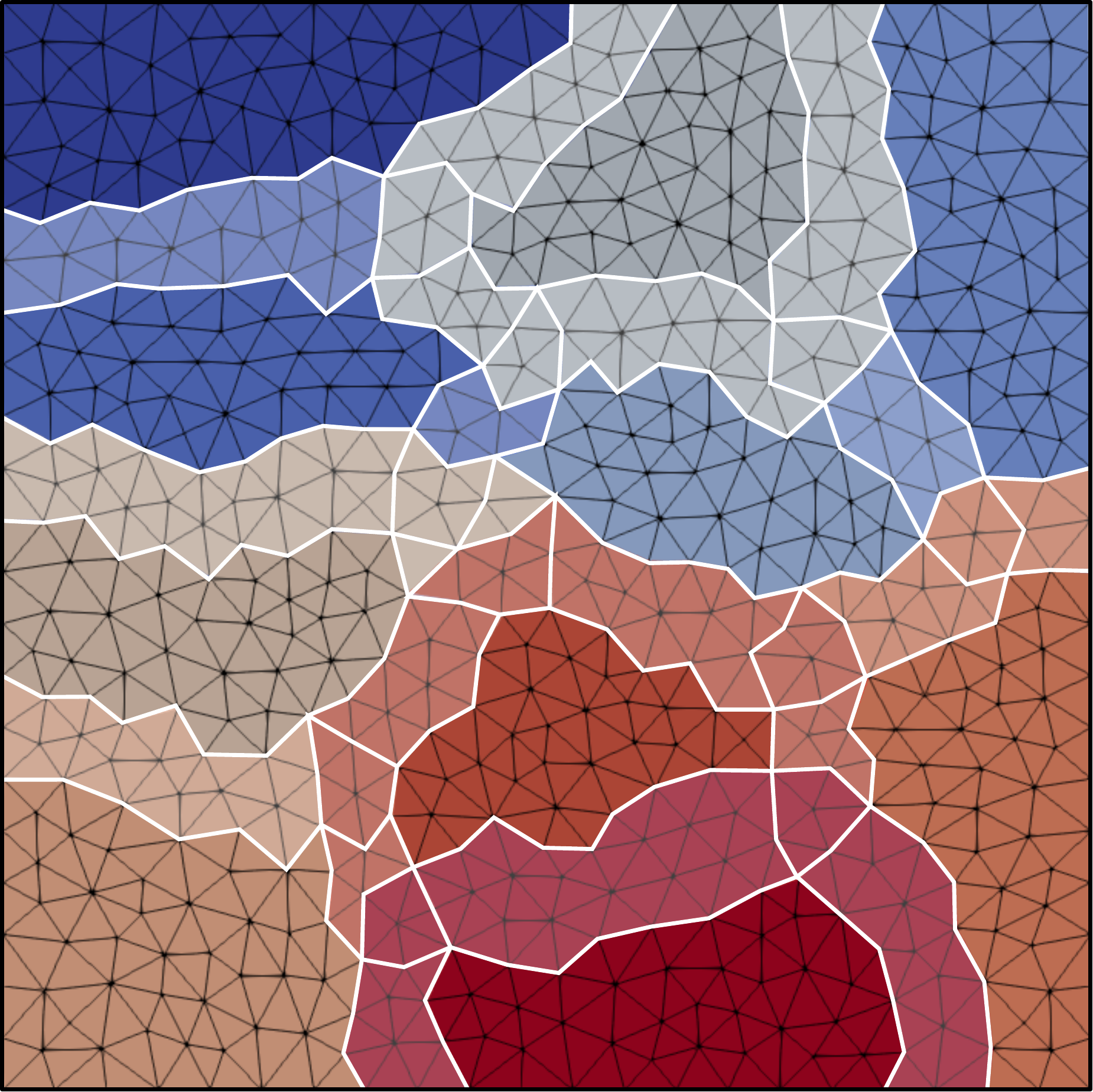} 
        \caption{Coarse grid with $10$ subdomains}
    \end{subfigure}   
  \caption{Finite element mesh with overlapping subdomains (overlap highlighted in white)}\label{fig:coarse_dd}
\end{figure}

One-level methods only communicate and exchange information with adjacent subdomains. This strategy is effective to reduce the high frequency component of error. When the number of subdomains is large, one-level methods converge slowly due to significant low-frequency components of error. The efficient global exchange of information among subdomains can be used to reduce the low-frequency component of error which enhance scalability. This is achieved using two-grid methods through coarse corrections \cite{victorita_book}. 

A coarse space correction is constructed to remove the low-frequency component of the error due to small eigenvalues in the preconditioned one-level matrix. Intuitively, these components represent constant functions for a Poisson problem or rigid body modes for elasticity \cite{victorita_book}. For a complex problem, where simple representation of the coarse space components is not available, a simple solution is to use a coarser triangulation from which the fine grid is constructed by refinement leading to a grid coarse space. In what follows, we will explain how the coarse correction matrix is built for a grid coarse space.

\cref{fig:coarse_dd} shows a square domain with overlapping subdomains and underlying finite element mesh. We construct a rectangular matrix $\mathbf{Z}$ of size $n \times n_c$ which is an interpolation operator from coarse to fine grid, where $n$ is the total number of degrees of freedom and $n_c$ is the number of coarse degrees of freedom. A coarse matrix $\mathbf{A}_c = \mathbf{R}_{0} \mathbf{A} \mathbf{R}_{0}^T$, is constructed using the Galerkin projection where $\mathbf{R}_{0} = \mathbf{Z}^{T}$ or by directly assembling the given PDE in the coarse grid. An additive coarse correction constructed from this coarse grid is applied to the one-level preconditioner in \cref{Eq.RAS} as \cite{victorita_book},
\begin{equation}
\mathbf{M}_2^{-1} = \mathbf{M}_{\rm{RAS}}^{-1}  + \mathbf{Q},
\end{equation}
where $\mathbf{Q} = \mathbf{R}_{0}^T \mathbf{A}_{c} ^{-1} \mathbf{R}_{0}$ is the global coarse correction. Other variants such as deflated and hybrid forms of corrections can also be applied to obtain favourable properties \cite{deflation_comparison}. Multiplicative corrections update the residual in between various levels providing better convergence rate than additive corrections \cite{dd_smith}. 

We utilize the multilevel parallel implementational architechture from PETSc \cite{petsc} (more implementational details in \cref{subsec.PC}) to construct a two-grid preconditioner with multiplicative corrections. A three-step correction is applied with a one-level RAS preconditioner as pre-smoother, post-smoother, and a coarse grid correction in between them. A combination of these three corrections applied at different levels (fine-coarse-fine) can be combined to construct the two-grid preconditioner as below.

Consider three preconditioners $\bf{P_1, P_2, P_3}$ applied to the system as pre-smoother, coarse correction, and post-smoother respectively as \cite{deflation_comparison},

\begin{align} \label{Eq.presmoother}
    \mathbf{u}^{i+ 1/3} &= \mathbf{u}^{i} + \mathbf{P_1} ( \mathbf{f} - \mathbf{A} \mathbf{u}^{i}) \\ \label{Eq.coarsecorrection}
    \mathbf{u}^{i+ 2/3} &= \mathbf{u}^{i+1/3} + \mathbf{P_2} ( \mathbf{f} - \mathbf{A} \mathbf{u}^{i+1/3}) \\ \label{Eq.postsmoother}
    \mathbf{u}^{i+ 1} &= \mathbf{u}^{i+2/3} + \mathbf{P_3} ( \mathbf{f} - \mathbf{A} \mathbf{u}^{i+2/3}).
\end{align}
Substituting \cref{Eq.presmoother} into \cref{Eq.coarsecorrection} and further \cref{Eq.coarsecorrection} into \cref{Eq.postsmoother} gives us,

\begin{align}
    \mathbf{u}^{i+ 2/3} &= \mathbf{u}^{i} + \underbrace{(\mathbf{P_1} + \mathbf{P_2} - \mathbf{P_2} \mathbf{A} \mathbf{P_1} )}_{\mathbf{P_4}} ( \mathbf{f} - \mathbf{A} \mathbf{u}^{i}) \\
    \mathbf{u}^{i+ 1} &= \mathbf{u}^{i} + \underbrace{(\mathbf{P_4} + \mathbf{P_3} - \mathbf{P_3} \mathbf{A} \mathbf{P_4} )}_{\mathbf{P_5}} ( \mathbf{f} - \mathbf{A} \mathbf{u}^{i}),
\end{align}
where $\mathbf{P_5}$ is our desired preconditioner expanded as,

\begin{equation}
    \mathbf{P_5} = \mathbf{P_1} + \mathbf{P_2} + \mathbf{P_3} - \mathbf{P_2} \mathbf{A} \mathbf{P_1} - \mathbf{P_3} \mathbf{A} \mathbf{P_1} - \mathbf{P_3} \mathbf{A} \mathbf{P_2} + \mathbf{P_3} \mathbf{A} \mathbf{P_3} \mathbf{A} \mathbf{P_1}.
\end{equation}
Using $\mathbf{P_1} = \mathbf{P_3} = \mathbf{M}_{\rm{RAS}} ^{-1}$ and $\mathbf{P_2} = \mathbf{Q} $, the final two-grid preconditioner can be written as,

\begin{align} \label{Eq.RAS_2L}
\mathbf{M}_2^{-1} = \mathbf{M}_{\rm{RAS}}^{-1} \mathbf{P} + \mathbf{Q}\mathbf{P}\mathbf{Q}^{-1} \mathbf{M}_{\rm{RAS}}^{-1} + \mathbf{Q} - \mathbf{M}_{\rm{RAS}}^{-1} \mathbf{P} \mathbf{A} \mathbf{M}_{\rm{RAS}}^{-1},
\end{align}
where $\mathbf{P} = (\mathbf{I} - \mathbf{AQ})$ is the projection matrix.
The residual is recalculated at each level using an updated solution which results in multiplicative corrections. Resriction and interpolation operators are $\mathbf{R}_0$ and $\mathbf{R}_0 ^{T}$ respectively as defined earlier.

For the two-grid preconditioner above, it is evident that the coarse problem has the same structure as the original problem and hence it can be constructed just by using an interpolation operator from the original system matrix. It also couples all the subdomains enabling global error propagation. However, for very large meshes, the system size grows rendering the solution of the coarse problem computationally expensive using a direct solver. In these cases, it is useful to adopt a preconditioner to iteratively solve the coarse problem $\mathbf{A}_{c}^{-1}$ in the coarse correction $\mathbf{Q}$ effectively to reduce the execution time and memory requirement.
To this end, we adopt two choices: $(i)$ a one-level RAS preconditioner as in \cref{Eq.RAS}, $(ii)$ an algebraic multigrid (AMG) V-cycle preconditioner \cite{mg_briggs}. Through numerical experiments, we demonstrate that the AMG preconditioner is more effective than the one-level RAS preconditioner. Next, we briefly explain the multigrid method relevant to the problem. 

Multigrid methods offer hierarchy of grids (geometric or algebraic) on which errors are iteratively reduced \cite{mg_briggs,mg_vassilevski,book_strang_CSE,mg_trottenberg}. The error in the iterative solution is decomposed into geometrically oscillatory and non-oscillatory parts. By utilizing a simple and cheap solver/smoother, highly oscillatory components of the errors are removed through a smoothing/relaxation step. The remaining non-oscillatory errors in the fine grid are corrected using a coarse grid. The advantage lies in the fact that geometrically smooth errors in fine grid become oscillatory in coarse grid which are then removed easily. After solving for the error in coarse grid they are interpolated back to fine grid to correct the solution \cite{book_strang_CSE}. Following the same steps repeatedly in cycles on these grids reduces the errors to required precision. Algebraic multigrid works on the same principle but does not depend on geometric information on the grid for error reduction \cite{mg_briggs}. This is useful for complex domains with unstructured meshes where construction of coarse grid is difficult. In AMG, each entries in the system matrix are analyzed to find the strength of dependence/strength of influence among elements which allows construction of coarse levels \cite{multigrid_xu,multigrid_Falgout}. Using such multi-level preconditioner for coarse grid solvers improves the scalability of the two-grid RAS solver of the original system.

\subsection{Preconditioners}\label{subsec.PC}

The linearized system at each  Picard iteration is solved using the GMRES iterative solver and associated preconditioners. Next we point out the different variants of one-level and two-grid preconditioners and their notations used in this work.

\begin{itemize}  
    \item One-level RAS: one-level RAS preconditioner as in \cref{Eq.RAS}.
    \item Two-grid RAS: two-grid RAS preconditioner with a coarse problem solved by the direct solver (LU factorization). 
    \item Two-grid RAS - V2: two-grid RAS  preconditioner with a coarse problem solved using GMRES iterative solver equipped with RAS preconditioner.
    \item Two-grid RAS - V3: two-grid RAS preconditioner with a coarse problem solved using GMRES with an AMG preconditioner. 
\end{itemize}

Note that the application of a two-grid preconditioner for the coarse solver in the two-grid preconditioners is  avoided due to  the complexity of constructing  coarser levels and associated interpolation operators between them. The time taken for construction of preconditioners in this case may not be able to balance the reduction in solution time achieved. In contrast, the two-grid RAS - V3 efficiently handles this by avoiding the use of grids in construction of the coarse level preconditioner as evident in numerical experiments later.

All numerical experiments are carried out in FreeFEM \cite{FreeFEM} integrated with PETSc \cite{petsc}. The two-grid variants of preconditioners differ only in the coarse solver. The preconditioner architecture is implemented using a single V-cycle multigrid framework from PETSc. Two-grid RAS-V3 makes use of HYPRE \cite{hypre} available through PETSc which runs BoomerAMG \cite{boomeramg} as preconditioner for coarse solve $A_{c}^{-1}$. This multilevel coarse solver adopts a  V-cycle with Jacobi smoother and Gaussian elimination for coarse correction \cite{mg_briggs,petsc}. This is specifically chosen to address the convergence bottleneck for high resolution models. A minimum overlap is used for RAS algorithms in all cases as depicted in \cref{fig:coarse_dd}. All preconditioners are used within the GMRES solver with the right preconditioning since it calculates the true residual norm in contrast to the preconditioned residual norm as in the left preconditioned systems. This permits a direct comparison of residuals among different methods \cite{preconditioning_wathen,petsc}. Stopping criteria for the Krylov solver (GMRES) follow PETSc implementations \cite{petsc} which are based on (a) the absolute residual norm, $atol$ (set as $10^{-50}$), (b) the decrement of the residual $l_2$ norm relative to the $l_2$ norm of the right-hand side, $rtol$ (set as $10^{-5}$), and (c) the relative increment in residual, $dtol$ (set as $10^{5}$). The convergence and divergence in any iteration $j$ is established respectively as: 
\begin{align}
\parallel r_{j} \parallel_2 &< max\;(rtol \;\times {\parallel b \parallel_2},\; atol) \\
\parallel r_{j} \parallel_2 & >  max\;(dtol \;\times {\parallel b \parallel_2}) 
\end{align}
where $r_{j}$ denotes the residual at the $j^{th}$ iteration of the GMRES solver and the right-hand  vector $b$. Apart from this, maximum number of outer and inner (coarse) Krylov iterations for two-grid solvers are restricted to $200$, $100$ respectively. In cases where an inner GMRES solver is used, the outer solver is modified to use the Flexible GMRES (FGMRES) algorithm \cite{FGMRES} which permits changes in the preconditioning at every step.
A detailed comparison of preconditioned Krylov methods for nonsymmetric systems relevant to this study can be found in \cite{PC_compare}. 

\section{Numerical Experiments}\label{numerical}

This section applies the preconditioners mentioned in the previous section to solve the linearized system in \cref{Eq.weak_vec_implicit} at each Picard iteration for all time steps. With five compartments, the total number of unknowns  becomes five times the number of degrees of freedom for the finite element mesh. The coarse grid is nested inside the fine grid, with a splitting ratio of two for fine-to-coarse grid. One-level and two-grid preconditioners are compared using their average number of Picard, GMRES iterations and total time-to-solution for a square domain. Numerical and parallel scalabilities for each case are studied to select the most suitable preconditioner. 
This preconditioner is then applied to a large geographical domain of Southern Ontario to demonstrate the scalibility of the solver in a realistic setting.  The parameter values used in each of the numerical experiment are shown in \cref{tab:parameter_all}. We denote the units of population as `people', time in `days' and length in `km'.
The initial ratio of exposed to infected population is assumed to be $1:1$ (50 \% detection is assumed).

\begin{table}[!htbp]
    \centering
    \caption{Parameter values used in numerical experiments.}
        \begin{tabular}{ | m{25mm}  |  m{35mm} |  m{30mm} |  m{25mm} |}
    \hline
      Parameter                &  Square domain  & Southern Ontario &   Units \\
                   & (Section \ref{s41})) & (Section \ref{s43}) &  \\
      \hline
      $A$                     					& $500$                  				& $8.9 \times 10^{-3}$  & $\frac{\rm{people}}{\rm{km^{2}}}$ \\ 
      $\beta_{I},\beta_{E}$            		& $3.78\times10^{-4}$	& \cref{Eq.beta_ie}  & $\frac{\rm{km^{2}}}{\rm{people} \times \rm{days}}$\\
      $\nu_{S},\nu_{R}$	& $3.94\times10^{-6}$  	&  $4.5\times 10^{-7}$  & $\frac{\rm{km^4}}{\rm{people} \times \rm{days}}$ \\
      $\nu_{E}$	& $3.94\times10^{-6}$  	&  $4.5\times 10^{-7}$  & $\frac{\rm{km^4}}{\rm{people} \times \rm{days}}$ \\
      $\nu_{I}$                			& $10^{-8}$    		&   $10^{-9}$ 		 &  $\frac{\rm{km^4}}{\rm{people} \times \rm{days}}$\\
      $\gamma_R$              	& 1/24          							& 1/11   									 & $\frac{1}{\rm{days}}$\\
      $\gamma_D$              	& 1/160         							&   1/750									&$\frac{1}{\rm{days}}$\\
      $\sigma$                			& 1/7		          							& 1/5										 		    & $\frac{1}{\rm{days}}$\\
     $\gamma_E$               		& 1/6          								& 1/15   						&$\frac{1}{\rm{days}}$\\
      \hline
    \end{tabular}
    \label{tab:parameter_all}
\end{table}

\subsection{Comparison of preconditioners}\label{s41}
\begin{figure}[H]
    \centering
    \begin{subfigure}[b]{0.475\textwidth}
      \centering
        \includegraphics[width=3.3in,height=2.9in]{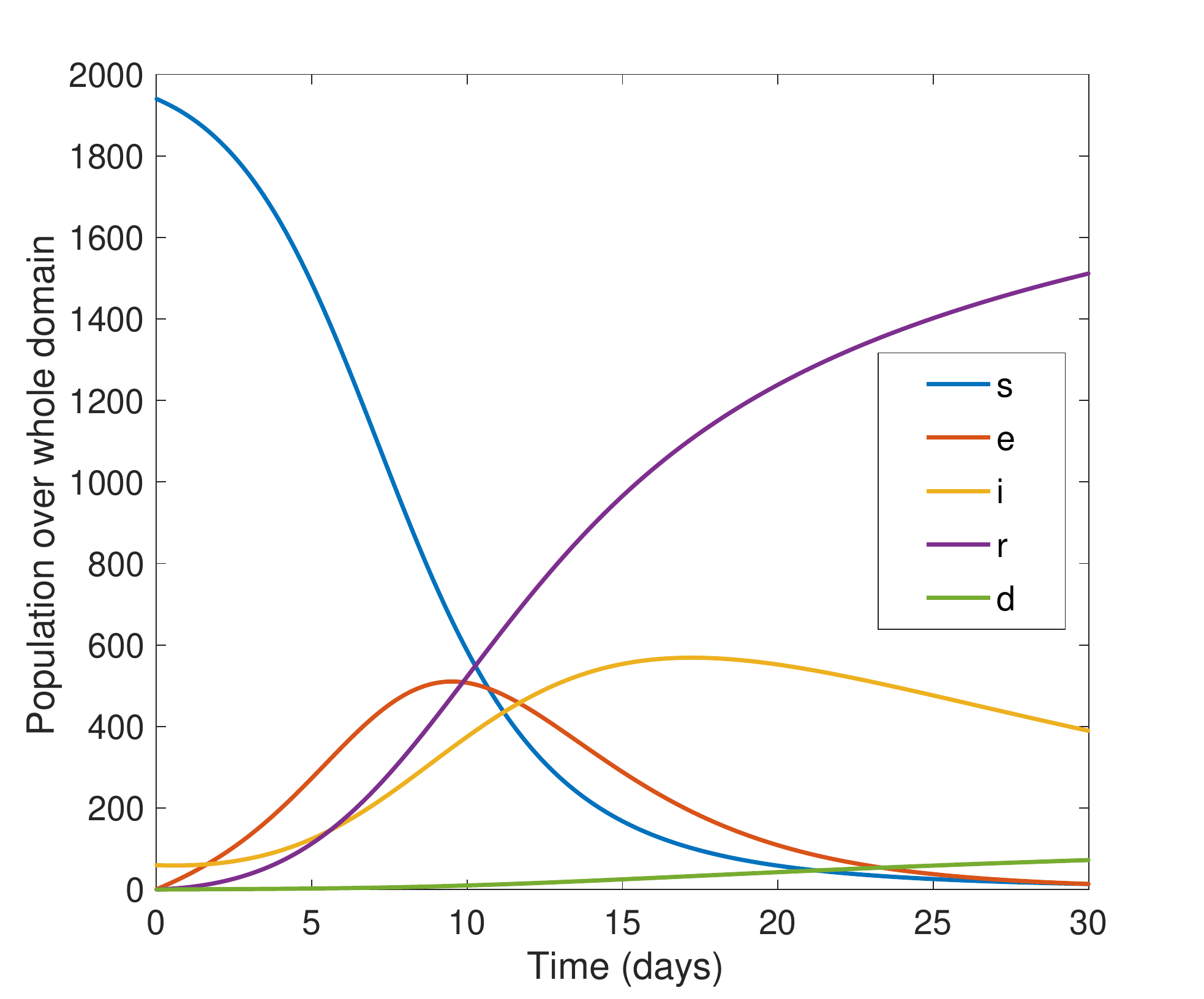} 
        \caption{Integrated solution over entire domain with time}
    \end{subfigure}
    \centering
    \begin{subfigure}[b]{0.475\textwidth}
      \centering
        \includegraphics[width=3.3in,height=3in]{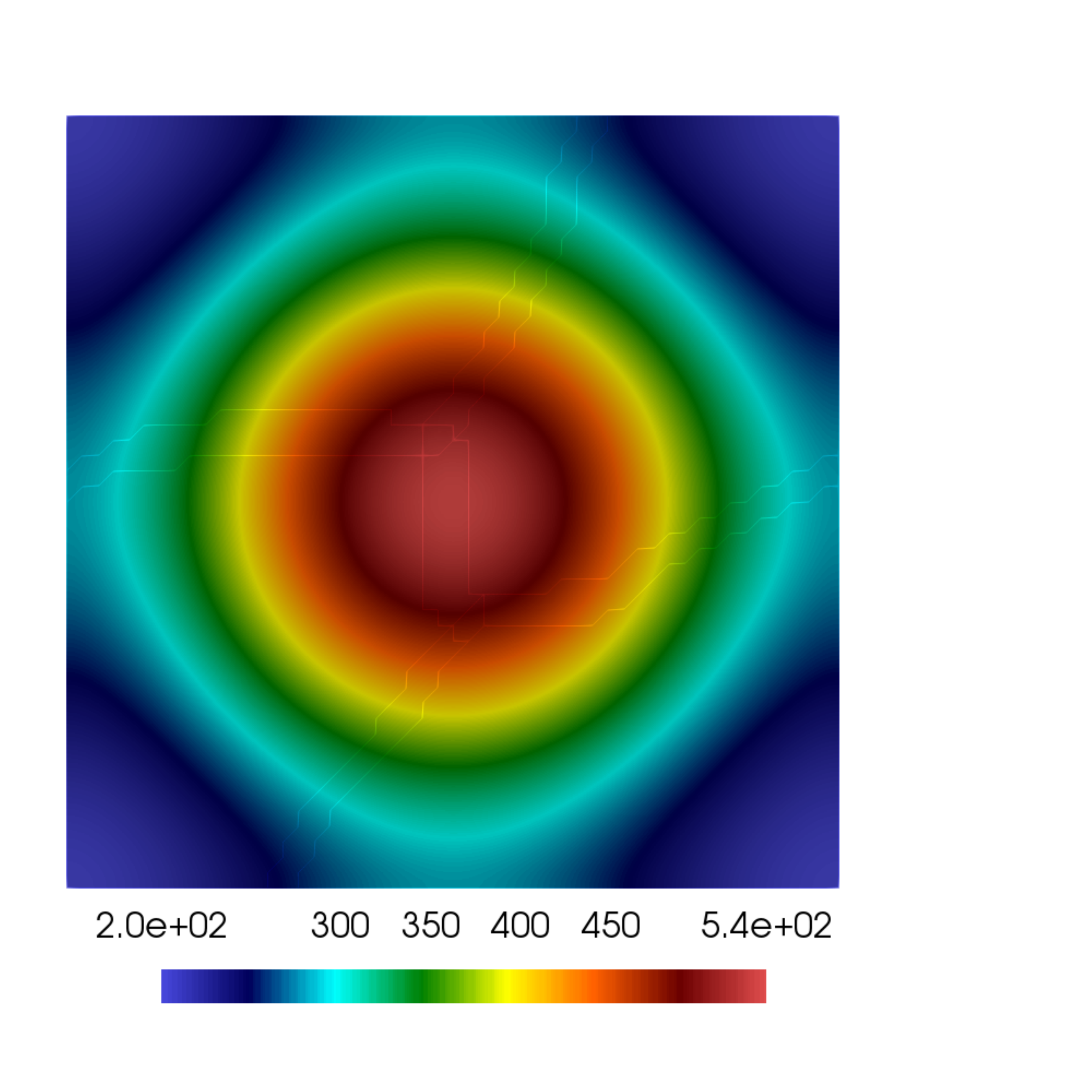} 
        \caption{Infected density at $10$ days}
    \end{subfigure}
    \caption{}\label{fig:NT}
\end{figure}

For numerical investigation, a unit square domain with a uniform total population density of $N= 2000 \; \frac{people}{km^2}$ is selected. A Gaussian function models the initial infected population at the center of domain as:

\begin{equation}
    i(x,y) = 0.1  N \exp \left(-10 \left[\left(x-0.5\right)^2 + \left(y-0.5\right)^2\right]\right),
\end{equation}
where $N$ represents the total population density. The densities of compartments $e$, $r$ and $d$ are initially zero. The susceptible density $s$ is calculated by subtracting the known densities from total density. The domain is discretized in space using triangular elements and backward Euler discretization is used for temporal discretization. The error tolerance for the Picard iteration is set to $10^{-8}$. The time traces of all compartments averaged over the entire domain and the contour plot for infected density at $10$ days is shown in \cref{fig:NT}.

\cref{tab:PC_2000_600cores} shows the time taken and iteration counts for various cases using a fine mesh of size $4000 \times 4000$ solved using 600 processes. We report the time-to-solution for ten time steps for comparison. In the decoupled approach, since all compartments are solved separately inside Picard iterations, the Krylov solver  iteration counts are calculated as the sum of all five compartments. The Picard iteration counts and the Krylov solver iteration counts are reported as the averaged value over all time steps. The Picard iteration counts does not change from one level and two-grid versions of RAS preconditioners. However, the number of Krylov iterations reduces drastically for two-grid methods. As the system size grows, the convergence rate decreases for one-level methods which can only be compensated by a coarse solve enabling global error propagation. Comparing two-grid methods, the time taken to solve ten steps is lowest for RAS-V3, followed closely by RAS-V2. The direct factorization of the two-grid RAS consumes more time than its counterparts which increases the execution time. It is interesting to note the smaller time for one-level RAS than two-grid RAS which demonstrates the importance of optimizing the coarse solver. We also note that the reported parameters for two-grid solvers do not change in time significantly and the selected duration represents the average behaviour.

\begin{table}[h!]
    \centering
    \caption{Comparison of various precondtioners for a fine mesh size of $4000 \times 4000$ .}
    \begin{tabular}{ | m{50mm}  |  m{20mm} |  m{20mm} |  m{20mm} |}
     \hline
      {Variants} & \specialcell{Average \\ Number of \\ Krylov \\iterations} & \specialcell{Average \\ Picard \\ iteration \\count} & \specialcell {Time taken\\ for 10 steps\\(s)}\\
      \hline
      \hline
      \cline{2-4}
        \hline
       \specialcell{ One-level  RAS}  & 3399 & 4 & 868 \\
        \hline
         \specialcell{Two-grid RAS} & 21 & 4 & 1308 \\
        \hline
         \specialcell{Two-grid RAS - V2 }  & 26 & 4 & 530 \\
        \hline
         \specialcell{Two-grid RAS - V3 } & 21 & 4 & 434 \\
        \hline
    \end{tabular}
    \label{tab:PC_2000_600cores}
\end{table}

\begin{figure}[H]
    \centering
    \begin{subfigure}[b]{0.473\textwidth}
      \centering
        \includegraphics[width=\textwidth]{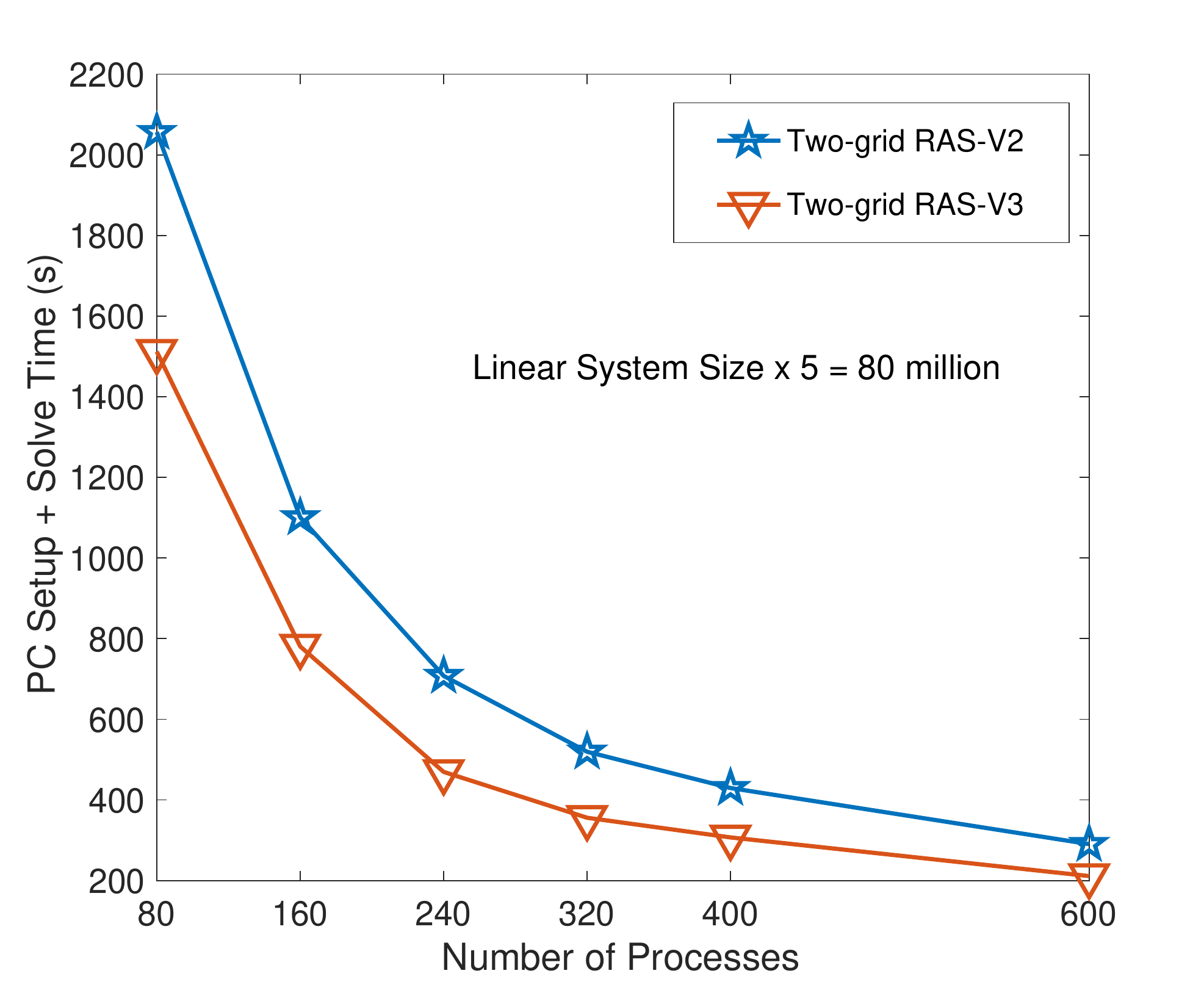} 
        \caption{Strong parallel scalability}\label{Fig:scalabilitya}
    \end{subfigure}
    \centering
    \begin{subfigure}[b]{0.475\textwidth}
      \centering
        \includegraphics[width=\textwidth]{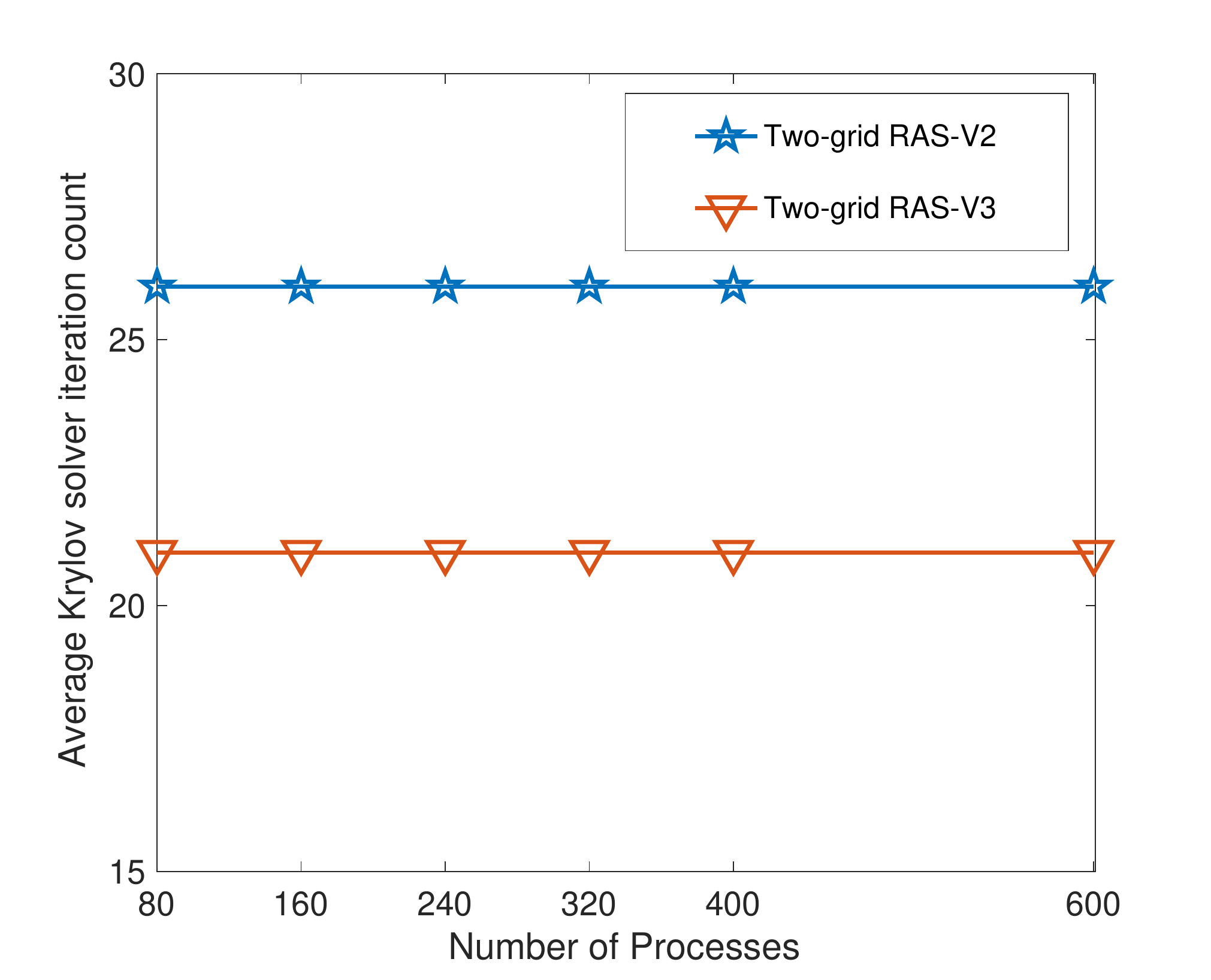} 
        \caption{Strong numerical scalability}\label{Fig:scalabilityb}
    \end{subfigure}
    \centering
    \begin{subfigure}[b]{0.475\textwidth}
      \centering
      \includegraphics[width=\textwidth]{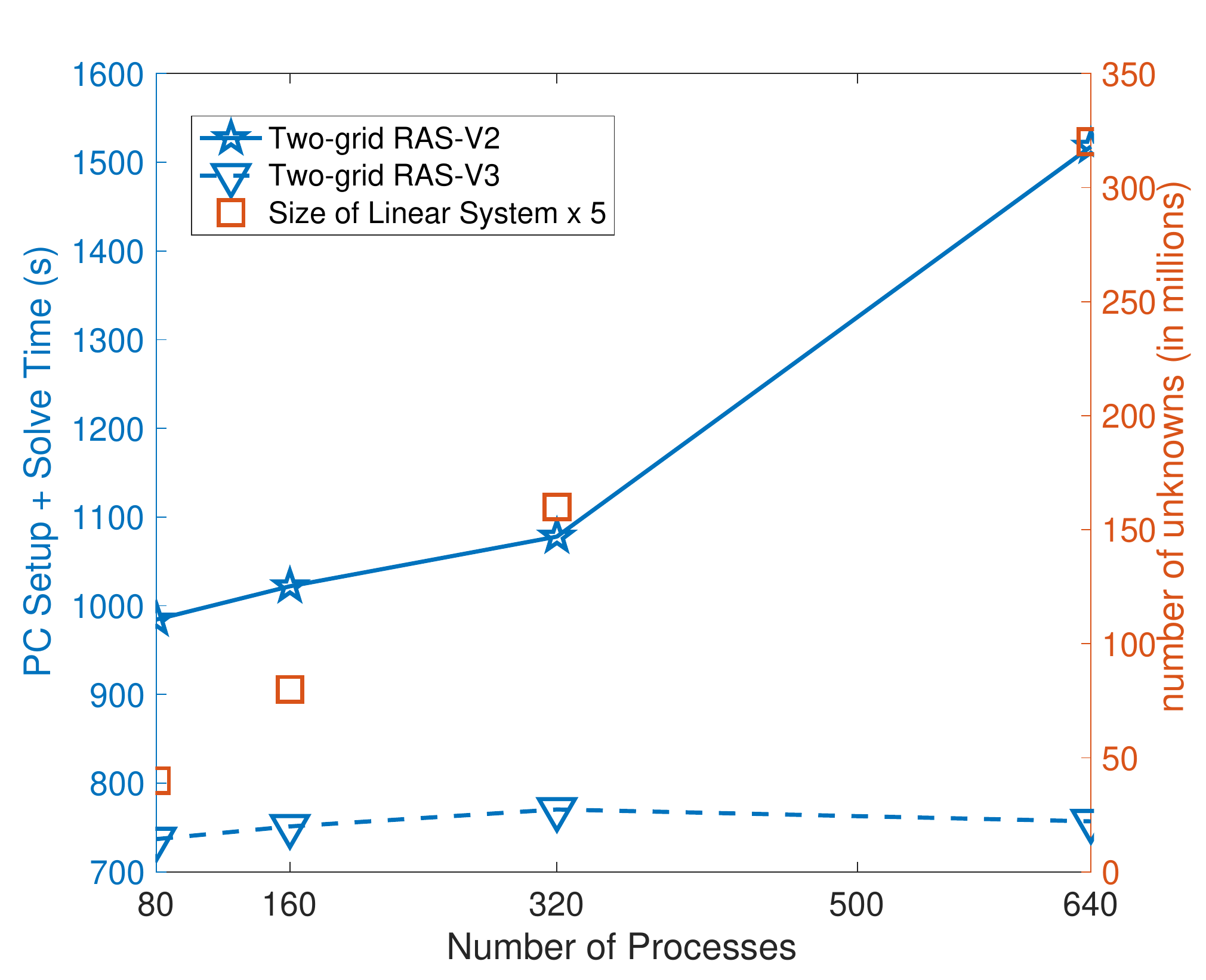} 
        \caption{Weak parallel scalability}\label{Fig:scalabilityc}
    \end{subfigure}
    \centering
    \begin{subfigure}[b]{0.475\textwidth}
      		 \centering
        \includegraphics[width=\textwidth]{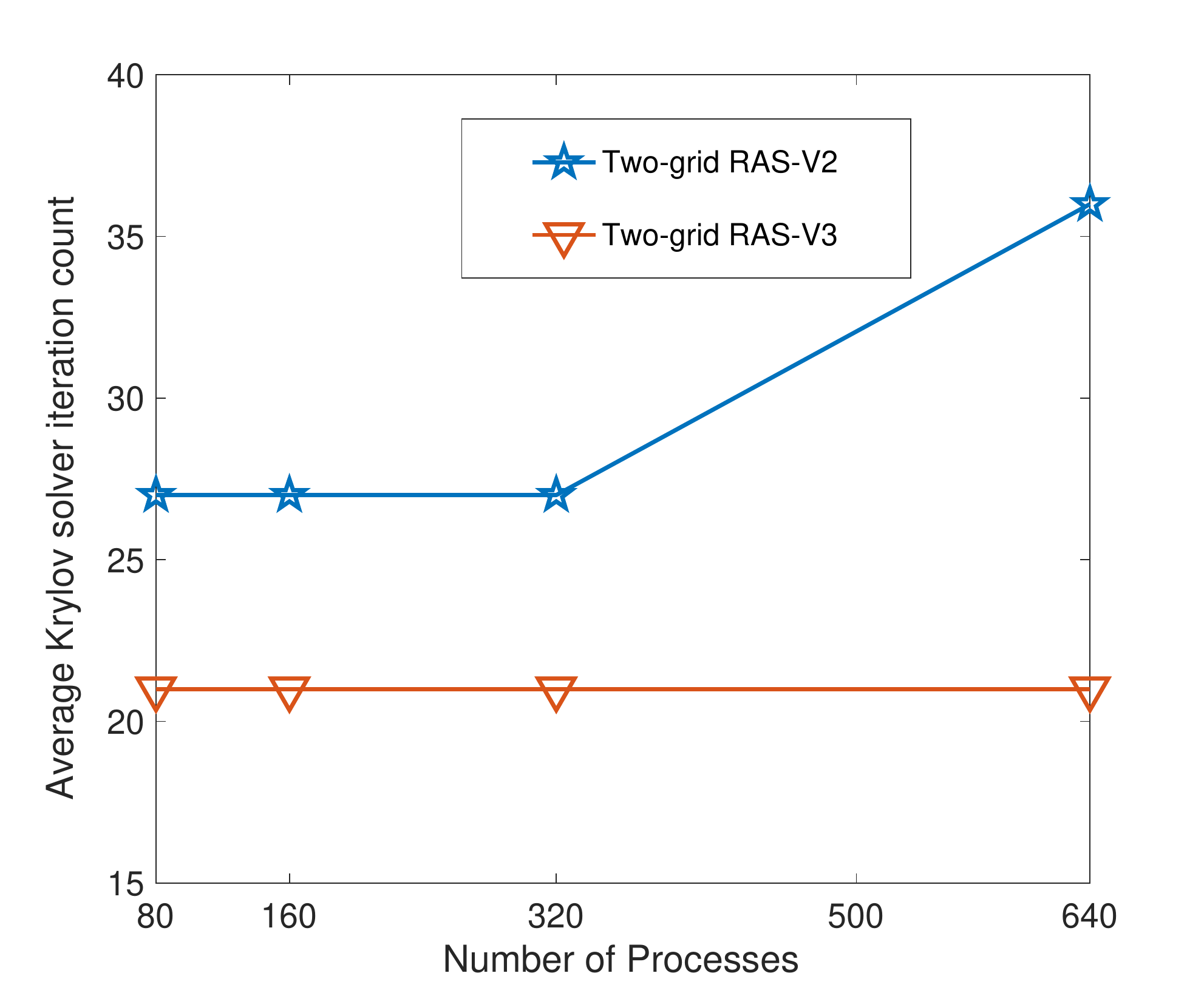} 
        \caption{Weak numerical scalability}\label{Fig:scalabilityd}
    \end{subfigure} 
  \caption{Scalability for two-grid RAS versions.}\label{Fig:scalability}
\end{figure}

After preliminary analysis, we select the two-grid variants RAS-V2 and RAS-V3 to perform further scalability studies using the same square domain and the same model parameters. The strong parallel and strong numerical scalabilities are measured by the execution time and the Krylov solver iteration counts respectively for solving a fixed problem with an increasing number of subdomains. A higher level of parallelization decreases the execution time demonstrating strong parallel scalability. Constant iteration counts with increasing processes shows the numerical scalability. However, a stagnation point is reached when the interprocessor communication time overwhelms the floating point operation time and it is no longer suitable to increase the number of processes. For the fixed fine mesh, the number of processes/subdomains are increased from 80 to 600. We plot the preconditioner setup time and system solve time which are fully parallelizable. The simulation involves ten time steps.  \cref{Fig:scalabilitya} shows the reduction in time for variants RAS-V2 and RAS-V3 with increasing processes. For a large number of subdomains/processes, both RAS-V2 and  RAS-V3 have minimal difference in time-to-solution. Even though iteration counts are constant with increasing processes for both RAS-V2 and RAS-V3, RAS-V2 takes higher iterations for convergence (see \cref{Fig:scalabilityb}).

Weak parallel and weak numerical scalabilities relate to a nearly constant execution time and constant Krylov iteration counts respectively for a constant problem size per subdomain with the increasing number of processes. Hence the total problem size is increased along with number of processes. \cref{Fig:scalabilityc,Fig:scalabilityd}   show the parallel and numerical scalabilities of both two-grid preconditioners. The system size is shown on the right y-axis of the \cref{Fig:scalabilityc}. As the problem size increases, the execution time for RAS-V2 increases significantly, but the execution time for RAS-V3 remains nearly constant in \cref{Fig:scalabilityc}. From the perspective of numerical scalability, note that outer iteration counts are the same for RAS-V3 and RAS-V2 for all processes except a slight increase observed for RAS-V2 (see \cref{Fig:scalabilityd}). 

The degraded performance of RAS-V2 can be attributed to the coarse solver. It is found that the coarse solver for RAS-V2 preconditioner requires more than $100$ iterations to reach a specified tolerance for some compartments (e.g. susceptible, exposed) while RAS-V3 converges in less than $10$ iterations. This increased coarse solver iterations at each outer Krylov iteration drastically increases the execution time for RAS-V2. This performance degradation of RAS-V2 becomes severe with an increasing size of coarse grid.  On the other hand, RAS-V3 shows excellent scalability having constant iteration counts at fine and coarse levels. This is due to the multiple levels of error reduction in the coarse solver using AMG preconditioner, leading to better convergence behaviour.

\begin{figure}[H]
    \centering
    \begin{subfigure}[b]{0.473\textwidth}
      \centering
        \includegraphics[width=\textwidth]{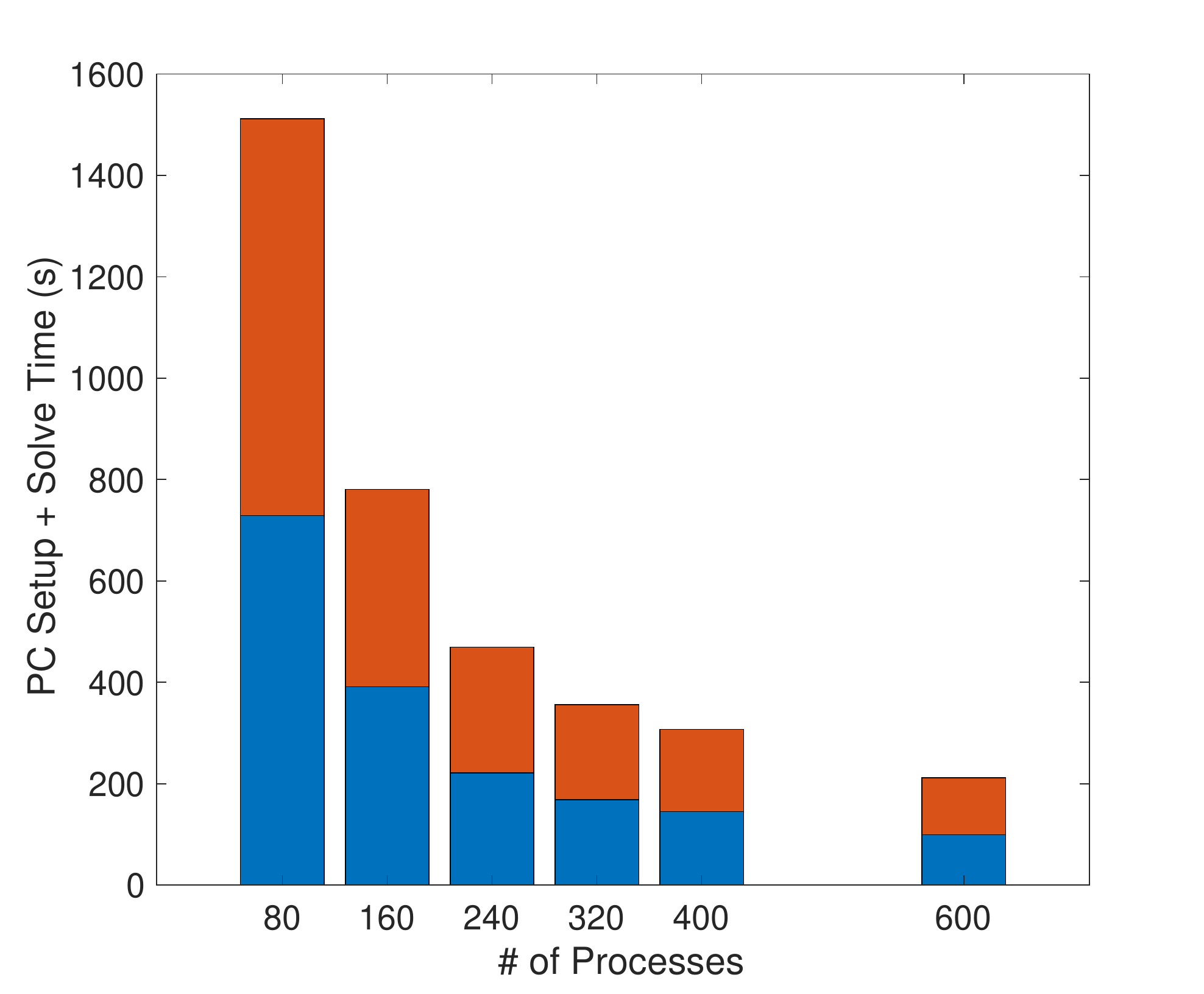} 
        \caption{Strong parallel scalability}
    \end{subfigure}
    \centering
    \begin{subfigure}[b]{0.475\textwidth}
      \centering
        \includegraphics[width=\textwidth]{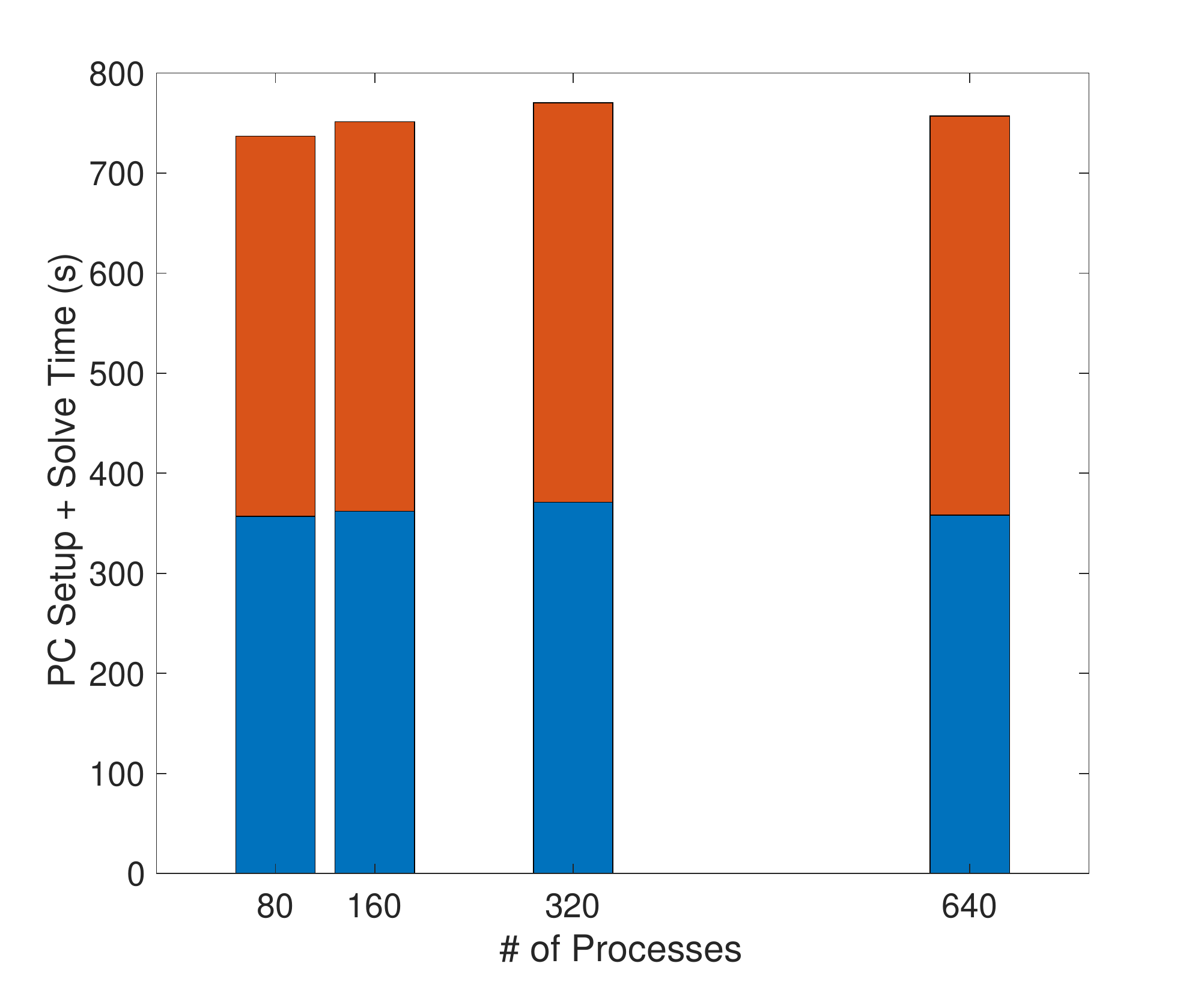} 
        \caption{Weak parallel scalability}
    \end{subfigure}
     \caption{PC setup (blue) and solve times (orange) - Two-grid RAS-V3.}\label{Fig:setup_hypre}
\end{figure}

\cref{Fig:setup_hypre} shows the preconditioner (PC) setup and solve times associated to two-grid RAS-V3. For strong scalability, the setup and solve time reduces with the increasing number of cores. For weak scalability, the setup and solve times remain nearly constant as the problem size increases. These facts clearly demonstrate the scalability of the RAS-V3 preconditioner against problem sizes and processes.

\subsection{Application: Southern Ontario}\label{s43}
\begin{figure}[H]
      \centering
        \includegraphics[width=0.5\textwidth]{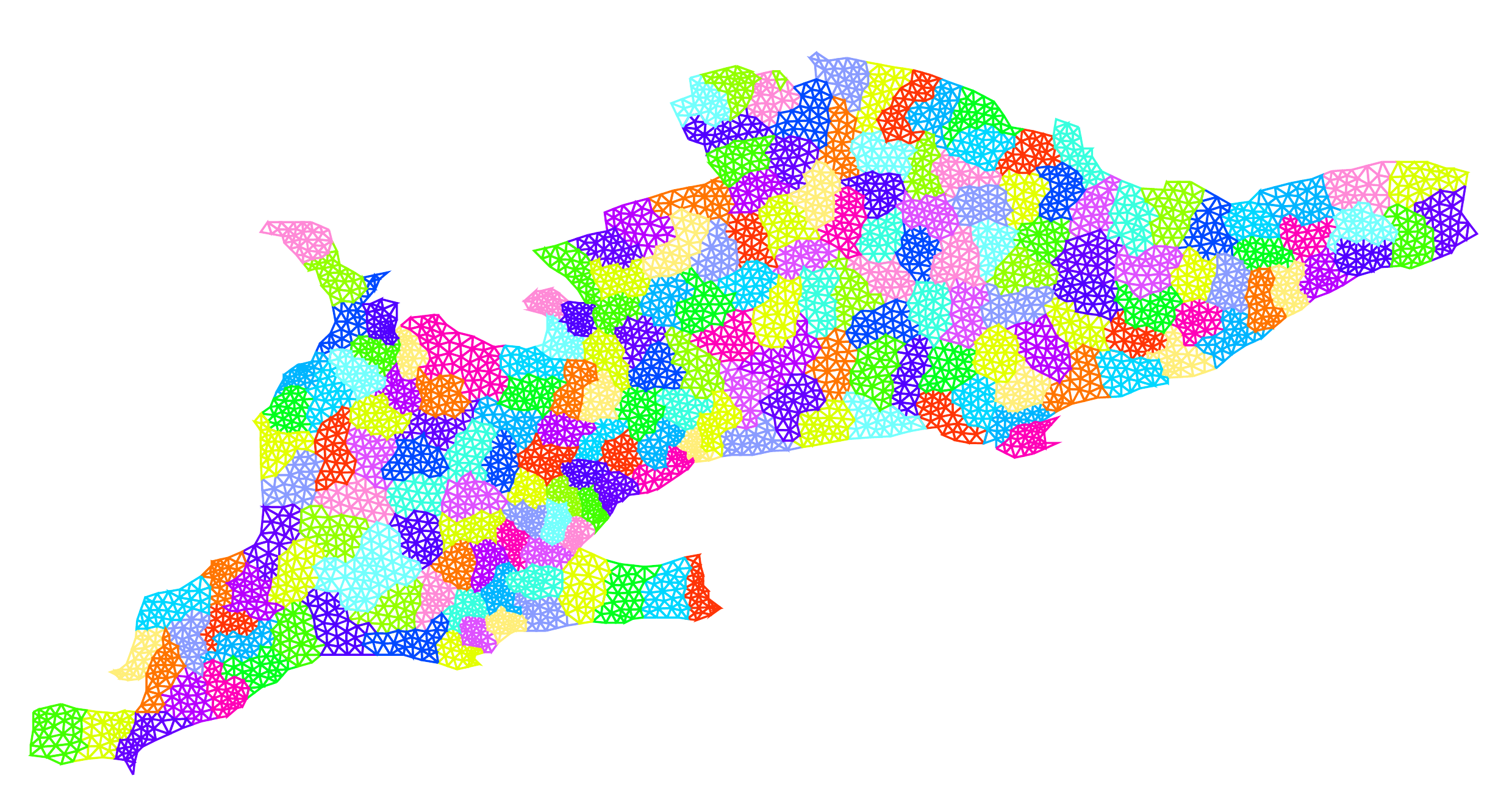} 
        \caption{Southern Ontario subdivided in to 200 subdomains}
        \label{Fig:south}
\end{figure}

In this section we apply two-grid RAS - V3 preconditioner to the region of Southern Ontario. This region, consisting of $27$ public health units (PHUs), is densely populated, accounting for 95\% of the population of the province of Ontario, while only occupying approximately 13\% of area \cite{phu}.
 The public data on  COVID-19 statistics from these PHUs can help to infer the status of COVID-19 infections in time and space. We utilize the open source software QGIS \cite{QGIS_software} to define the geographical boundaries and generate a mesh file \cite{phu_shpfile}. The meshed domain is then subdivided into different subdomains for the domain decomposition solver. The mesh of Southern Ontario subdivided into $200$ subdomains is shown in \cref{Fig:south} (overlapping parts not shown for clarity). Note these subdomains do not correspond to the aforementioned PHUs. The initial conditions are obtained from publicly reported testing data on September 1, 2020. We generate densities of each compartment as a sum of 27 Gaussian pulses, centered at main city of each PHU (following similar approach as \cite{oden_seird}),

\begin{equation}
    s(\mathbf{x},0) = \sum _{i=1} ^{27} A_i \exp \Big[ \frac{(x - x_i)^2 + (y - y_i)^2}{2B_i^2} \Big],
\end{equation}
where each $(x_i,y_i)$ are the coordinates of the cities, and $B_i$ represents the radius around ($x_i,y_i$) which captures 95 \% of the population of the $i^{th}$ PHU. The value of $A_i$ is a constant which ensures that the integral of the two dimensional function over the entire domain equals to the total number of people in the respective compartment for that PHU. 

\begin{figure}[H]
    \centering
    \begin{subfigure}[b]{0.475\textwidth}
      \centering
        \includegraphics[width=\textwidth]{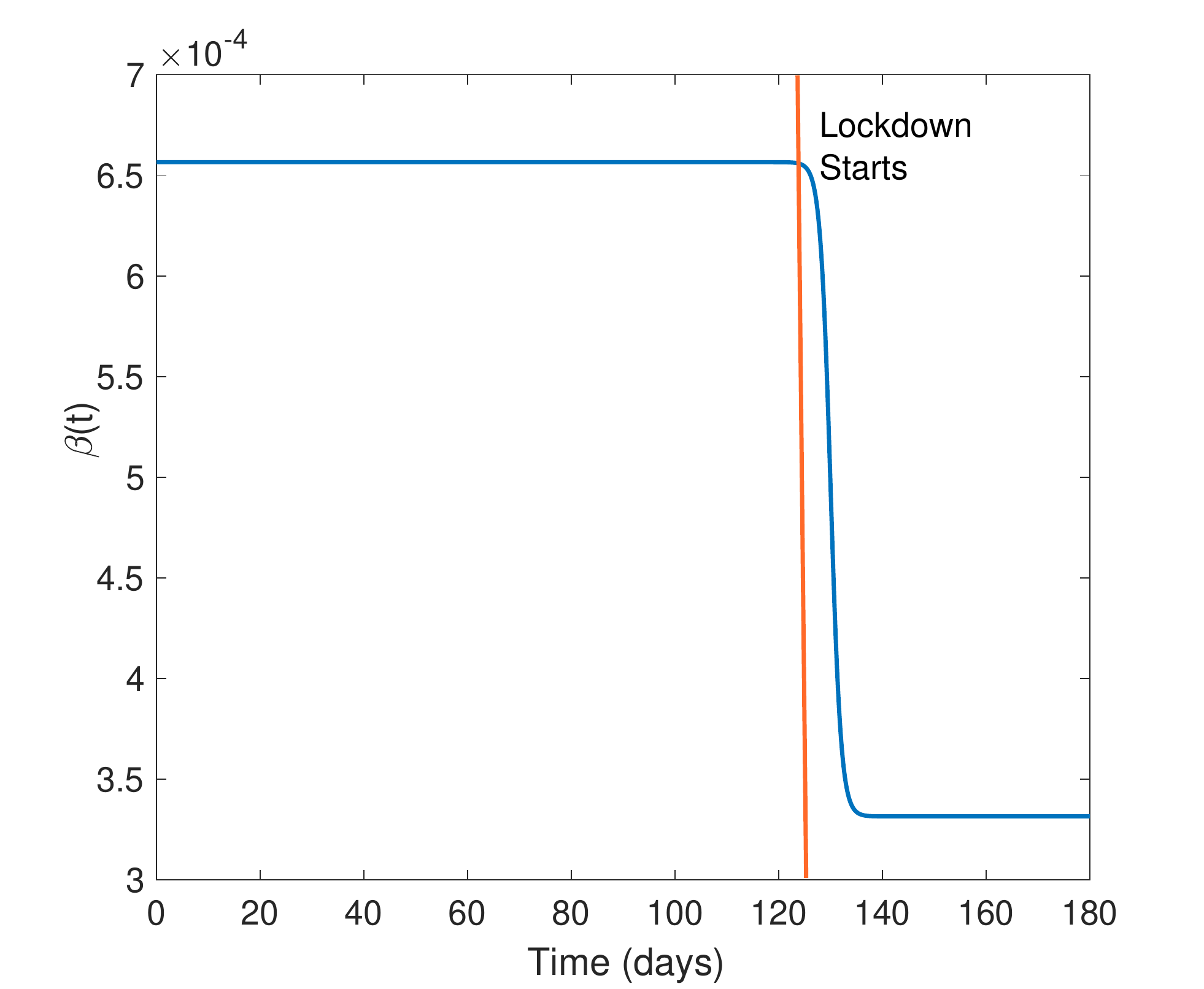} 
        \caption{Temporal variation of beta from $1^{st}$ Sep $2020$ to $28^{th}$ Feb 2021.}
    \end{subfigure}
    \centering
    \begin{subfigure}[b]{0.475\textwidth}
      \centering
        \includegraphics[width=\textwidth]{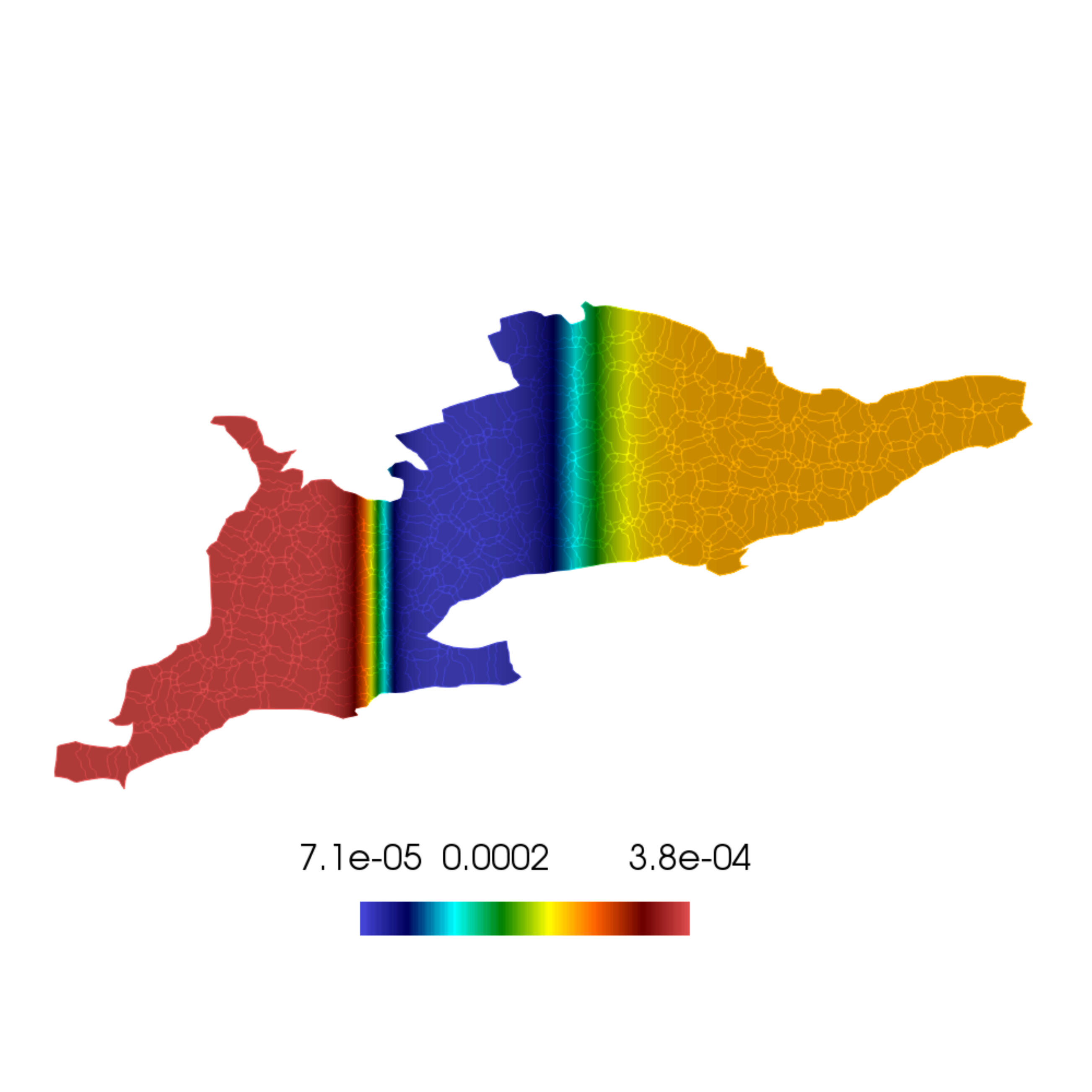} 
        \caption{Spatial variation of $\beta(\mathbf{x})$ on three regions}
    \end{subfigure}
    \caption{}\label{Fig:south_beta}
\end{figure}

The time-invariant rate parameters of the SEIRD model ($\gamma_E$, $\sigma$ and $\gamma_R$) are based on clinically relevant information regarding average duration of infection, average time to detection, and an assumed detection rate ($80\%$) \cite{Tuite,oden_seird}. The time-varying infection rate parameters $\beta_E$, $\beta_I$, rate of death $\gamma_D$ and the diffusion coefficients $\nu_S,\nu_E,\nu_I, \nu_R$ are estimated using the testing and death statistics data collected during the six-month period from September 1, 2020 to February 28, 2021, representing the second wave of infection in the province \cite{Ontario}. 
We define the infection rate parameter as a sigmoid function to mimic the sudden reduction in disease transmission following  the provincial lockdown after the 2020 holiday season as shown in \cref{Fig:south_beta}($a$). 

Furthermore, we subdivide the domain of Southern Ontario into western, central and eastern regions as in \cref{Fig:south_beta}($b$). The central region contains many PHUs with large populations in close proximity to one another. This region is separated from the larger population centres in the eastern and western regions of the province by large areas of low population density. In order to capture this spatio-temporal variation of infection rate, we assume,
\begin{equation}\label{Eq.beta_ie}
\beta_{i}(\mathbf{x},t) = \beta_{e}(\mathbf{x},t) = \underbrace{\left(0.101 - \frac{0.05}{1 + \exp^{ (130-t)} } \right)}_{\beta (t)} \; \beta (x),
\end{equation}
where,

\begin{equation}\label{Eq.beta_x}
\beta (x) =  \beta_{\rm{central}}   + \frac{\beta_{\rm{eastern}} - \beta_{\rm{central}}} {(1 + \exp ^{-5 (x - x_{\rm{eastern}})})} + \frac{\beta_{\rm{western}} - \beta_{\rm{central}}} {(1 + \exp ^{10 (x - x_{\rm{western}})})},
\end{equation}
with $x_{\rm{eastern}}$ and $x_{\rm{western}}$ denoting the longitudinal boundaries separating the eastern from the central region and the central from the western region of the domain, respectively. Similarly $\beta_{\rm{eastern}}$, $\beta_{\rm{central}}$ and $\beta_{\rm{western}}$ are the infection rates in the corresponding regions.
In \cref{Eq.beta_x}, we model the spatial variation for $\beta_{i}(\mathbf{x},t)$ and $\beta_{e}(\mathbf{x},t)$ along the longitude $x$, but they are invarient along the latitude $y$.

We adopt a simple calibration step by which the averaged results of PDE case with negligible diffusion is initially matched with the aggregated case counts for all PHUs in Southern Ontario. 
Next, the diffusion coefficients  $\nu_S,\nu_E,\nu_I$ and $\nu_R$ are changed to model the localized spatial interaction/spread of the different compartments. These parameters are calibrated through simulations to ensure that they encourage sufficient mixing of the population within the city without causing significant inter-city population movements. Finally, we determine unique values of $\beta_{\rm{eastern}}, \beta_{\rm{central}}$, and $ \beta_{\rm{western}}$ by concurrently matching the reported data both at the provincial level (globally) and at the major population centres (locally).

\begin{figure}[H]
    \centering
    \begin{subfigure}[b]{0.475\textwidth}
      \centering
        \includegraphics[width=\textwidth]{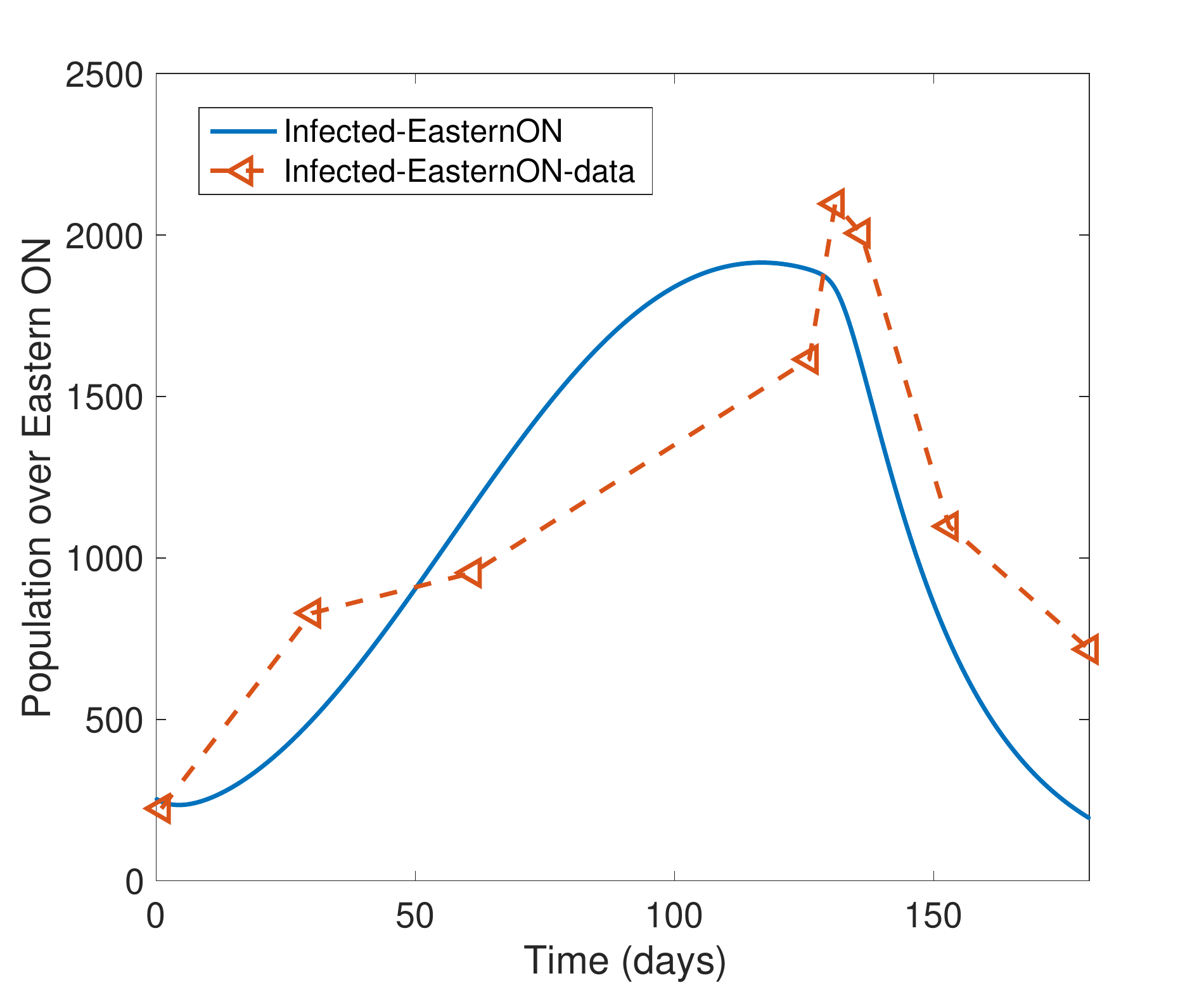} 
        \caption{Eastern Ontario}
    \end{subfigure}
    \centering
    \begin{subfigure}[b]{0.475\textwidth}
      \centering
        \includegraphics[width=\textwidth]{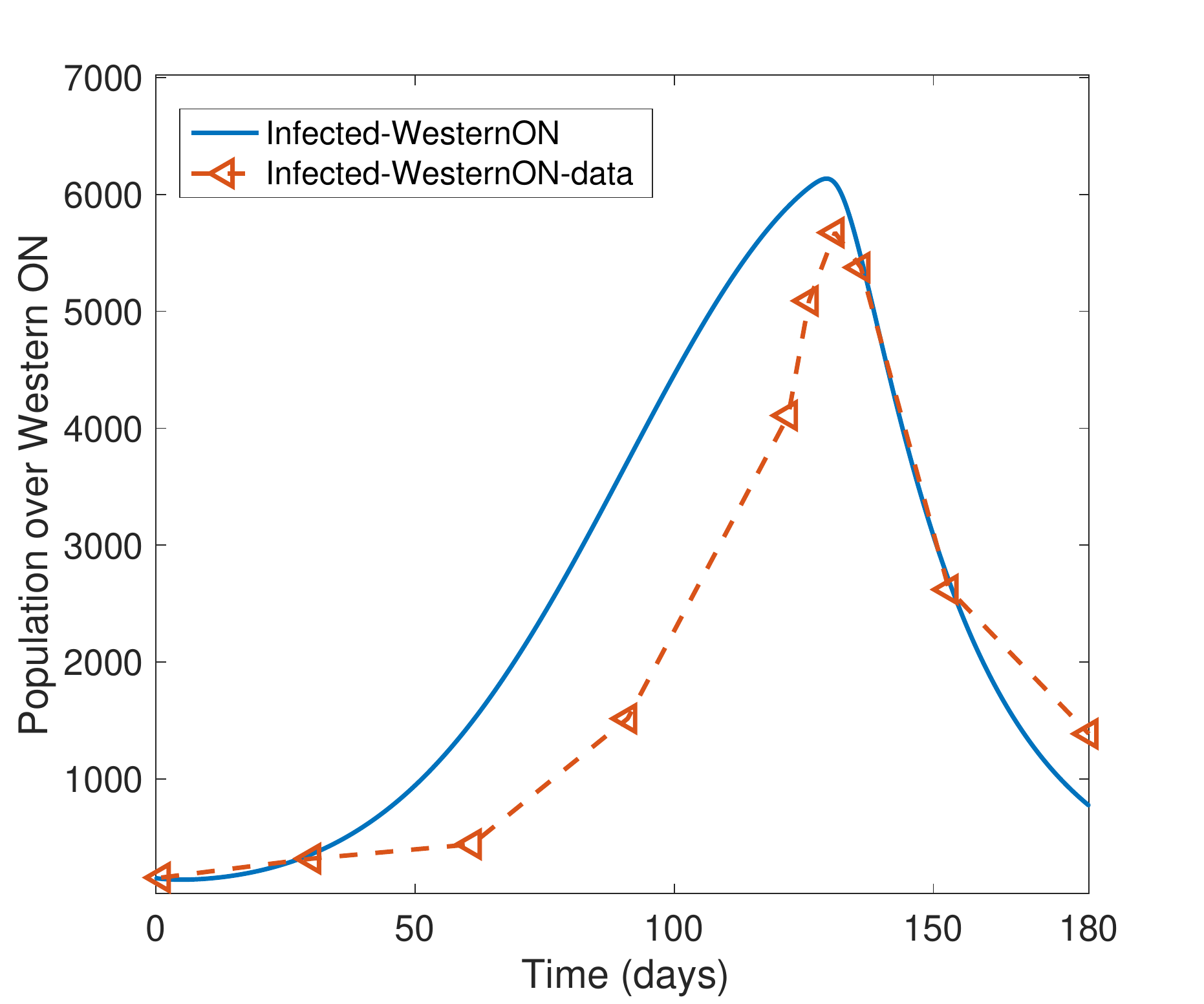} 
        \caption{Western Ontario}
    \end{subfigure}
    \centering
    \begin{subfigure}[b]{0.475\textwidth}
      \centering
        \includegraphics[width=\textwidth]{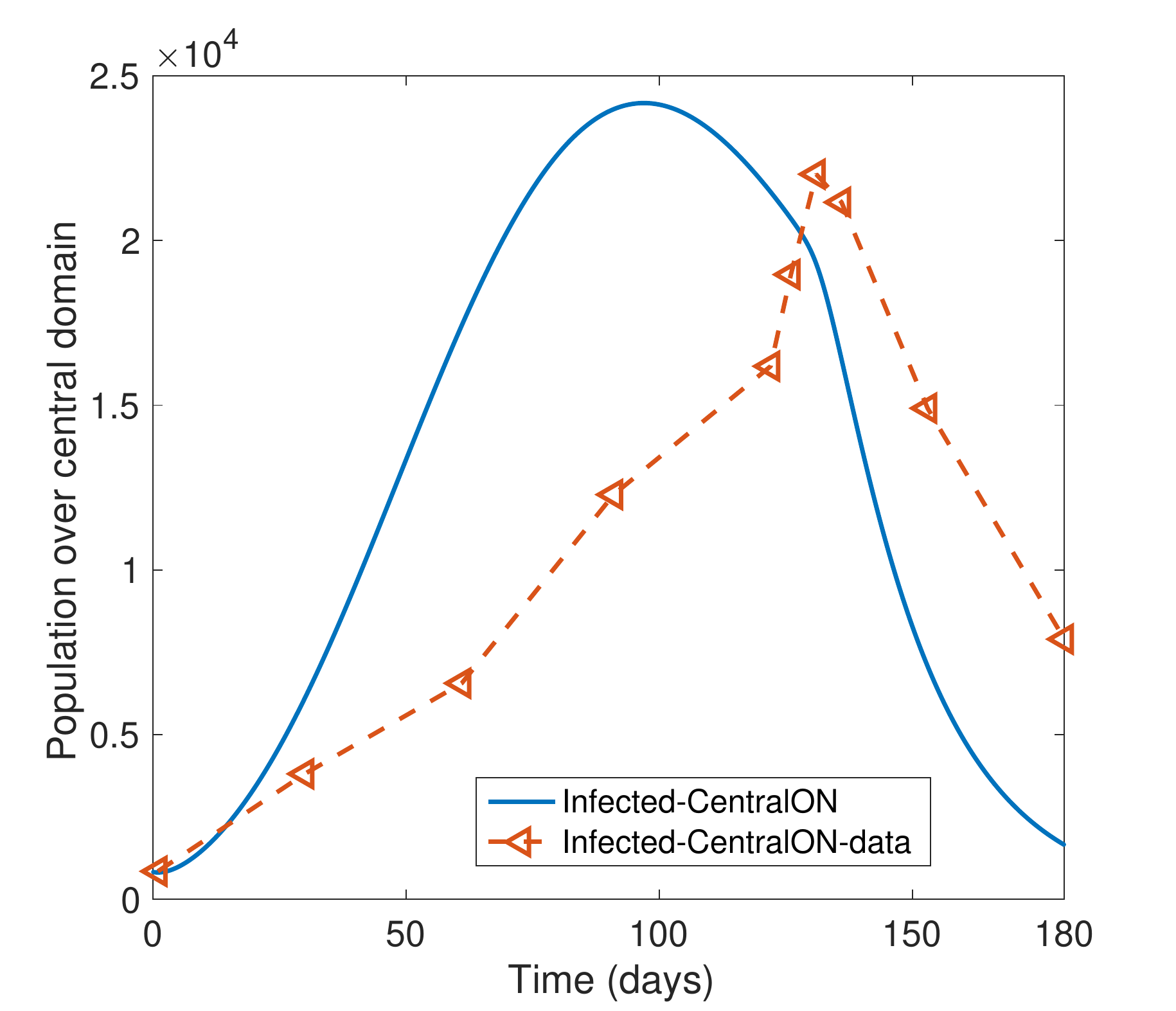} 
        \caption{Central Ontario}
    \end{subfigure}
    \centering
    \begin{subfigure}[b]{0.475\textwidth}
      \centering
        \includegraphics[width=\textwidth]{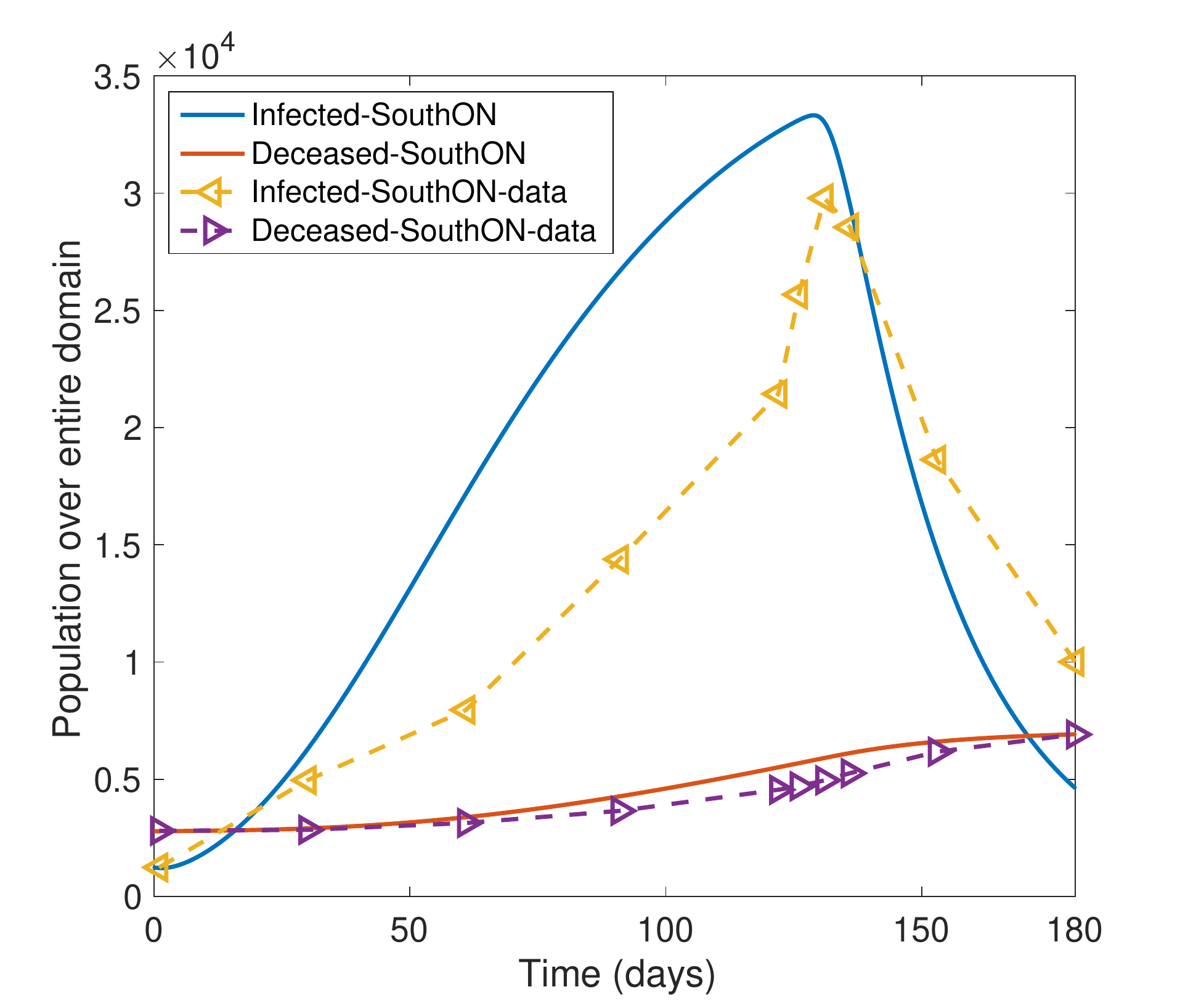} 
        \caption{Full Southern Onatrio}
    \end{subfigure}
    \caption{Comparison of simulation results and reported case counts from $1^{st}$ Sep $2020$ to $28^{th}$ Feb 2021.}\label{Fig:south_regions}
\end{figure}

\cref{Fig:south_regions} shows the averaged infected population counts for different regions against the reported field data. There is reasonable agreement between the trends in the observed data and the numerical solutions. As described above, a full scale statistical (e.g., Bayesian) inference of all the model parameters were not pursued currently as the  objective is to illustrate the performance (e.g., scalabilities) of the parallel solvers. To improve the agreement between the numerical simulation and the observed case counts, the Bayesian calibration method is currently being implemented by the authors and its findings will be reported in future publications.

\begin{figure}[H]
    \centering
    \begin{subfigure}[b]{0.225\textwidth}
      \centering
        \includegraphics[width=1.05\textwidth]{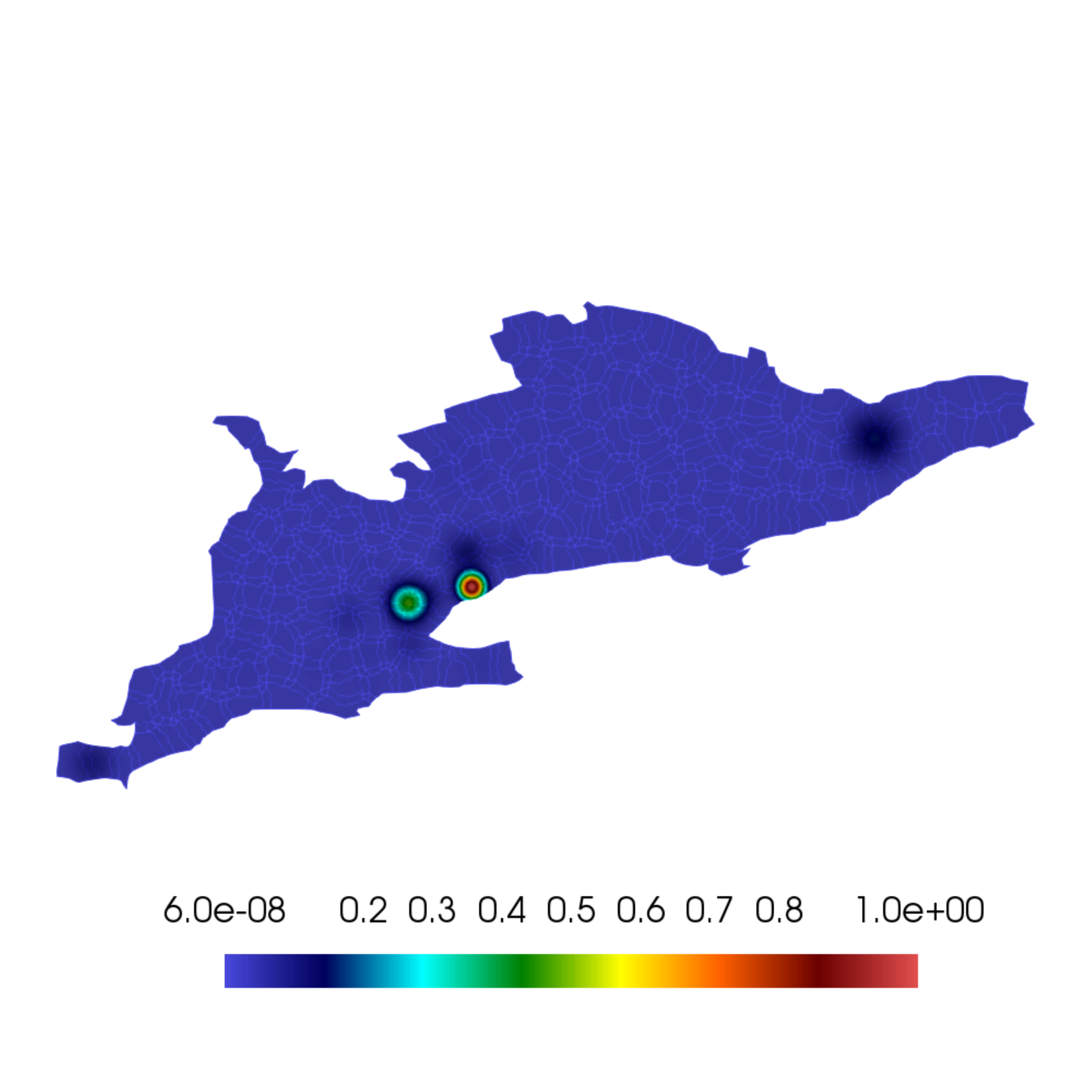} 
        \caption{Infected - initial state}
    \end{subfigure}
    \centering
    \begin{subfigure}[b]{0.225\textwidth}
      \centering
        \includegraphics[width=1.05\textwidth]{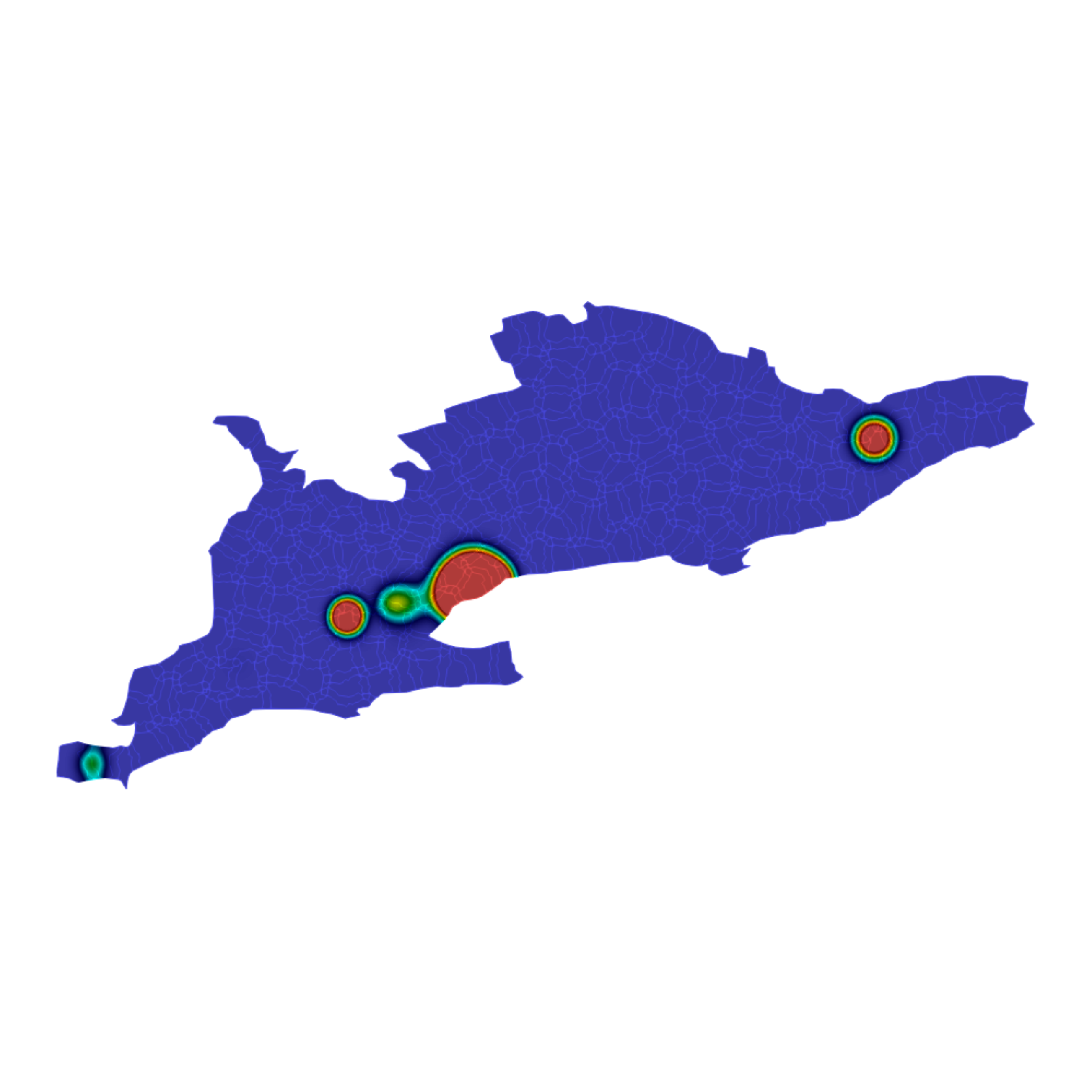} 
        \caption{Infected - 50 days}
    \end{subfigure}
\centering
    \begin{subfigure}[b]{0.225\textwidth}
      \centering
        \includegraphics[width=1.05\textwidth]{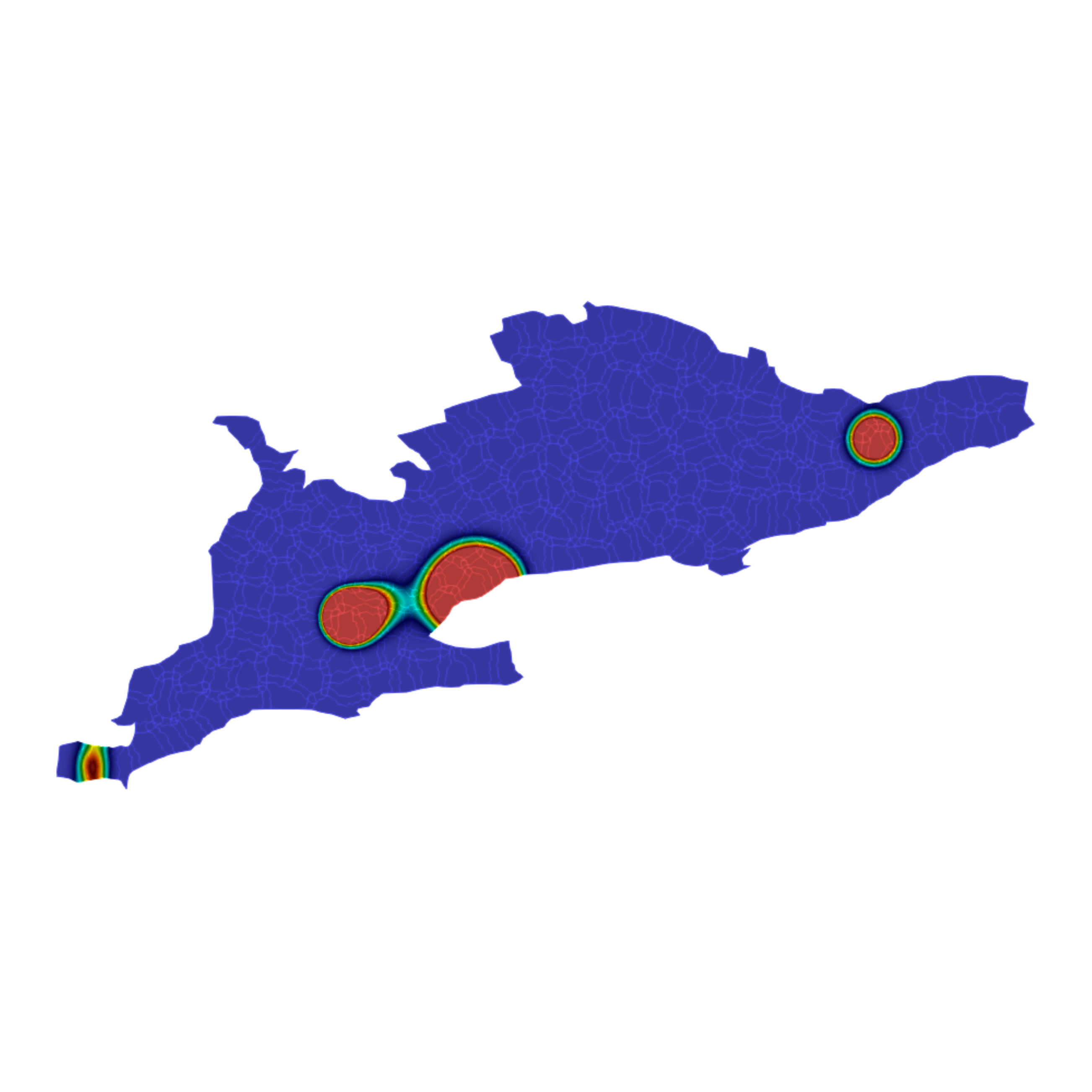} 
        \caption{Infected - 120 days}
    \end{subfigure}
     \centering
    \begin{subfigure}[b]{0.225\textwidth}
      \centering
        \includegraphics[width=1.05\textwidth]{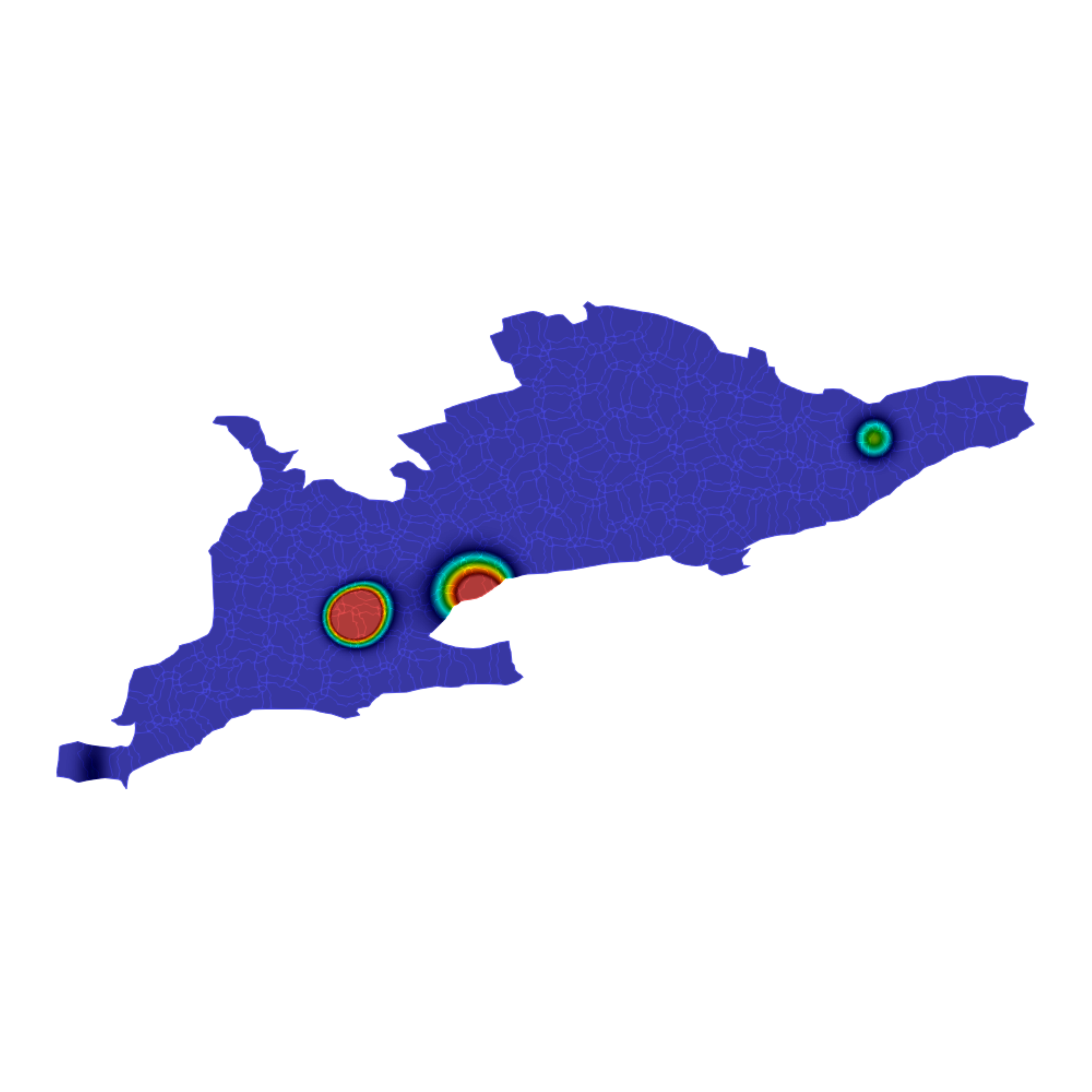} 
        \caption{Infected - 180 days}
    \end{subfigure}  
  \caption{Infected densities in Southern Ontario}\label{Fig:south_contour_inf}
\end{figure}

\begin{figure}[H]
    \centering
 \begin{subfigure}[b]{0.225\textwidth}
      \centering
        \includegraphics[width=1.05\textwidth]{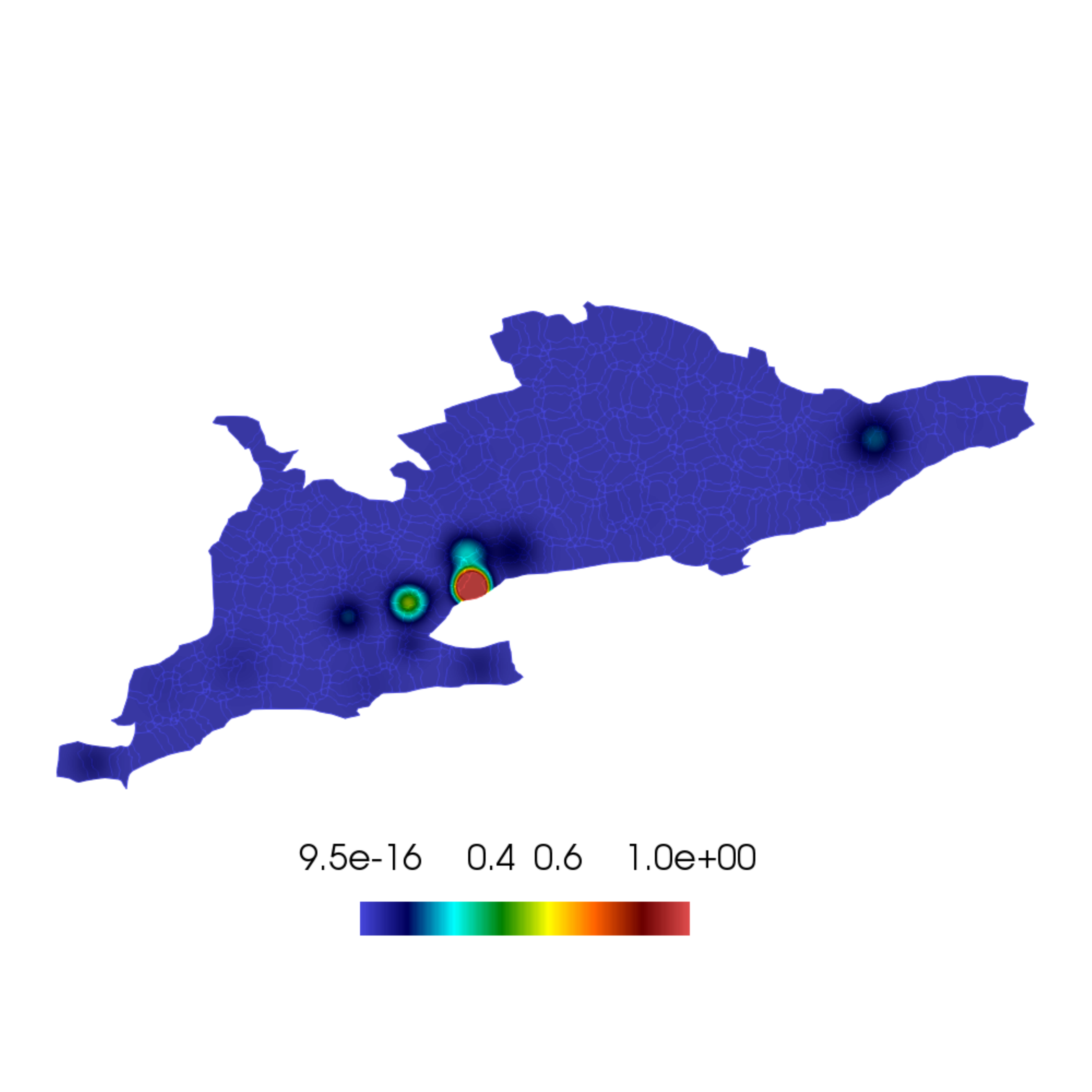} 
        \caption{Deceased - initial state}
    \end{subfigure}
     \centering
    \begin{subfigure}[b]{0.225\textwidth}
      \centering
        \includegraphics[width=1.05\textwidth]{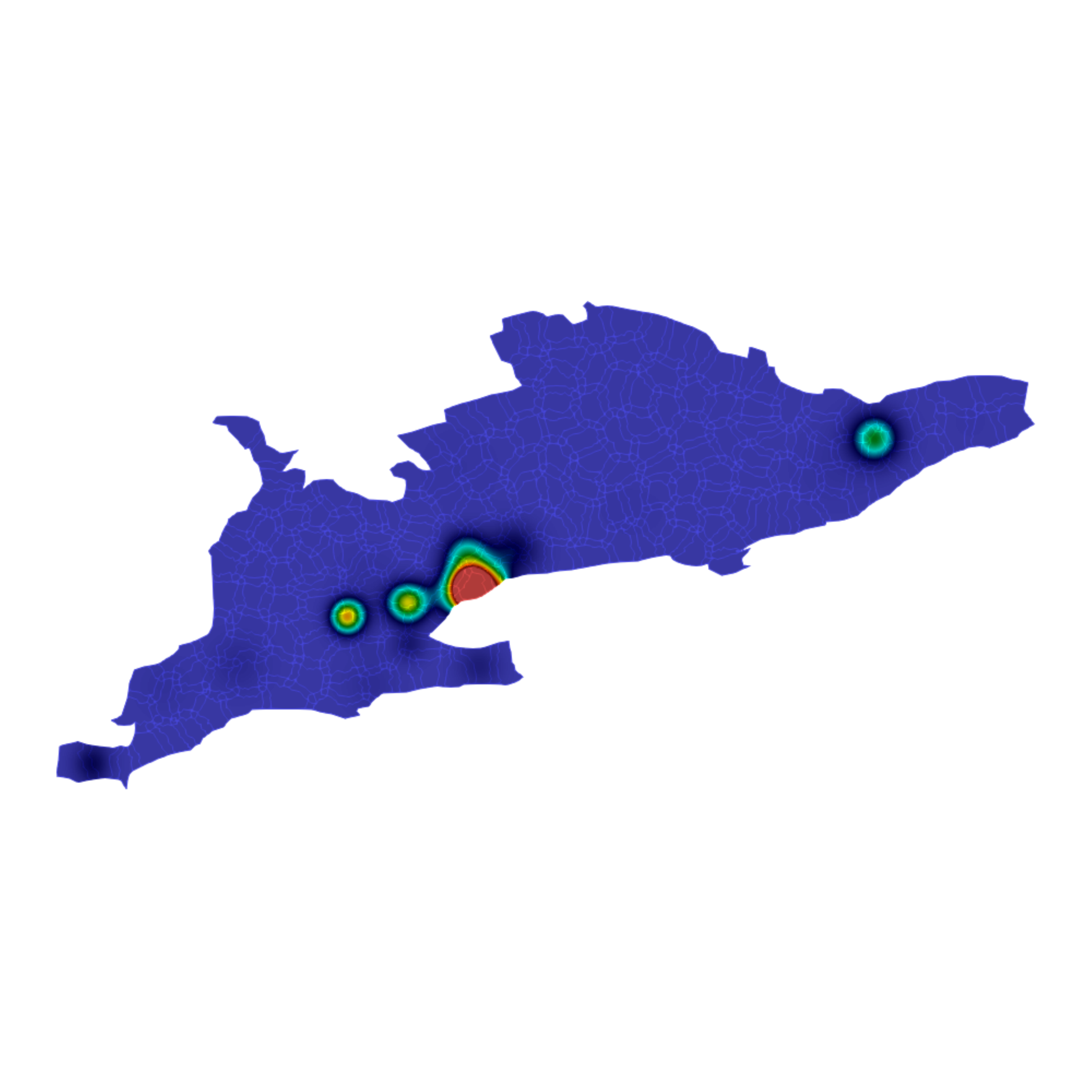} 
        \caption{Deceased - 100 days}
    \end{subfigure}
 \centering
    \begin{subfigure}[b]{0.225\textwidth}
      \centering
        \includegraphics[width=1.05\textwidth]{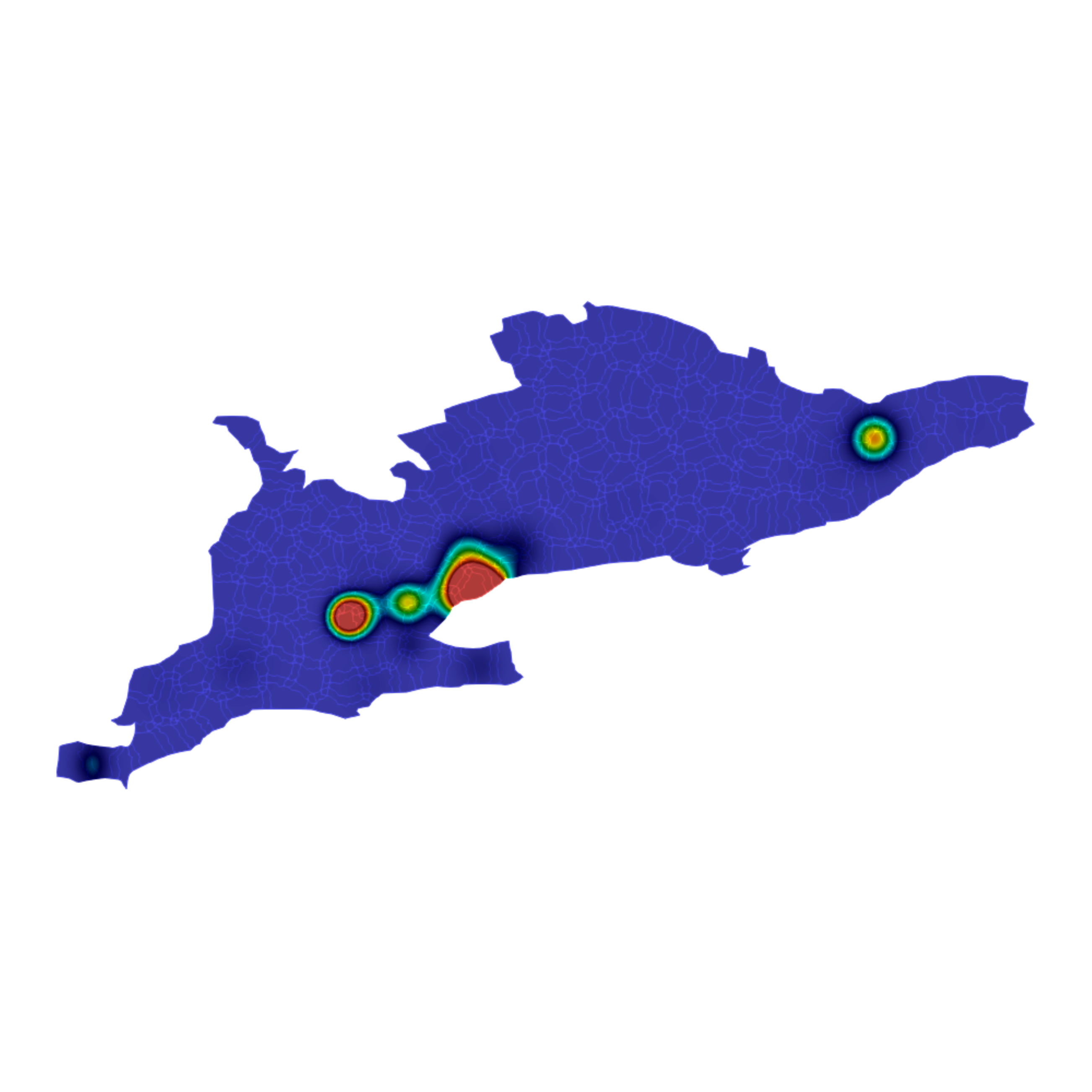} 
        \caption{Deceased - 150 days}
    \end{subfigure}
     \centering
    \begin{subfigure}[b]{0.225\textwidth}
      \centering
        \includegraphics[width=1.05\textwidth]{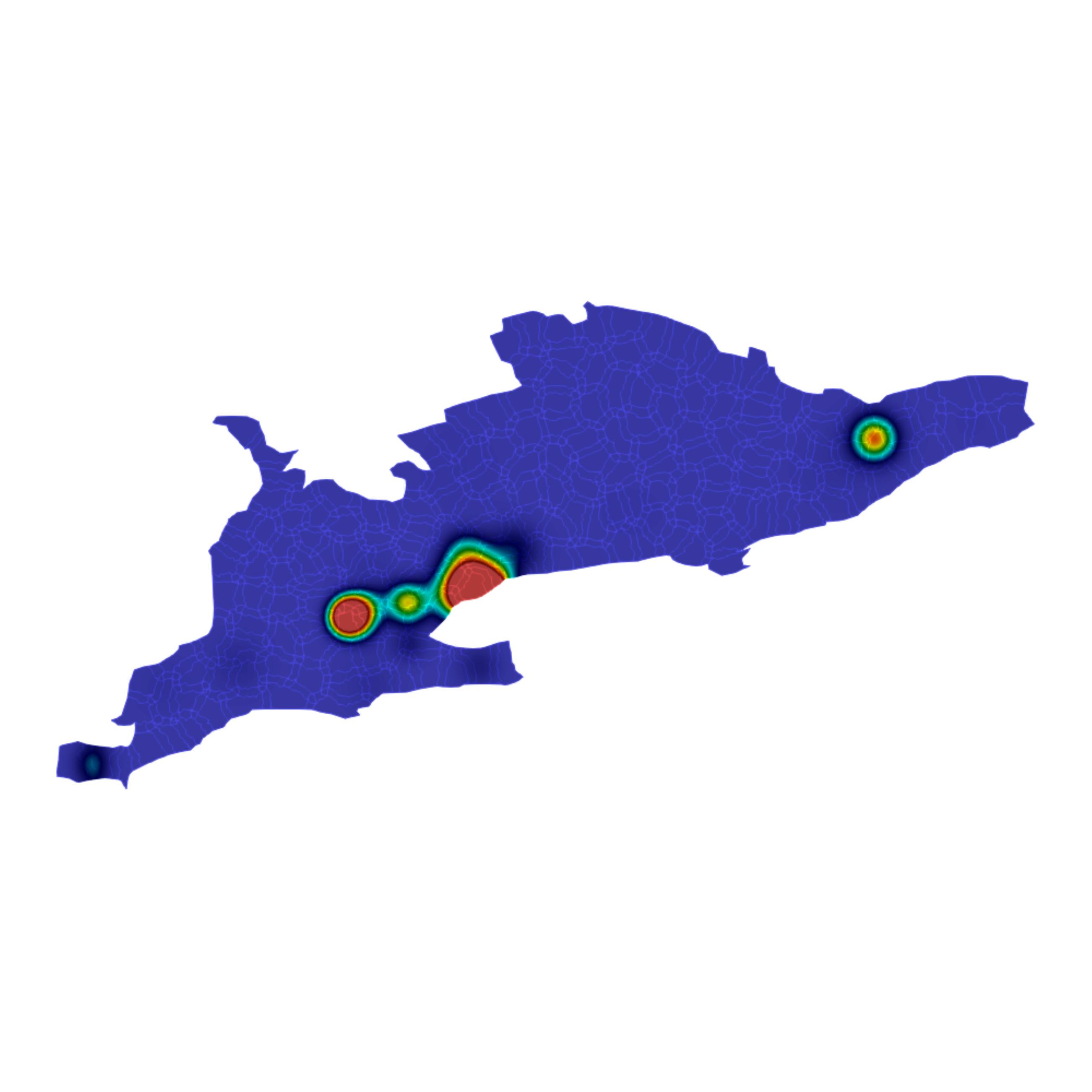} 
        \caption{Deceased - 180 days}
    \end{subfigure}
    \caption{Deceased densities in Southern Ontario}\label{Fig:south_contour_dec}
\end{figure}

The contour plots in \cref{Fig:south_contour_inf} and \cref{Fig:south_contour_dec} show the interaction of the populations of adjacent PHUs in the central region (containing the densely populated Greater Toronto and Hamilton areas) as the case counts increase. We observe similar growth in the number of infections in the more isolated regions of Windsor in the western region, and Ottawa in the eastern region. The use of a heterogeneous population density-dependent diffusion coefficients permit the mixing of the population at a relatively local scale within individual PHUs and with adjacent PHUs in high density regions. This does not permit for a realistic account of the spread of the disease among the population centres that are separated by regions of low-density. In this case, more sophisticated  methods such as in \cite{kinetic_uncertain} which uses kinetic transport equations to model commuters and diffusion to model non-commuters, or \cite{rabies_spatial} which defines an anisotropic diffusion coefficient according to various geographical features (such as highway networks, rivers and mountains) would be required. A more rigorous framework would permit for the effects of human mobility patterns (e.g., \cite{long2022associations,google_mobility}) to reflect the spatial spread. Note that the spatial variability in COVID-19 incidence rates can be related to various social and economic parameters (e.g., \cite{GIS} and \cite{long2022associations}). These variations can be captured in the model by including much finer stratifications of age and socio-economic status as described in \cite{Robinsone052681}. 
Although the critical demands of intensive care unit (ICU) beds, increased infections to co-morbid people or children, implications of post-acute COVID-19 (long covid) etc., could also be modelled by introducing appropriate compartments (see \cref{ac}),  these aspects are considered in the current investigation.

\begin{figure}[H]
    \centering
    \begin{subfigure}[b]{0.475\textwidth}
      \centering
        \includegraphics[width=\textwidth]{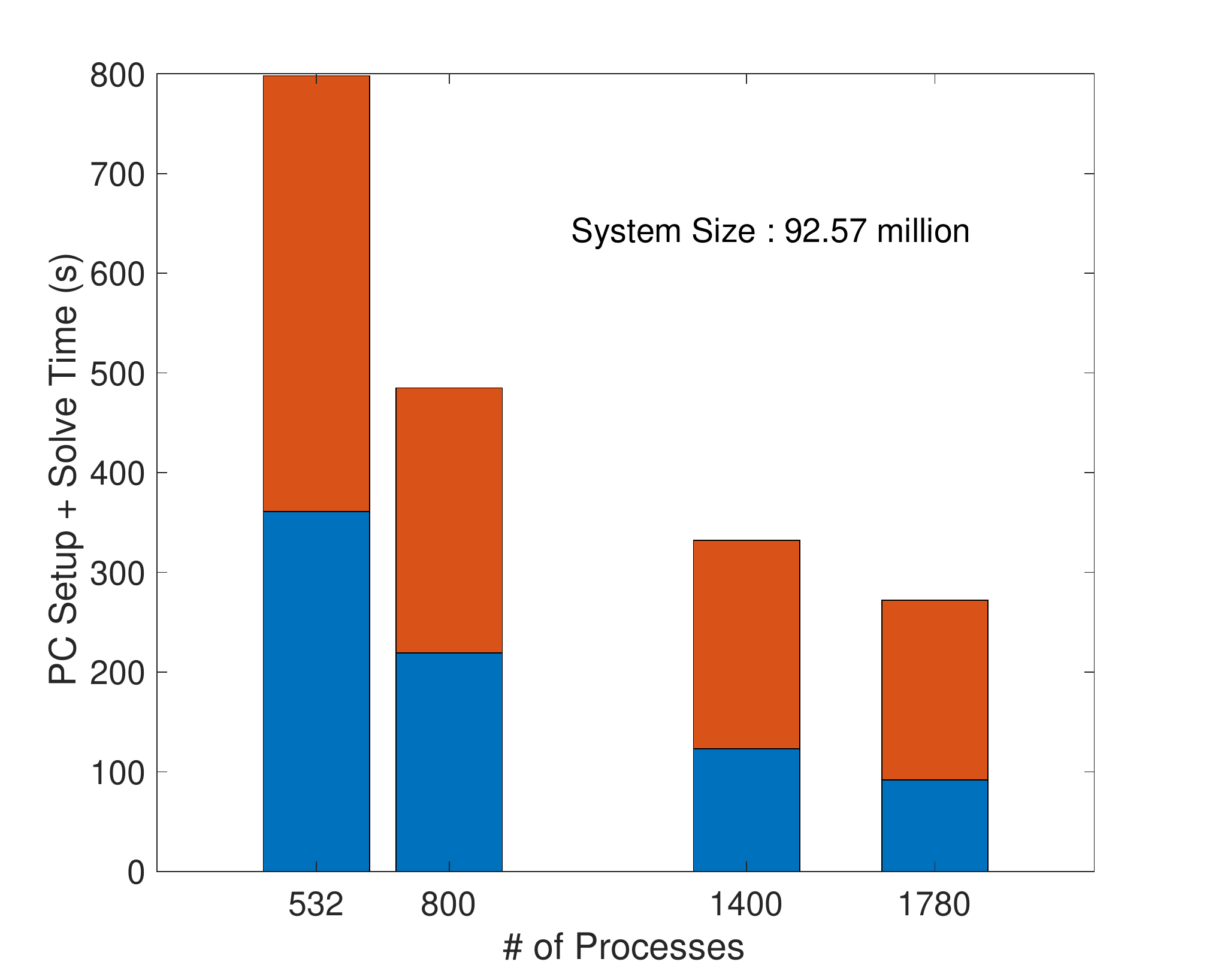} 
        \caption{Strong scalability}
    \end{subfigure}
    \centering
    \begin{subfigure}[b]{0.475\textwidth}
      \centering
        \includegraphics[width=\textwidth]{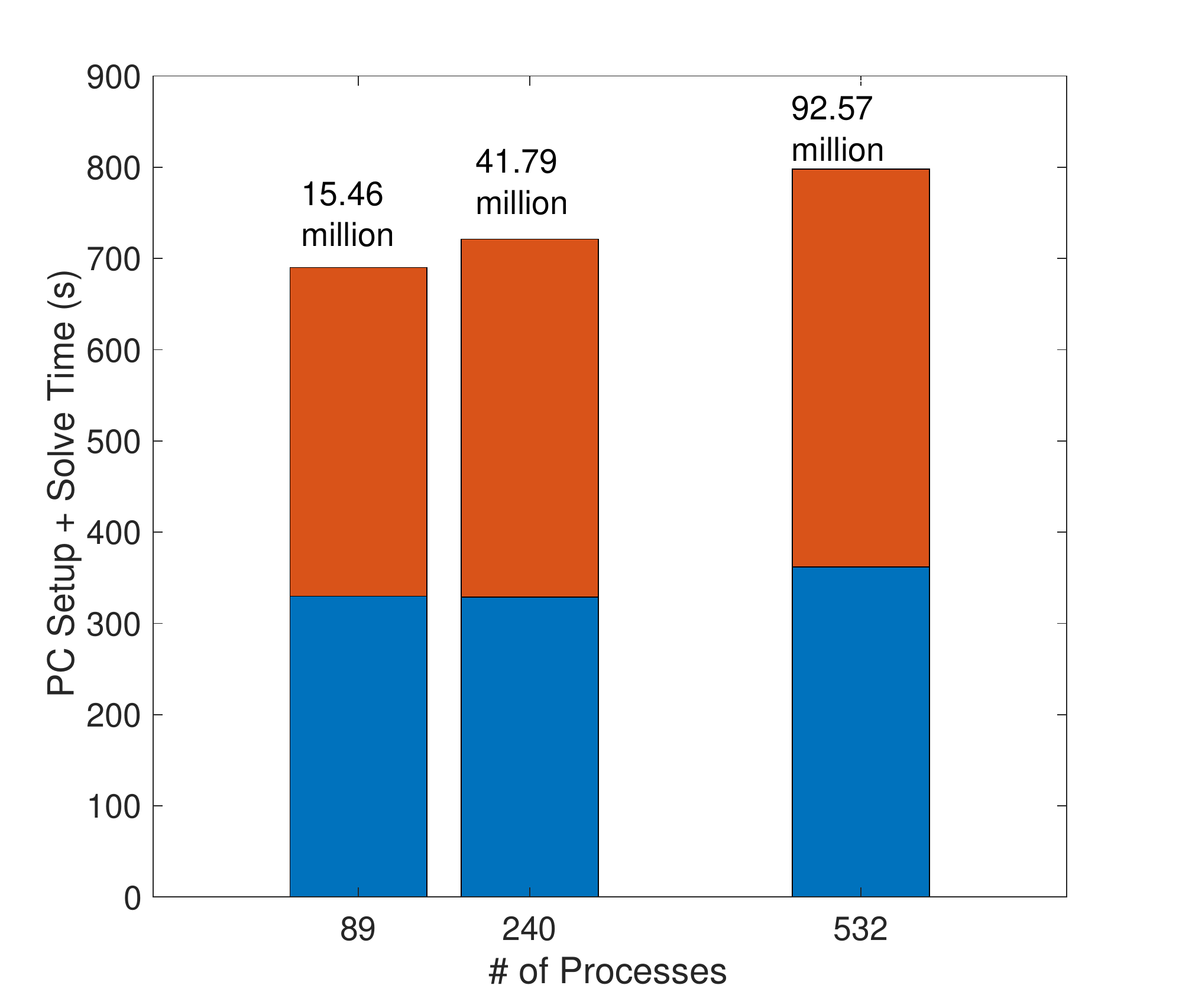} 
        \caption{Weak scalability}
    \end{subfigure}
    \caption{Scalability using domain of Southern Ontario for 10 time steps (PC setup in blue and solve in red)}
  \label{Fig:south_scalability}
\end{figure}

The strong and weak scalabilities of the proposed solver is tested for the case of Southern Ontario with increasing number of processes. \cref{Fig:south_scalability} shows the preconditioner setup and solve times. A fixed system size of $92.57$ million is tested with upto $1780$ processes in order to demonstrate strong scaling (evidenced by the reduction of execution time with more processes). Weak scalability plot shows only small increase in solve time with the increase in total problem size from $15.46$ to $92.57$ million. From the scalability plots, we can extrapolate that forecasts of infections for three months can be obtained for a model size of $92.57$ million within $7$ hours using $1780$ processes. 

Currently, we have only employed deterministic modelling with constant parameters showing the excellent scalability of DD-based preconditioners. However, model parameters like infection rate, diffusion coefficient, recovery rates etc., and initial conditions of the system are not precisely known. Moreover, error in modelling and noise in the data could also be accounted with a stochastic error term. It is thus important to consider a stochastic model to calibrate and reliably predict the infections with uncertainty bounds. The increased dimensionality and increase in time taken for likelihood evaluations with the sampling approach could be effectively reduced using the sampling-free approaches and domain decomposition based solvers \cite{inverse_najm,inverse_najm2,khalil_thesis}. Details on future works of extending this five-compartment model to 22-compartment model and applying state of the art Bayesian inference algorithms for reliable predictions are provided in \cite{Robinsone052681}.

\section{Conclusion}
A PDE-based compartmental model for COVID-19 is essential for continuous space-time trace of infections. For  high resolution meshes, highly stratified compartmental models  can drastically increase the computational requisite needed to accurately capture the disease dynamics. In this investigation, a DD-based parallel scalable iterative solver is developed to enhance the computational efficacy  of these complex models. A two-grid RAS preconditioner equipped with an algebraic multigrid preconditioner for the coarse solver provides excellent scalabilities. A five-compartment SEIRD model of COVID-19 for a large geographical domain of Southern Ontario is used to demonstrate the  scalabilities of the solver in a realistic setting. The solver is  capable of predicting infections up to three months for a system size of 92 million within 7 hours using 1780 processes, thereby saving an order of magnitude computational time required for conventional sequential solvers.

\section*{Acknowledgments}

First author acknowledges the support of Ontario Trillium Scholarship for International Doctoral Students.

\section*{Conflict of interest}

The authors declare there is no conflict of interest.


\providecommand{\href}[2]{#2}
\providecommand{\arxiv}[1]{\href{http://arxiv.org/abs/#1}{arXiv:#1}}
\providecommand{\url}[1]{\texttt{#1}}
\providecommand{\urlprefix}{URL }

\begin{appendices}
\renewcommand\thefigure{A.\arabic{figure}}    
\setcounter{figure}{0}   

\section{The details of weakform }\label{aa}

The bilinear form in \cref{Eq.weak_vec_implicit} are explicitly written for each of the compartments using a decoupled approach below. This approximation of handling each PDE seperately works best for weakly coupled systems. This approach has advantages for a stochastic extension of a sampling-free method for uncertainty quantification \cite{sarkar_stochasticdd} whereby each scalar PDE is transformed into a coupled set of deterministic PDEs. More elaborate studies on this aspect is the subject of future research. proposed in future works. Note that we have ignored $\alpha$ and $\mu$ terms in the formulations below since their values are considered to be zero for the current study, although their inclusion is straightforward. We define the different indices as,

\begin{enumerate}
    \item $n+1$: current time step,
    \item $n$: previous time step,
    \item $k+1$: current Picard iteration number and
    \item $k$: previous Picard iteration number.
\end{enumerate}

Weak form for the SEIRD compartmental model can be written as,
\vspace{3pt}

\underline{\textbf{Susceptible}}\\

\begin{multline} 
    (s^{n+1,k+1},\phi_S) + \Delta t \Big( (1- \frac{A}{N^{n+1,k}}) \beta_I \; s^{n+1,k+1} i^{n+1,k},\phi_S \Big ) 
    + \Delta t \Big ( (1- \frac{A}{N^{n+1,k}}) \beta_E \; s^{n+1,k+1} e^{n+1,k}, \phi_S \Big )  \\
    + \Delta t \Big( N^{n+1,k} \bar{\nu}_S \nabla s^{n+1,k+1}, \nabla \phi_S \Big ) = ( s^{n},\phi_S)
\end{multline}

\underline{\textbf{Exposed}}\\

\begin{multline} \label{Eq. exposed_A1}
    (e^{n+1,k+1},\phi_E) - \Delta t \Big( (1- \frac{A}{N^{n+1,k}}) \beta_E \; s^{n+1,k+1} e^{n+1,k+1},\phi_E \Big ) 
    + \Delta t \Big(  (\sigma + \gamma_E) e^{n+1,k+1},\phi_E \Big ) \\
    + \Delta t \Big( N^{n+1,k} \bar{\nu}_E \nabla e^{n+1,k+1}, \nabla \phi_E \Big ) 
    = ( e^{n},\phi_E) + \Delta t \Big ( (1- \frac{A}{N^{n+1,k}}) \beta_I \; s^{n+1,k+1} i^{n+1,k}, \phi_E \Big ) 
\end{multline}

\underline{\textbf{Infected}} \\

\begin{multline}
    (i^{n+1,k+1},\phi_I) + \Delta t \Big( (\gamma_D + \gamma_R) i^{n+1,k+1},\phi_I \Big ) 
    + \Delta t \Big( N^{n+1,k} \bar{\nu}_I \nabla i^{n+1,k+1}, \nabla \phi_I \Big ) 
    = ( i^{n},\phi_I) + \Delta t \Big ( \sigma e^{n+1,k+1},\phi_I \Big ) 
\end{multline}

\underline{\textbf{Recovered}} \\

\begin{multline}
    (r^{n+1,k+1},\phi_R)  + \Delta t \Big( N^{n+1,k} \bar{\nu}_R \nabla r^{n+1,k+1}, \nabla \phi_R \Big ) 
    = ( r^{n},\phi_R) + \Delta t \Big ( \gamma_R i^{n+1,k+1} ,\phi_R \Big ) + \Delta t \Big ( \gamma_E e^{n+1,k+1} ,\phi_R \Big ) 
\end{multline}

\underline{\textbf{Deceased}} \\

\begin{equation}
    (d^{n+1,k+1},\phi_D) = ( d^{n},\phi_D) + \Delta t \Big ( \gamma_D\; i^{n+1,k+1} ,\phi_D \Big ),
\end{equation}
where $\phi_S, \phi_E, \phi_I, \phi_R$ and $\phi_D$ are the test functions for each compartments.
Note that once the solution is obtained for a particular (e.g. susceptible) compartment at iterate $k+1$, 
its updated value can be used in the calculations of the other compartments. This has been used to improve the convergence rate of the algorithm \cite{oden_seird}.

\section{Model verification}\label{ab}

The SEIRD compartmental model is verified using the process of method of manufactured solutions. Convergence of the finite element system with increasing discretizations in space and time are studied. This process is applied to both one-dimensional and two-dimensional models.  Validation of the model is conducted by comparing the spatially averaged PDE solution to an ODE model which provides a time trace of aggregated infection over a region.

\subsection{Method of manufactured solutions}
\label{subsec:sample:appendix}
Method of manufactured solutions (MMS) is a process of generating analytical solutions to mathematical models of a system to verify computational simulations. This approach can help to detect errors in numerical implementations and solution accuracy. Mathematical models of many physical processes does not have exact analytical solutions and hence numerical methods are used to obtain solutions. The MMS can verify the numerical solution through the manufactured solution. A compatible forcing to generate the manufactured solution is then found by solving the model backwards. This forcing is used to generate the numerical solutions and their accuracy can be verified through the manufactured solutions.

Typically, three different acceptance criteria for the test can be used, namely the percentage error, consistency, and order of accuracy. Consistency ensures that the discretization error decreases to zero as the grid size tends to zero \cite{MMS_sandia}. However, it may not be possible to test this aspect since reducing the grid size to zero is computationally impractical. The order of accuracy criterion checks for consistency and calculates the order in which the error decreases. The theoretical order of accuracy $p$ for spatial refinement is calculated as \cite{MMS_sandia},
\begin{align}
E_{\rm{grid}1} &\approx Ch^p \\
E_{\rm{grid}2} &\approx C(h/r)^p \\
p &\approx \frac{\log\Big(\frac{E_{\rm{grid}1}}{E_{\rm{grid}2}} \Big)}{\log(r)},
\label{Eq.MMS_order}
\end{align}
where $E_{\rm{grid}1}, E_{\rm{grid}2}$ are the error in discretization for grid $1$ and $2$, $h$ is the grid spacing, $r$ is the refinement ratio which dictates the amount of refinement from coarse to fine grid and $C$ is a constant independant of $h$. Theoretical order of accuracy for a triangular element with linear interpolation functions can be estimated to be two\cite{FEM_zienkowicz}. Similarly, the temporal order of accuracy for the backward Euler implicit scheme can be computed as  one \cite{hughes}. We test our numerical solutions to retain the theoretical order of accuracy as obtained from interpolation functions and temporal discretization scheme. 
The $L_2$ norm of the relative error $\epsilon$ for any given compartment is calculated as:-
\begin{align}
\epsilon &=\frac{  \parallel  u-u_h \parallel_2 }{\parallel  u \parallel_2},
\end{align}\label{Eq.MMS_error}
where $u, u_h$ are the true solution and its finite element counterpart for a compartment. The total error at any time step is then calculated as the sum of all five compartments.

\subsection{One-dimensional model}
The order of accuracy in spatial and temporal dimensions for a simple one-dimensional SEIRD model is examined in this section. Manufactured solutions for each compartment is expressed as,

\begin{align}
s &= B \; \sin(10x + 0.2t) + A_{S} \\
e &= B \; \sin(10x + 0.2t) + A_{E} \\
i &= B \; \sin(10x + 0.2t) + A_{I} \\
r &= B \; \sin(10x + 0.2t) + A_{R} \\
d &= B \; \sin(10x + 0.2t) + A_{D},
\end{align}
where $B = 25$, $A_{S} = 500$, $A_{E}= 300$, $A_{I} = 200$, $A_{R} = 100$ and $ A_{D} = 80 $ are the constant parameter values.

\begin{table}[!htbp]
    \centering
    \caption{Parameter values used for one dimensional model.}
        \begin{tabular}{ | m{25mm}  |  m{35mm} |  m{25mm} |}
    \hline
      Parameter                &  Value &   Units \\
      \hline
      $A$                     					& $0$  & $\rm{people}$ \\ 
      $\beta_{I},\beta_{E}$            			& $0.01$  & $\frac{1}{\rm{people} \times \rm{days}}$\\
      $\nu_{S},\nu_{R}$		&  $4.5\times 10^{-5}$  & $\frac{1}{\rm{people} \times \rm{days}}$ \\
      $\nu_{E}$		&  $10^{-3}$  & $\frac{1}{\rm{people} \times \rm{days}}$ \\
      $\nu_{I}$                			&   $10^{-10}$ 		 &  $\frac{1}{\rm{people} \times \rm{days}}$\\
      $\gamma_R$              			& 1/24   									 & $\frac{1}{\rm{days}}$\\
      $\gamma_D$              	&   1/160									&$\frac{1}{\rm{days}}$\\
      $\sigma$                				& 1/8									 		    & $\frac{1}{\rm{days}}$\\
     $\gamma_E$               			& 1/6   						&$\frac{1}{\rm{days}}$\\
      \hline
    \end{tabular}
    \label{tab:parameter_1d}
\end{table}

The model parameters used are given in \cref{tab:parameter_1d}. Note that the domain is normalized by dividing with original length as $\hat{X} = x/L$ which is also reflected in units. Initial conditions and Dirichlet boundary on both ends are derived from the manufactured solution.  Since the model involves both spatial and temporal discretizations, it is important to remove the error caused by one discretization on the other. Thus we chose a sufficiently small time step of $10^{-5}$ days to remove any error propagating from temporal scale. We fix the spatial element length as $2 \times 10^{-4}$ and calculate the error at $5$ days to estimate the order of accuracy in temporal scale.

\begin{figure}[H]
    \centering
    \begin{subfigure}[b]{0.475\textwidth}
       \centering
        \includegraphics[width=\textwidth]{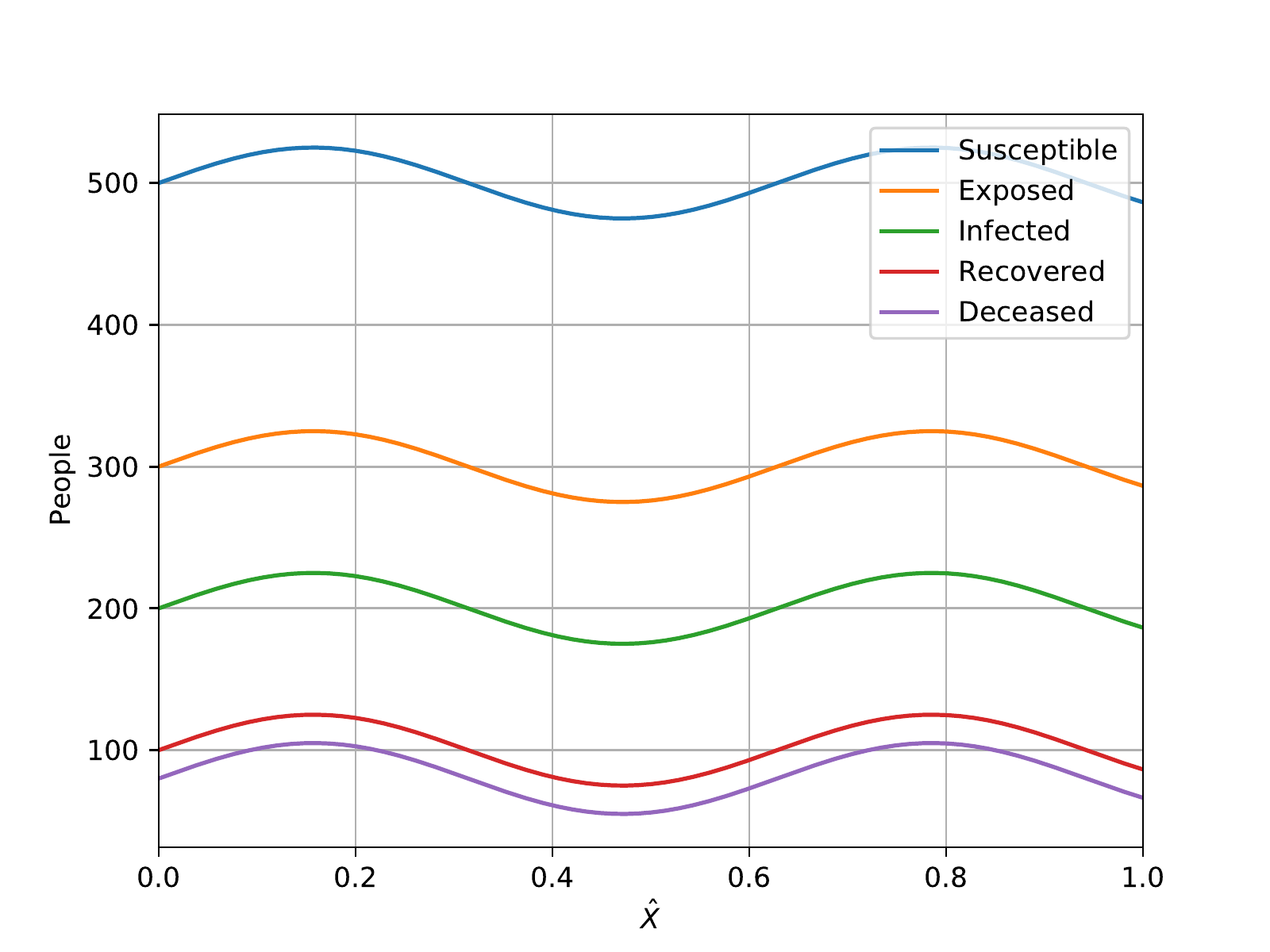} 
        \caption{Initial condition for infected compartment.}
     \end{subfigure}
         \begin{subfigure}[b]{0.475\textwidth}
      		 \centering
        \includegraphics[width=\textwidth]{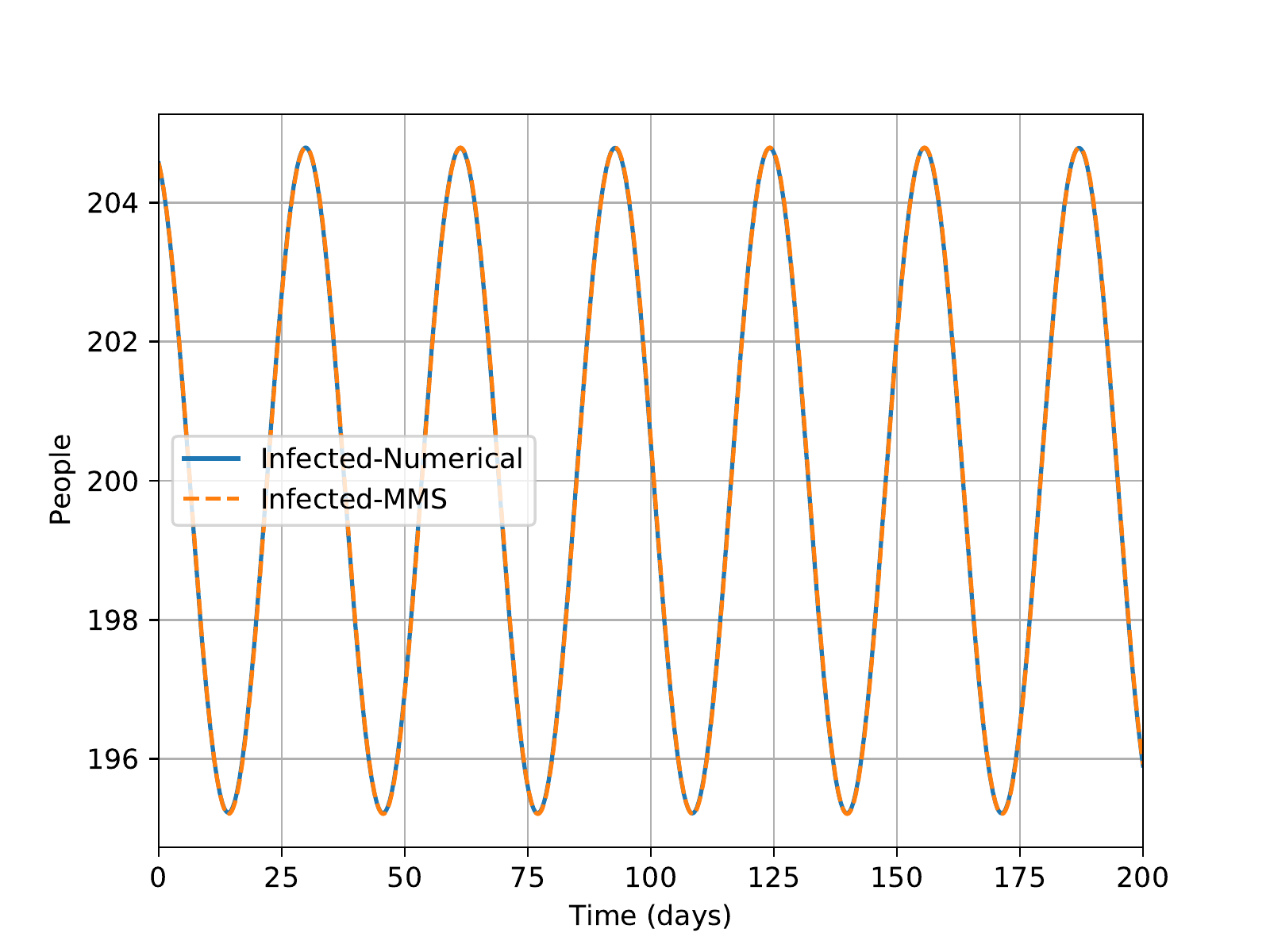} 
        \caption{Infected people averaged over entire domain.}
        \end{subfigure} 
          \hfill     
        \begin{subfigure}[b]{0.475\textwidth}
       \centering
        \includegraphics[width=\textwidth]{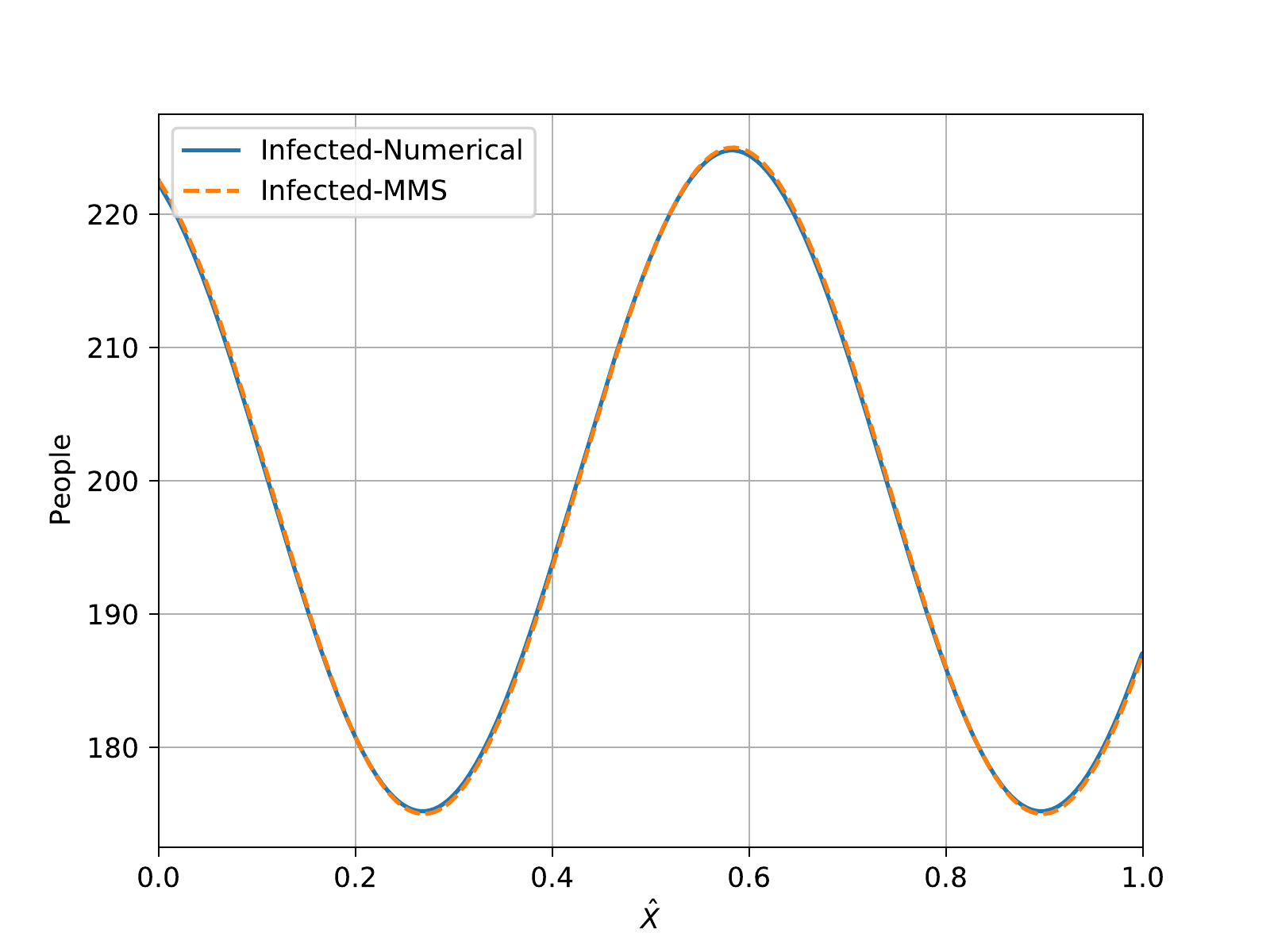} 
        \caption{Infected people in space at $10$ days.}
 \end{subfigure}
 
  \caption{Spatial and temporal variations of infected people}\label{Fig.mms_1d_comp}
\end{figure}

Note from \cref{tab:ord_spatial_2} and \cref{tab:order_time}, the estimated accuracies of spatial and temporal discretizations are close to $2$ and $1$ as expected from theoretical considerations. Figures \ref{Fig.mms_1d_comp} demonstrates that the numerical solutions and manufactured solutions match closely both in space and time with a characteristic length of $0.0005$ and time step of $0.1$ days.

\begin{table}[!h]
    \centering
      \caption{Order of accuracy in spatial discretization for one-dimensional model.}
    \begin{tabular}{|c|c|c|}
    \hline
      $\Delta h$ (Characteristic Length)   & Error at $0.002$ days & Order of Accuracy ($p$) \\
      \hline
      \hline
       $0.05$  & $0.01289$ &  \\
       $0.02$  & $0.00208$ &  1.9920 \\
       $0.01$  & $0.00052$ & 1.9985 \\
       $0.002$  & $2.0809\times10^{-5}$ & 1.9998 \\
       $0.001$  & $5.2025\times10^{-6}$ & 1.9999 \\
       $0.0005$  & $1.3008\times10^{-6}$ & 1.9998 \\
       $0.0002$  & $2.0842\times10^{-7}$ & 1.9985 \\
      \hline

    \end{tabular}
    \label{tab:ord_spatial_2}
\end{table}

\begin{table}[!h]
    \centering
    \caption{Order of accuracy in temporal discretization for one-dimensional model.}
    \begin{tabular}{|c|c|c|}
    \hline
      $\Delta t$ (Temporal Discretization)   & Error at $5$ days  & Order of Accuracy ($p$) \\
      \hline
       $0.1$  & $0.00374$ &  \\
       $0.01$  & $0.00037$ &  0.9995 \\
       $0.005$  & $0.00019$ & 0.9994 \\
       $0.001$  & $3.7553\times10^{-5}$ & 0.9982 \\
       $0.0005$  & $1.8847\times10^{-5}$ & 0.9946 \\
       \hline

    \end{tabular}
    \label{tab:order_time}
\end{table}

\subsection{Two-dimensional model}

The sinusoidal functions similar to those in the one-dimensional case are used as the manufactured solutions for two-dimensional case as:

\begin{align}
s &= B \; \sin(10xy + 0.2t) + A_{S} \\
e &= B \; \sin(10xy + 0.2t) + A_{E} \\
i &= B \; \sin(10xy + 0.2t) + A_{I} \\
r &= B \; \sin(10xy + 0.2t) + A_{R} \\
d &= B \; \sin(10xy + 0.2t) + A_{D},
\end{align}\label{Eq. MMS2d}
where $B = 25$, $A_{S} = 500$, $A_{E}= 300$, $A_{I} = 200$, $A_{R} = 100$ and $ A_{D} = 80 $. Other model parameters 
are given in \cref{tab:parameter_all} for square domain except the value of $A = 100$. We use Neumann boundary conditions that varies in time derived from manufactured solution as:

\begin{align}
g_{\rm{N,left}} &= -10 By \; \cos(0.2t + 10xy) \\
g_{\rm{N,right}} &= 10 By \; \cos(0.2t + 10xy) \\
g_{\rm{N,top}} &= 10 Bx \; \cos(0.2t + 10xy) \\
g_{\rm{N,bottom}} &= -10 Bx \; \cos(0.2t + 10xy)
\end{align}

\begin{figure}[H]
    \centering
    \begin{subfigure}[t]{0.475\textwidth}
       \centering
        \includegraphics[width=3in,height=2.1in]{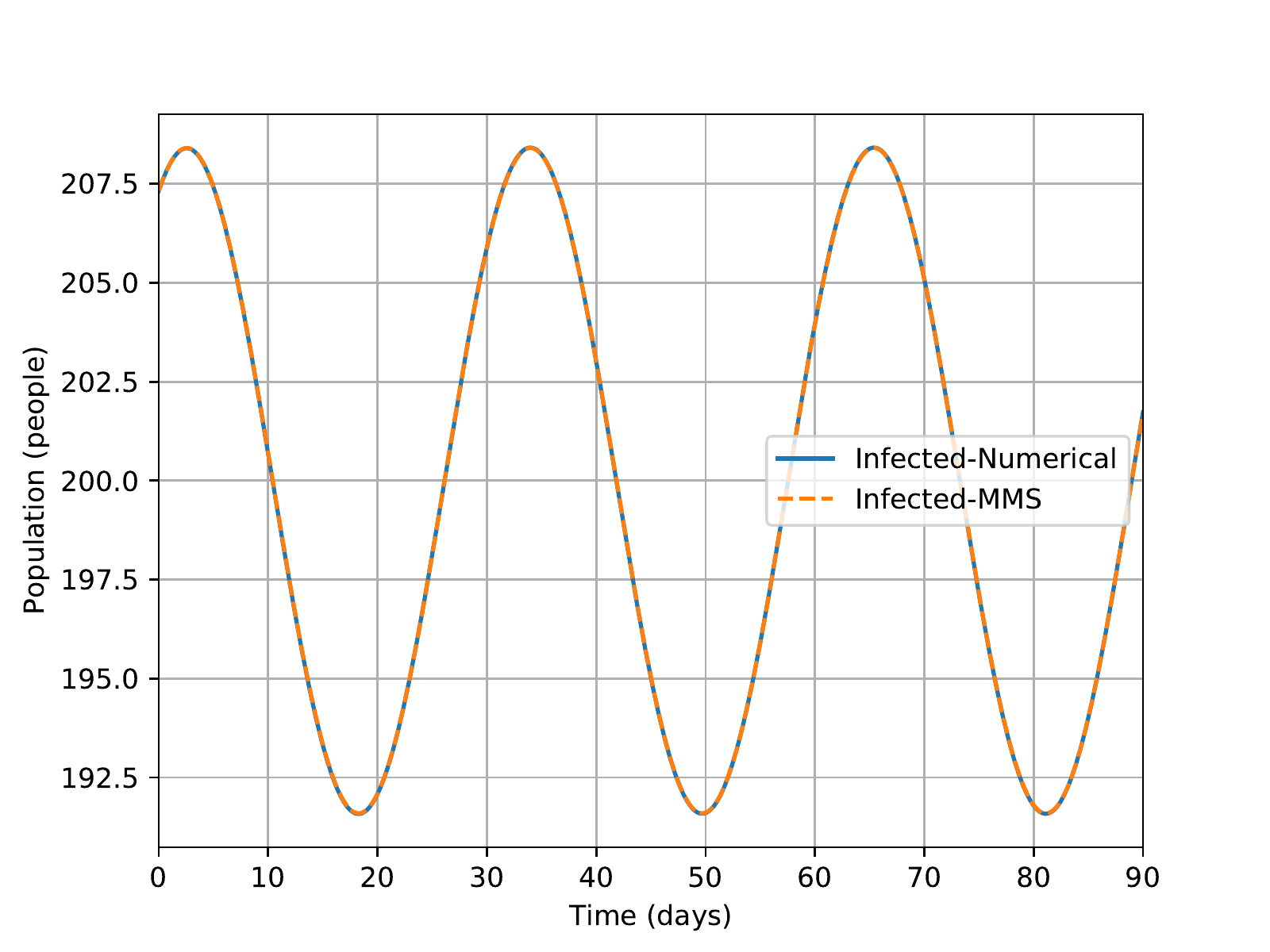} 
        \caption{\centering{Infected compartment averaged over the entire domain.}}
    \end{subfigure}
     \centering
    \begin{subfigure}[t]{0.475\textwidth}
      		 \centering
        \includegraphics[width=3in,height=2.1in]{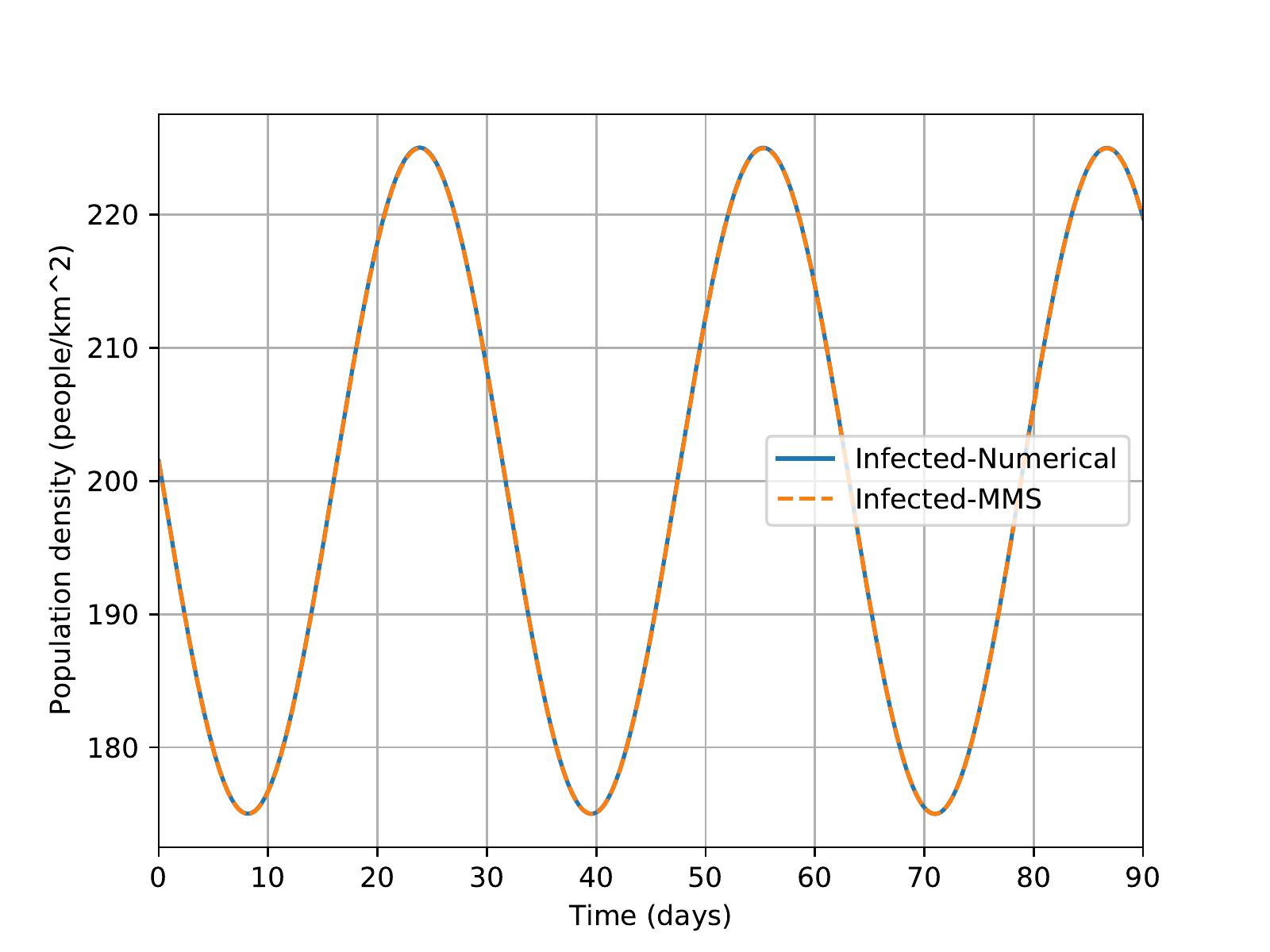} 
        \caption{\centering{Infected compartment at $(0.7,0.3)$.}}
    \end{subfigure} 
    \centering
        \begin{subfigure}[t]{0.475\textwidth}
       \centering
        \includegraphics[width=3in,height=2.1in]{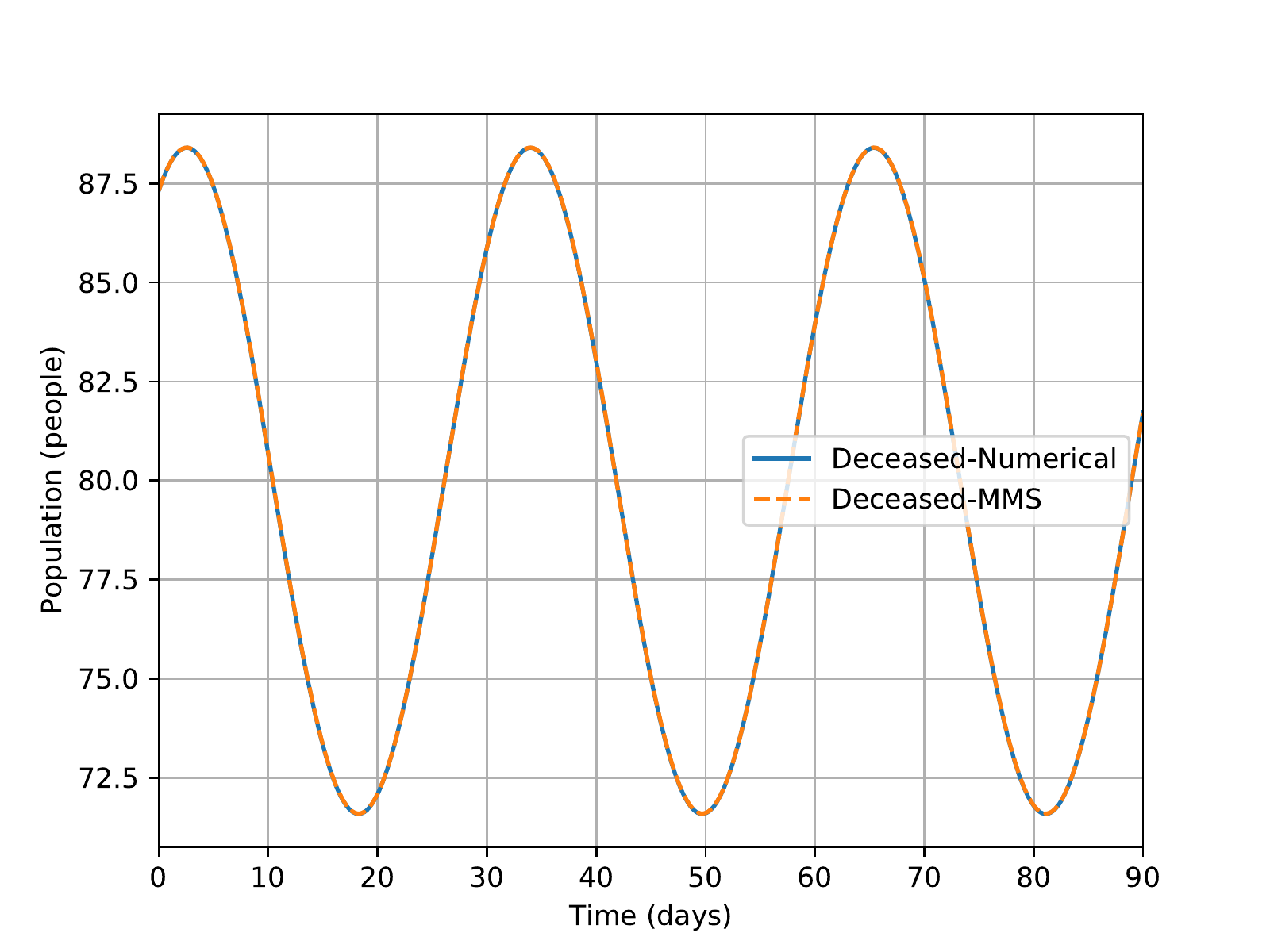} 
        \caption{\centering{Deceased compartment averaged over the entire domain.}}
    \end{subfigure}
     \centering
    \begin{subfigure}[t]{0.475\textwidth}
      		 \centering
        \includegraphics[width=3in,height=2.1in]{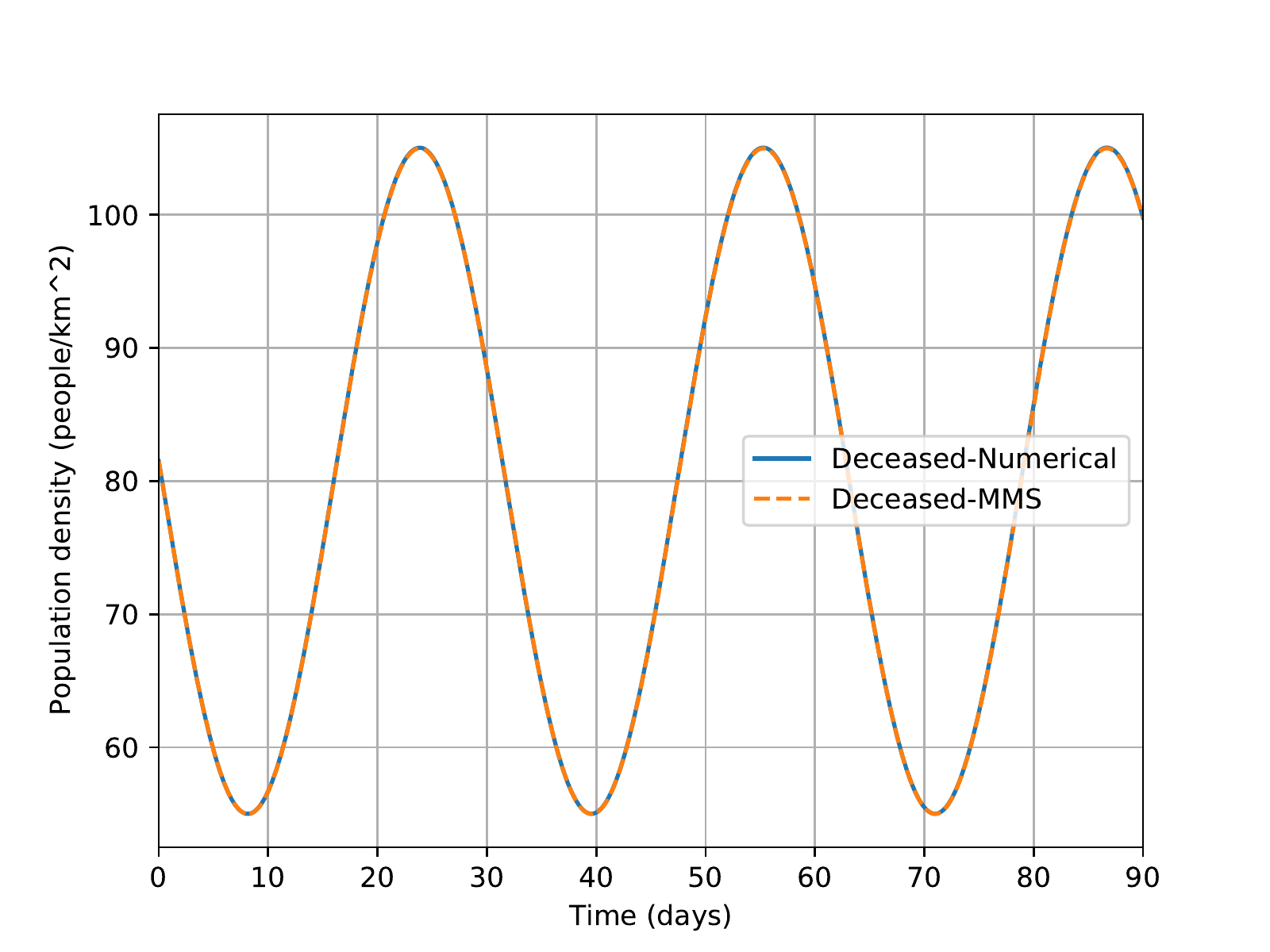} 
        \caption{Deceased compartment at $(0.7,0.3)$.}
    \end{subfigure} 
    \centering
    \begin{subfigure}[b]{0.475\textwidth}
       \centering
        \includegraphics[width=3.1in,height=2.1in]{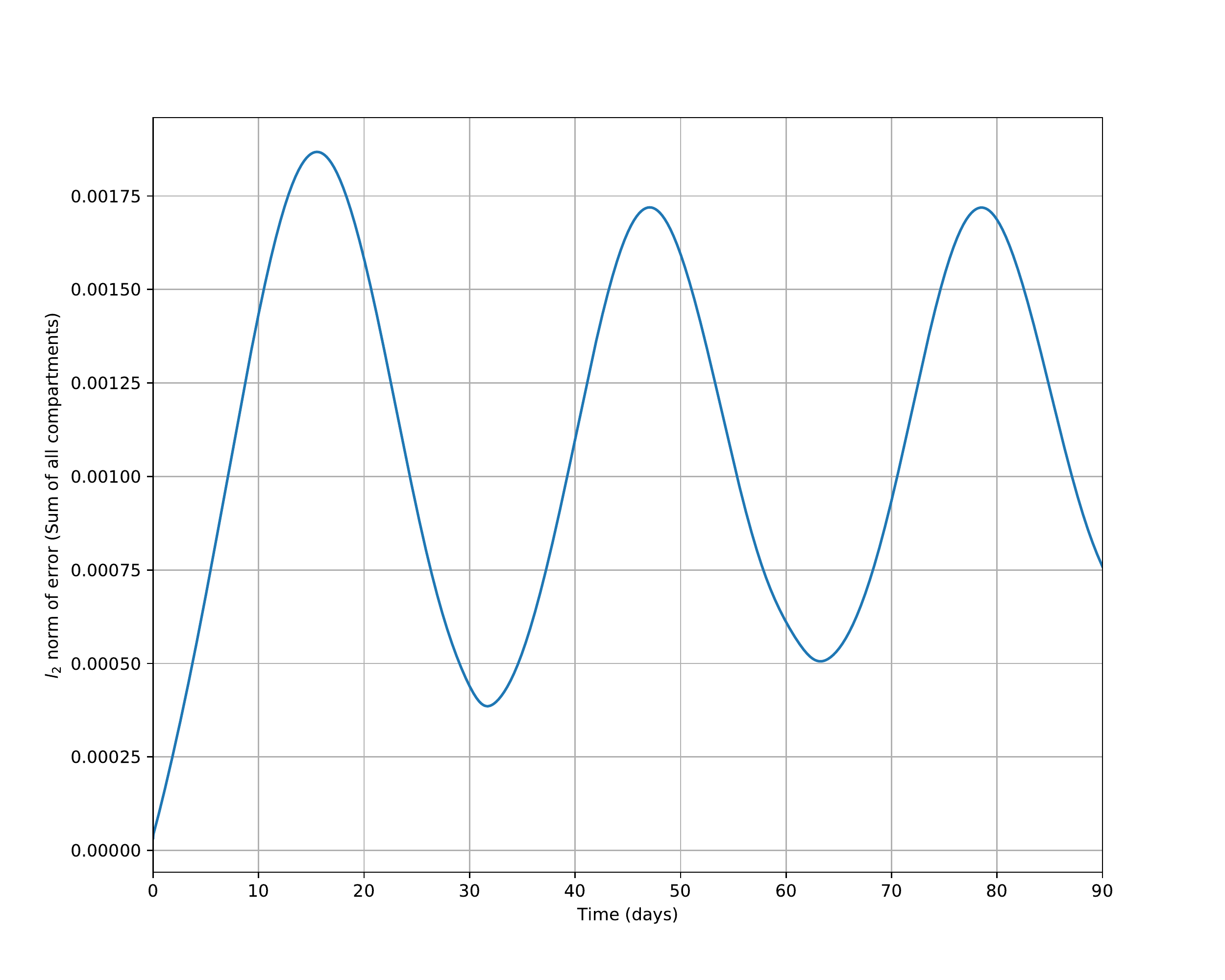} 
        \caption{$l_2$ norm of error over time.}
    \end{subfigure}
        
  \caption{Temporal variation of infected and deceased compartments.}\label{Fig:MMS_2D}
\end{figure}

A square domain with a mesh size containing $13470$ vertices and a time step of $0.01$ days is chosen to verify the solutions. A fully implicit approach of time discretization is implemented. The tolerance of Picard  iteration is chosen to be $10^{-10}$. The GMRES solver with a one-level RAS preconditioner is used for the linear system solve. The integrated solution over the entire domain and the solution at a point against time are shown in \cref{Fig:MMS_2D}. The $l_2$ norm of the error for the sum of all compartments is plotted which reduces over time in an oscillatory fashion.

\subsection{Comparison with the ordinary differential equation  model}
\begin{figure}[H]
    \centering
    \begin{subfigure}[b]{0.475\textwidth}
       \centering
        \includegraphics[width=\textwidth]{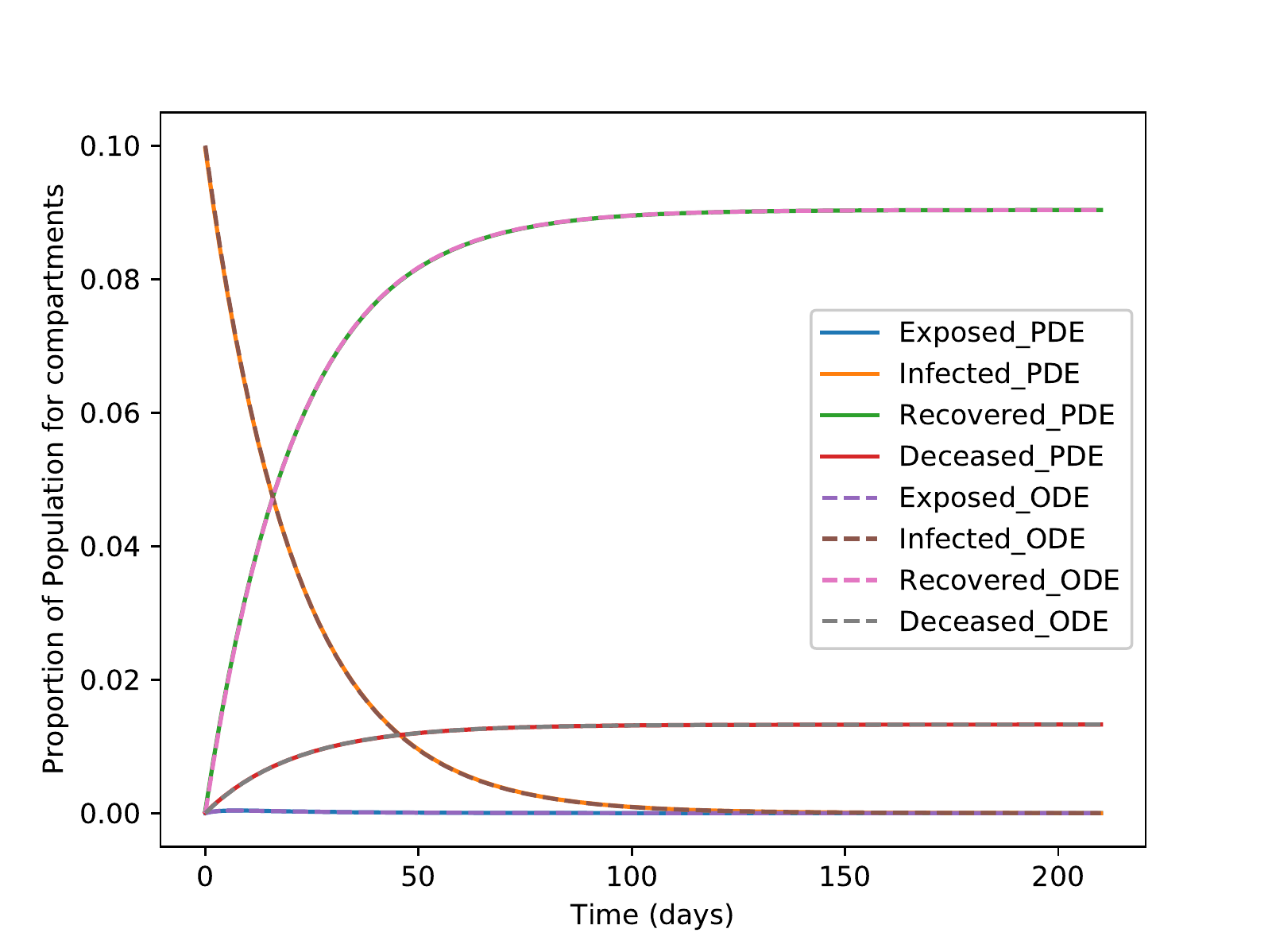} 
        \caption{Total population of $10$}
    \end{subfigure}
    \centering
    \begin{subfigure}[b]{0.475\textwidth}
       \centering
        \includegraphics[width=\textwidth]{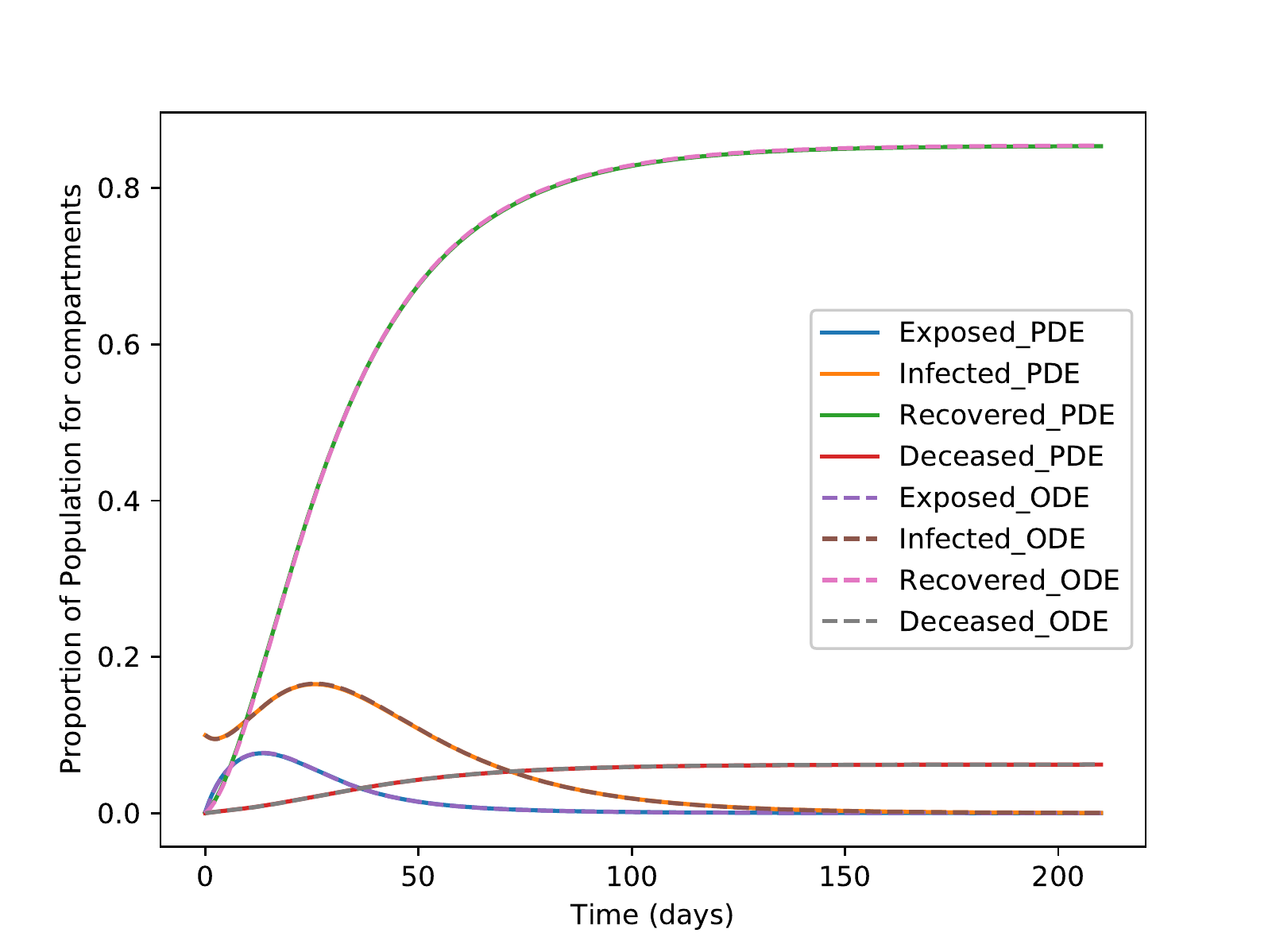} 
        \caption{Total population of $1000$}
    \end{subfigure}
  \caption{Comparison of ODE model with integrated and normalized PDE model output}\label{Fig:ODE_compare}
\end{figure}

The PDE-based SEIRD model captures the variation of infection in space through the diffusion term. By decreasing the diffusion to a very low value and integrating the solution over the entire domain, the PDE model can reduce to an equivalent ODE system. An ODE compartmental model describing the same dynamics as the PDE model can be expressed as:
\begin{align}
\frac{ds}{dt} & = -(1-A)\beta_I  s i - (1-A)\beta_E  s e\\\label{Eq.ODE_s}
\frac{de}{dt} & = (1-A)\beta_I s i + (1-A)\beta_E s e - \sigma e - \gamma_E e\\
\frac{di}{dt} & = \sigma e - (\gamma_R + \gamma_D) i \\
\frac{dr}{dt} & = \gamma_E e + \gamma_R i \\
\frac{dd}{dt} & = \gamma_D i,
\label{Eq.ODE_d}
\end{align}
where $s$, $e$, $i$, $r$, and $d$ denote the susceptible, exposed, infected, recovered and deceased proportion of population where sum of all compartments $s+e+i+r+d = 1$. Model parameters are observed from the square domain case in \cref{tab:parameter_all} except the diffusion coefficients are now $10^{-20}$. As the initial condition, $10\%$ of the total population is chosen to be in the infected compartment. The initial value of the exposed, recovered and deceased compartments are set to be zero, and the susceptible compartment is calculated as 0.9. For the PDE model, all compartment densities are initially assumed to be uniform over the entire domain. For the square domain, we consider the population sizes of $10$ and $100$ respectively for numerical investigation. The tolerance of the Picard iterations is set at $10^{-10}$ and the integration time step of $0.1$ days is used for a total duration of $210$ days. The PDE model is integrated over the entire domain and normalized by total population for comparison with the ODE model.  Results in \cref{Fig:ODE_compare} demonstrates that the solutions of PDE and ODE models match closely when the diffusion coefficients are very small.

\section{22-compartment model}\label{ac}
This 22-compartment  PDE-based SEIRD model becomes computationally expensive which inspired the development of the scalable solvers reported in this paper. Next we describe the extension of a 22-compartment ODE-based SEIRD model to the PDE-based system to consider the geo-spatial spread of the disease dynamics. If properly calibrated by field data, such comprehensive model can more realistically represent the disease dynamics in order to assist clinical and public health decision makers. To this end, we modify the 22-compartment ODE-based model proposed by Robinson et al. \cite{Robinsone052681}, to a system of PDEs, representing the model states as population densities in space and time. In the system of equations below, we use capital letters to denote model compartments (see \cref{tab:compartments}) to maintain consistency with the format presented in \cite{Robinsone052681}, and use Greek letters to denote model parameters (see \cref{tab:parameters}).

The following ten compartments involves a diffusion term, modeling spatio-temporal population movement : $S$, $V$, $E$, $F$, $A$, $B$, $C$, $P$, $R1$, $R2$. 
Publicly available data may not permit construction and calibration of stratified SEIRD models based on age, co-morbidity, sex, socio-economic status etc \cite{Robinsone052681}. However, private health care database (e.g. COVID-19 database from ICES \cite{ices}) can permit such detailed stratified SEIRD-based modeling.

\begin{figure}[htbp!]
    \centering
\includegraphics[width=\textwidth]{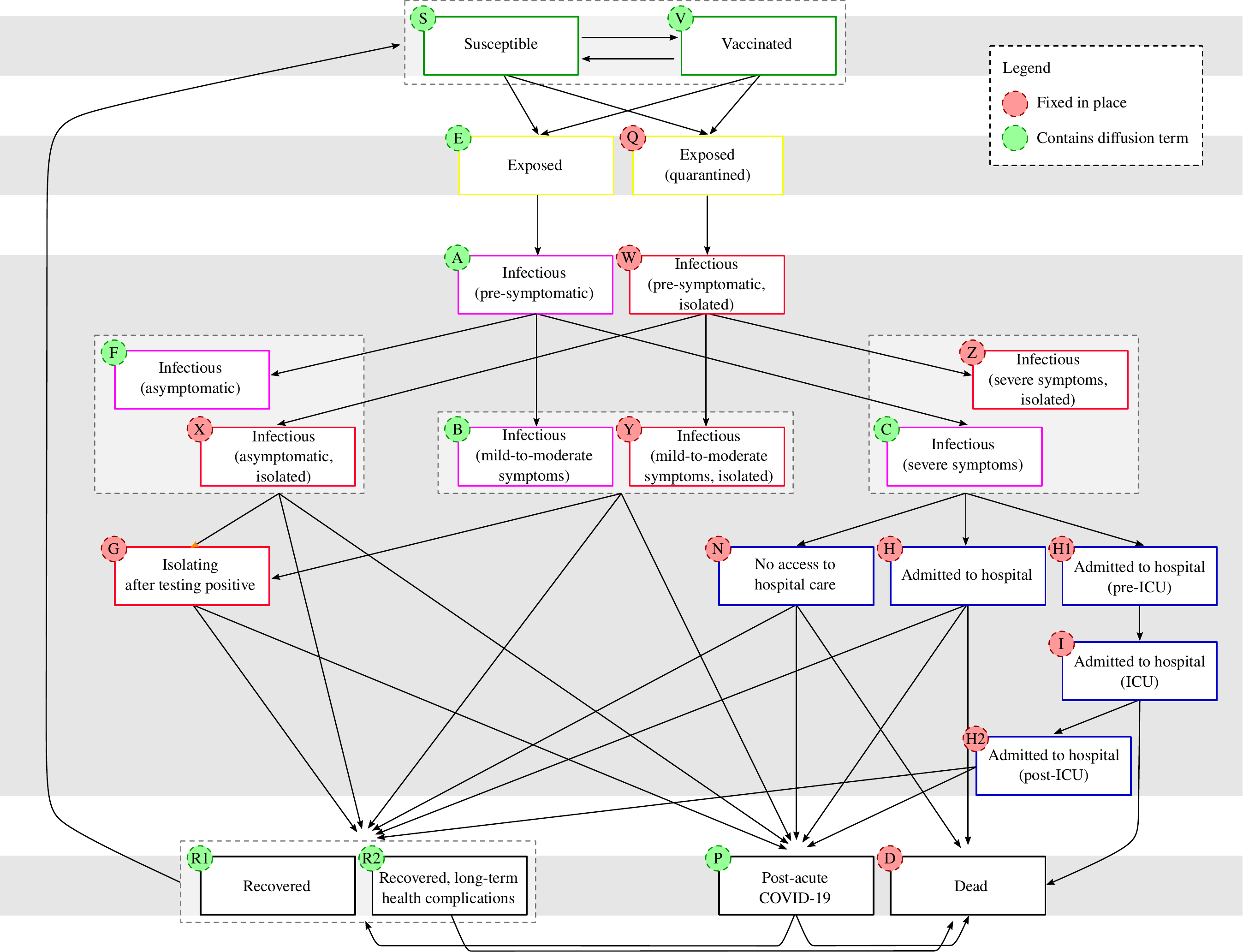}
    \caption{22-compartment model (adapted from \cite{Robinsone052681})}
    \label{fig:21state}
\end{figure}

{\small
\begin{align}
& \text{Susceptible:} \nonumber\\
 \frac{\partial S}{\partial t} &= -\lambda  S   -\gamma_V S   + \gamma_T V + \gamma_R (R1  + R2) + \nabla \cdot (\nu_S\nabla S)\label{eq:S}\\[8pt]
&  \text{Vaccinated:}\nonumber\\
\frac{\partial V }{\partial t} & =-(1-r_V)\lambda  V  + \gamma_V S  - \gamma_TV + \nabla \cdot (\nu_V\nabla V) \label{eq:V}\\[8pt] 
&  \text{Exposed:}\nonumber\\
 \frac{\partial E }{\partial t} &= (1-{\delta_S})\lambda  S  +  (1-{\delta_V})(1-r_V)\lambda  V  - \gamma_E E  + \nabla \cdot (\nu_E\nabla E)\label{eq:E}\\[8pt]  
& \text{Exposed, isolating:}\nonumber\\
 \frac{\partial Q }{\partial t}& = {\delta_S}\lambda  S  +  {\delta_V}(1-r_V)\lambda  V   - \gamma_E Q  \\[8pt] 
& \text{Infectious, presymtomatic:}\nonumber\\
 \frac{\partial A }{\partial t}& =  \gamma_E E  -\gamma_P A  + \nabla \cdot (\nu_A\nabla A)  \label{eq:A}\\[8pt] 
& \text{Infectious, pre-symptomatic, isolating:}\nonumber\\
 \frac{\partial W }{\partial t} &=\gamma_E Q  -\gamma_P W \\[8pt] 
& \text{Infectous, asymptomatic:}\nonumber\\
\frac{\partial F }{\partial t}&  = \sigma_{A}\gamma_P A  -\gamma_A F  - \gamma_{DA} F   + \nabla \cdot (\nu_F\nabla F) \\[8pt] 
& \text{Infectious, mild-to-moderate symptoms:}\nonumber\\
 \frac{\partial B }{\partial t} &= (1-{\sigma_{A}})(1-{\sigma_S})\gamma_P A  -\gamma_M B  - \gamma_{DM} B   + \nabla \cdot (\nu_B\nabla B)\\[8pt]  
& \text{Infectious, severe symptoms:}\nonumber\\
 \frac{\partial C }{\partial t} &= (1-{\sigma_{A}})\sigma_{S}\gamma_p A  -\gamma_{S1} C   + \nabla \cdot (\nu_C\nabla C)\\[8pt] 
& \text{Infectious, asymptomatic, isolating:}\nonumber\\
\frac{\partial X }{\partial t}& = {\sigma_A}\gamma_P W  -\gamma_A X  - \gamma_{DA} X  \\[8pt] 
& \text{Infectious, mild-to-moderate symptoms, isolating:}\nonumber\\
 \frac{\partial Y }{\partial t} &= (1-{\sigma_A})(1-{\sigma_S})\gamma_PW  -\gamma_M Y  - \gamma_{DM} Y  \\[8pt] 
& \text{Infectious, severe symptoms, isolating:}\nonumber\\
 \frac{\partial Z }{\partial t} &=(1-{\sigma_A})\sigma_{S}\gamma_P W  -\gamma_{S1} Z  \\[8pt] 
& \text{Infectious, isolating after testing positive:}\nonumber\\
 \frac{\partial G }{\partial t} &= \gamma_{DA} (F  + X ) + \gamma_{DM}( B  + Y  ) -\gamma_I G \\[8pt] 
& \text{Inadequate access to health care resources:}\nonumber\\
\frac{\partial N }{\partial t} &  =(1-{\sigma_H})\gamma_{S1}(C  + Z ) -\gamma_{S2} N \\[8pt] 
& \text{Hospital:}\nonumber\\
\frac{\partial H }{\partial t}&  ={\sigma_H}(1-{\sigma_C})\gamma_{S1}(C  + Z ) -\pi_H H\\[8pt] 
& \text{Pre-ICU:}\nonumber\\
\frac{\partial H1 }{\partial t}&  = \sigma_H\sigma_C\gamma_{S1}(C  + Z )  -\pi_A H1\\[8pt] 
& \text{ICU:}\nonumber\\
 \frac{\partial I }{\partial t}& = \pi_A H1  -\pi_B I \\[8pt] 
& \text{Post-ICU:}\nonumber\\
 \frac{\partial H2 }{\partial t}& = (1-\kappa_I)\pi_B I  - \pi_C H2  \\[8pt] 
& \text{Recovered:}\nonumber\\
\frac{\partial R1 }{\partial t} & = (1-\phi_C)\Big((1-\phi_M)\left(\gamma_I G  + \gamma_A(F +X ) + \gamma_M(B +Y )\right) \\& \qquad   + (1-\phi_S)\left((1-\kappa_N)\gamma_{S2} N  + (1-\kappa_H)\pi_H H  + \pi_C H2 \right)\Big)  + (1-\phi_P)(1-\kappa_P)\gamma_C P  - \gamma_R R1  + \nabla \cdot (\nu_{R1}\nabla R1)  \nonumber \\[8pt]
& \text{Recovered with long-term health complications:}\nonumber\\ 
\frac{\partial R2 }{\partial t}& = (1-\phi_C)\Big(\phi_M \left(\gamma_I G  + \gamma_A(F +X ) + \gamma_M(B +Y )\right) \label{eq:R} \\& \qquad  + \phi_S \left((1-\kappa_N)\gamma_{S2} N   + (1-\kappa_H)\pi_H H  + \pi_CH2  \right)\Big) + \phi_P(1-\kappa_P)\gamma_C P  - \gamma_R R2   - \gamma_L R2    + \nabla \cdot (\nu_{R2}\nabla R2) \nonumber\\[8pt] 
& \text{Post-acute COVID-19:}\nonumber\\
\frac{\partial P }{\partial t} & = \phi_C\left(\gamma_I G  + \gamma_A(F +X ) + \gamma_M(B +Y )   + (1-\kappa_N)\gamma_{S2} N  + (1-\kappa_H)\pi_H H  + \pi_C H2 \right) - \gamma_C P  + \nabla \cdot (\nu_{P}\nabla P)  \\[8pt]
& \text{Death:} \nonumber \\
\frac{\partial D }{\partial t}&  = \kappa_H \pi_H H  + \kappa_I\pi_B I  +\kappa_N\gamma_{S2} N  + \gamma_L R2 + \kappa_P\gamma_C P \label{eq:D}
\end{align} 
}

\begin{table}[h!]
{\small
\centering
\begin{tabular}{ l  l }
\hline
Symbol & Definition \Tstrut \Bstrut\\
\hline
\textcolor{diffusion}{$S$}& Susceptible \Tstrut\\
\textcolor{diffusion}{$V$} & Vaccinated\\
\textcolor{diffusion}{$E$} & Exposed \\
\textcolor{nodiffusion}{$Q$} & Exposed, isolating \\
\textcolor{diffusion}{$A$} & Infectious, pre-symptomatic \\
\textcolor{nodiffusion}{$W$} & Infectious, pre-symptomatic, isolating \\
\textcolor{diffusion}{$F$} & Infectious, asymptomatic \\
\textcolor{diffusion}{$B$} & Infectious, mild-to-moderate symptomatic (i.e., symptoms not requiring hospitalization) \\
\textcolor{diffusion}{$C$} & Infectious, severe symptomatic (i.e., symptoms requiring hospitalization) \\
\textcolor{nodiffusion}{$X$} & Infectious, asymptomatic, isolating \\
\textcolor{nodiffusion}{$Y$} & Infectious, mild-to-moderate symptomatic, isolating\\
\textcolor{nodiffusion}{$Z$} & Infectious, severe symptomatic, isolating \\
\textcolor{nodiffusion}{$G$} & Infectious, mild-to-moderate symptomatic, isolating but not previously in isolation\\
\textcolor{nodiffusion}{$N$} & No access to hospital care\\
\textcolor{nodiffusion}{$H$} & Hospitalized, never to be admitted to the intensive care unit (ICU)\\
\textcolor{nodiffusion}{$H1$} & Hospitalized, to be admitted to the ICU\\
\textcolor{nodiffusion}{$I$} & Hospitalized, in the ICU\\
\textcolor{nodiffusion}{$H2$} & Hospitalized, after being discharged from the ICU\\
\textcolor{diffusion}{$R1$} & Recovered, without long-term health complications\\
\textcolor{diffusion}{$R2$} & Recovered, with long-term health complications\\
\textcolor{diffusion}{$P$} & Post-acute COVID-19\\
\textcolor{nodiffusion}{$D$} & Death \Bstrut\\
 \hline
\hline
\end{tabular} \\
\caption{Model states/compartments, indices omitted for brevity (reproduced from \cite{Robinsone052681})}\label{tab:compartments}
}
\end{table}

\begin{table}[h!]
{\small 
\centering
\begin{tabular}{ l  l}
\hline
Symbol & Definition  \Tstrut \Bstrut\\
\hline 
$r_V$ & Vaccine effectiveness (1 indicates 100\% immunity, 0 indicates no immunity)\\
${\delta_S}$ & Probability that a susceptible individual exposed to the virus will self-isolate (without prior testing) \\
${\delta_V}$ & Probability that a vaccinated individual exposed to the virus will self-isolate (without prior testing) \\
$\gamma_A$ & 1/the average duration of the infectious period for asymptomatic individuals \\
$\gamma_C$ & 1/the average duration of sub-acute COVID-19 \\
$\gamma_{DA}$ & Rate of detection among asymptomatic cases \\
$\gamma_{DM}$ & Rate of detection among mild-to-moderate cases \\
$\gamma_E$ & 1/the average incubation period \\
$\gamma_I$& 1/the average duration of self-isolation \\
$\gamma_L$ & Rate of deaths due to long-term health complications \\
$\gamma_M$ & 1/the average duration of the infectious period for individuals with mild-to-moderate symptoms \\
$\gamma_P$ & 1/the average duration of the pre-symptomatic infectious period \\
$\gamma_R$ & 1/the average effective duration of temporary immunity from having recovered from the virus\\
$\gamma_{S1}$ & 1/the average duration of severe symptoms before seeking hospitalization \\
$\gamma_{S2}$ & 1/the average remaining duration of symptomatic period for individuals with severe symptoms \\
$\gamma_T$ & 1/the average effective duration of temporary immunity from vaccination\\
$\gamma_V$ & Rate of vaccination \\
${\sigma_A}$ & Probability that an infectious individual is asymptomatic \\
${\sigma_C}$ & Probability that a hospitalized case will be admitted to the ICU \\
${\sigma_H}$ & Probability that an individual has access to hospital care \\
${\sigma_S}$ & Probability that a case displaying symptoms will require hospitalization \\
$\pi_A$ & 1/the average time in hospital prior to ICU \\
$\pi_B$ & 1/the average time in ICU \\
$\pi_C$ & 1/the average time in hospital following ICU \\
$\pi_H$ & 1/the average duration hospitalization (non-ICU track) \\
${\phi_A}$ & Probability of acute COVID \\
${\phi_M}$ & Probability of long-term complications for asymptomatic, mild-to-moderate cases \\
${\phi_S}$ & Probability of long-term complications for severe cases \\
${\phi_P}$ & Probability of long-term complications for post-acute COVID-19 cases \\
${\kappa_H}$ & Probability of death among hospital cases \\
${\kappa_I}$ & Probability of death among ICU cases \\
${\kappa_N}$ & Probability of death among cases without access to hospital care\\
${\kappa_P}$ & Probability of death among post-acute COVID-19 cases \\
\hline
\hline
\end{tabular}\\
\caption{Model parameters with definition and reference, indices omitted for brevity (reproduced from \cite{Robinsone052681})}\label{tab:parameters}
}
\end{table}

\end{appendices}

\end{document}